\documentclass[aps,prd,twocolumn,amsmath,amssymb,superscriptaddress,nofootinbib,longbibliography]{revtex4-2}
\usepackage{graphicx}
\usepackage{bm}
\usepackage{amssymb,amsmath}
\usepackage{mathrsfs} 
\usepackage{mathtools} 
\usepackage[normalem]{ulem} 
\usepackage{dcolumn}
\usepackage[colorlinks=true,citecolor=blue,urlcolor=blue]{hyperref}
\usepackage[usenames,dvipsnames]{xcolor}

\allowdisplaybreaks

\newcommand{\CGP}{\affiliation{Center for Gravitational Physics, University of Texas at Austin, Austin, TX 78712, USA}}

\begin{document}

\title{An approach to computing spectral shifts for black holes beyond Kerr}

\author{Asad Hussain}
\CGP
\author{Aaron Zimmerman}
\CGP
\date{\today}

\begin{abstract}
Recent measurements of gravitational-wave ringdown following the merger of binary black holes raise the prospect of precision black hole spectroscopy in the near future.
To perform the most sensitive tests of the nature of black holes using ringdown measurements, it is critical to compute the deviations to the spectrum of black holes in particular extensions of relativity.
These spectral shifts are also needed to interpret any violations of the predictions of relativity that may be detected during ringdown.
Here we present a first step towards computing the shifts to the spectrum of Kerr black holes with arbitrary spins, by deriving a modified Teukolsky equation governing the perturbations of black holes in theories beyond GR.
Our approach applies to a class of theories which includes dynamical Chern-Simons gravity and shift-symmetric scalar Gauss-Bonnet gravity, in the case where the deviations from relativity are small.
This allows for a perturbative approach to solving the equations of motion.
Further, we show how to use the modified equation to compute the leading-order spectral shifts of Kerr black holes, using eigenvalue perturbation methods.
Our formalism provides a practical approach to predicting ringdown for black holes in a range of promising extensions to relativity, enabling future precision searches for their signatures in black hole ringdown.
\end{abstract}

\maketitle
 
\section{Introduction}

The direct detection of gravitational waves~\cite{LIGOScientific:2016aoc,LIGOScientific:2018mvr,LIGOScientific:2020ibl,LIGOScientific:2021usb,LIGOScientific:2021djp,Nitz:2018imz,Nitz:2020oeq,Nitz:2021uxj,Nitz:2021zwj,Zackay:2019tzo,Venumadhav:2019tad,Venumadhav:2019lyq,Zackay:2019btq,Olsen:2022pin} by the Advanced LIGO~\cite{LIGOScientific:2014pky} and Virgo~\cite{VIRGO:2014yos} interferometers has opened a new window into strong-field and dynamical gravity.
These detections have enabled stringent tests of relativity, e.g.~\cite{LIGOScientific:2016lio,Yunes:2016jcc,LIGOScientific:2016dsl,LIGOScientific:2018dkp,LIGOScientific:2019fpa,LIGOScientific:2020tif,LIGOScientific:2021sio}, primarily in the form of null tests.
With the number of detections rapidly increasing, the strongest constraints on and the most sensitive searches for new physics require combining tests across many gravitational wave events.
For many tests, such as searches for parametrized deviations from inspiral, merger, and ringdown models, this requires either a hierarchical analysis~\cite{Zimmerman:2019wzo,Isi:2019asy,LIGOScientific:2020tif,LIGOScientific:2021sio} or a specific model from which to derive constraints on a common model parameter.

The ringdown following the merger of binary black holes is of especial interest from this perspective.
This final, exponentially decaying emission is the superposition of quasinormal modes (QNMs), which in the perturbative regime are determined by the mass and spin of the merged black hole~\cite{Berti:2009kk}.
The measurement of two or more ringdown modes allows for black hole spectroscopy~\cite{Dreyer:2003bv,Berti:2005ys,Berti:2018vdi}, probing both the properties of the black hole and allowing for tests of relativity targeting the merged remnant.
While deviations from the expected structure of Kerr black holes (violations of the no-hair theorem) also alter the gravitational waves produced during inspiral and merger, one attraction of ringdown tests is that are conceptually straightforward.
In addition, measurements of black hole mergers with total mass $\gtrsim 65 M_\odot$, together with recent advances in modeling~\cite{Giesler:2019uxc,Carullo:2019flw,Finch:2021qph,Isi:2021iql}, indicate that the measurement of multiple ringdown modes may already be within reach, e.g.~\cite{Isi:2019aib,Ghosh:2021mrv,Capano:2021etf} (see also~\cite{Cotesta:2022pci,Isi:2022mhy}). 

Perhaps more importantly, predicting deviations from ringdown in specific extensions to GR is tractable.
With a theory selected, the QNM spectrum can be computed by perturbing the metric and any additional fields around the equilibrium black hole solutions.
Shifts to the QNM spectrum have been computed in many cases for perturbations around Schwarzschild backgrounds, for example in quadratic gravity extensions~\cite{Cornelius:1938,Lovelock_1971,Deser:1981wh,Kanti:1995vq,Jackiw:2003pm,Alexander:2009tp} such as dynamical Chern-Simons (dCS) gravity~\cite{Cardoso:2009pk,Molina:2010,Cano:2020cao,Wagle:2021tam,Srivastava:2021imr}, scalar Gauss-Bonnet (sGB) gravity~\cite{Blazquez-Salcedo:2017txk,Blazquez-Salcedo:2016enn,Cano:2020cao}, and also in theories with even higher powers of curvature~\cite{Endlich:2017tqa,Cano:2020cao,Cano:2021myl}.
With spectral predictions from these theories in hand, direct searches for their signatures during ringdown are possible, and combining constraints across events is straightforward.
In addition, recent advances have allowed for numerical simulations of binary black holes in some of these beyond-GR theories~\cite{Okounkova:2017yby,Okounkova:2018pql,Okounkova:2019zjf,Witek:2018dmd,Okounkova:2020rqw,Silva:2020omi,Witek:2020uzz,Cayuso:2017iqc,Cayuso:2020lca,East:2020hgw,East:2021bqk}, and computation of the QNM spectra in these theories can contribute to future models covering inspiral, merger, and ringdown in these theories.

Most progress to date has been limited to the regime of slowly-spinning black holes, perturbing around a Schwarzschild background. 
Meanwhile, astrophysical merger remnants are expected to have dimensionless spin $\chi \sim 0.7$, e.g.~\cite{Hofmann:2016yih,Fishbach:2017dwv}.
Although expansions exist to very high orders in small spin~\cite{Cano:2019ore,Cano:2021myl}, an approach to computing deviations to the QNM spectra on Kerr backgrounds without any assumptions on the spin would be valuable.
This is true both for practical data analysis, where for Bayesian inference, predictions are needed across parameter space including for high spins $\chi \lesssim 0.99$, and for understanding how the unique features of the spectra of rapidly rotating black holes (e.g.~\cite{Andersson:1999wj,Yang:2013uba,Gralla:2016sxp,Compere:2017hsi,Gralla:2017lto}) carry over to other theories.

In this study we outline an approach to computing the deviations to QNMs in a broad class of beyond-GR theories. 
Our method is especially suited for dCS and sGB gravity when the coupling parameter that controls deviations from relativity is small and so the theories are in the decoupling limit.
This allows us to follow the same order-reduction scheme used in some recent simulations of binary black holes in these theories~\cite{Okounkova:2017yby,Okounkova:2018pql,Okounkova:2019zjf,Okounkova:2020rqw}. 
Our approach builds on previous work for computing the spectra of black hole backgrounds which are a small deformation away from Kerr~\cite{Zimmerman:2014aha}.
That method, based on standard eigenvalue perturbation (EVP) theory in quantum mechanics, has been used to make the first predictions of the QNMs of rapidly rotating, weakly-charged black holes~\cite{Mark:2014aja}, to understand parametric instabilities near the horizons of rapidly rotating black holes~\cite{Yang:2014tla}, to compute part of the QNM shifts in dCS for slowly rotating black holes~\cite{Srivastava:2021imr}, and has been discussed in the context of coupled-oscillator models of interacting QNMs~\cite{Yang:2015jja}.

In order to apply this method, we first derive a modified Teukolsky equation~\cite{Teukolsky:1973ha}, accounting for the deviations to the background of the Kerr black hole, the presence of additional non-minimally coupled background fields, and the changes to the dynamics of perturbations due to beyond-GR effects.
The result is a set of coupled equations for the metric perturbations and the additional fields.
We then show how these equations can be partially decoupled, allowing for an iterative approach to computing the dynamics of the fields, and then the QNM shifts of the gravitational perturbations.
This approach preserves the separability of the equations up to the final integrals required to compute the shifts.

Our derivation primarily takes place at the level of the field equations, and we only project onto the Newman-Penrose (NP) formalism~\cite{Newman:1961qr} at the last stages in order to take advantage of the separability of the Teukolsky equation.
Our approach, while compact and transparent, obscures any potential simplifications that may arise if the equilibrium black hole solution remains algebraically special even when accounting for the deviations from Kerr.
As such, we also provide a derivation of a modified Teukolsky equation entirely in the NP language, which may prove to be convenient in specific cases.

Following the initial stages of deriving our formalism, we became aware of an independent but equivalent effort for deriving a modified Teukolsky equation for dCS and similar theories in an NP language~\cite{Li2022InPrep}.
That work outlines additional choices of tetrad and gauge freedoms that further simplify the NP approach.
These two independent approaches serve as valuable cross-checks, and in the future can provide validation of technically challenging steps in the eventual computation of QNMs beyond Kerr. 
For example, both require metric reconstruction~\cite{Chrzanowski:1975wv,Wald:1978vm,Ori:2002uv,Keidl:2010pm} (in the form of tetrad reconstruction in the latter case) in order to compute QNM shifts, and both require an approach to solving for the dynamics of non-minimally coupled scalar fields (in our case this reduces to solving separable, sourced wave equations).

The remainder of this paper is as follows. 
In Sec.~\ref{sec:FieldEquations} we present the field equations for a broad class of models which are parametrically deformed away from relativity.
These include quadratic gravity models such as dCS and sGB gravity.
We then show how to partially decouple the field equations governing black hole ringdown.
Some further details on the operators arising in these equations are given in Appendix~\ref{sec:G2}.
A few example applications are given in Sec.~\ref{sec:Examples}, including further discussion of dCS and sGB gravity, as well as how the QNMs of weakly charged black holes fit into this formalism.
Additional details on the comparison of our approach to previous results on the QNMs of weakly charged black holes is in Appendix~\ref{sec:KNApp}.
We describe our modified Teukolsky equation and outline a practical approach to compute the leading shifts to the Kerr QNM spectra in these theories in Sec.~\ref{sec:MasterEqn}.
Section~\ref{sec:NPApproach} provides an alternative derivation of a modified Teukolsky equation, governing gravitational perturbations on a deformed background that is not Type D, with further details given in Appendix~\ref{sec:NPApp}.
This provides a convenient approach for cases where the deformed black hole remains algebraically special.
We discuss future directions and conclude in Sec.~\ref{sec:Conclusions}.

{\it Conventions}: 
In this paper, we set $c = 1$.
We use Latin indices from the beginning of the alphabet for spacetime quantities, while Latin indices from the middle of the alphabet generally index over sums.
We use capital subscripts $A, B$ as abstract indices over field quantities.
In our sections including NP quantities~\cite{Newman:1961qr}, we use $A$ to identify non-dynamical ``background fields'' and $B$ to denote dynamical degrees of freedom, as an extension of the notation of~\cite{Teukolsky:1973ha}.
In the same sections, $i,j$ index over miscellaneous collections of NP quantities as specified in the text.

\section{Field equations for QNMs beyond Kerr}
\label{sec:FieldEquations}

\subsection{Field equations}

Our goal is to create a formalism appropriate for quadratic gravity theories such as dCS and sGB gravity, in the decoupling limit where the modifications to relativity can be treated perturbatively, e.g.~\cite{Stein:2014xba}.
In such theories a scalar field $\vartheta$ is coupled to terms quadratic in the curvature, for example the Pontryagin density ${}^*R R$, such that a nontrivial geometry (specifically a black hole background) serves as a source term for the scalar field.
To tackle these theories, we consider actions of the more general form
\begin{align}
S & = S_{EH} + \int d^4 x \sqrt{-g} [\mathcal L_\vartheta + \epsilon \mathcal L_{\rm int} + \mathcal L_{\rm matter} ] \,.
\end{align}
Here $\mathcal L_{\vartheta}$ is the Lagrange density for a collection of fields we denote $\vartheta_A$, while $\mathcal L_{\rm matter}$ represents normal matter which is minimally coupled to gravity.
The new fields can be of any type, for example collections of scalar fields or vector fields, and $A$ is an abstract index running over all the field components.
We assume that $\mathcal L_\vartheta$ is at least quadratic in the new field degrees of freedom.
The term $\mathcal L_{\rm int}$ meanwhile provides a nontrivial coupling between the fields $\vartheta_A$ and the spacetime curvature, and we assume that it enters first at linear order in the fields $\vartheta_A$.
The parameter $\epsilon$ can be viewed as a small coupling term which governs the deviations to relativity.
Formally we treat it as a bookkeeping parameter, and match terms order by order in $\epsilon$.
Finally, the Einstein Hilbert action is
\begin{align}
S_{EH} & = \frac{1}{2\kappa_0} \int   d^4 x \sqrt{-g} R \,,
\end{align}
with $\kappa_0 = 8 \pi G$.

Varying the action and neglecting boundary terms as usual, the equations of motion for the field take the form\footnote{This assumes the equations of motion are second order in the field derivatives; these expressions can be extended to other cases.}
\begin{align}
\label{eq:WaveEqs}
\mathcal W_A(\vartheta,g) & =  \epsilon \rho_A(\vartheta,g)  \,, \\
\mathcal W_A(\vartheta,g) & \coloneqq \frac{\partial \mathcal L_\vartheta}{\partial \vartheta_A} - \nabla_a  \frac{\partial \mathcal L_\vartheta}{\partial \nabla_a \vartheta_A} \,,  \\
\rho_A(\vartheta,g) & \coloneqq - \frac{\partial \mathcal L_{\rm int}}{\partial \vartheta_A} + \nabla_a  \frac{\partial \mathcal L_{\rm int}}{\partial \nabla_a \vartheta_A} \,.
\end{align}
Here $\mathcal W_A$ is a collection of generalized wave equations for the fields, sourced by $\rho_A$.
For brevity, here and elsewhere we leave off the abstract indices of all fields when they arise in the arguments of operators.
Meanwhile, the gravitational field equations are
\begin{align}
\label{eq:EFE}
G_{ab}(g) & = \kappa_0 \left[ T^\vartheta_{ab}(\vartheta, g) + T^{\rm matter}_{ab} + \epsilon V^{\rm int}_{ab}(\vartheta, g) \right]\,,
\end{align}
with each stress-energy tensor defined as usual from variations with respect to the (inverse) metric. 
For example in a variational language we can write
\begin{align}
T^{\vartheta}_{ab} \coloneqq - \frac{2}{\sqrt{-g}} \frac{\delta (\sqrt{-g} \mathcal L_{\vartheta})}{\delta g^{ab}} \,.
\end{align}
Meanwhile, $V^{\rm int}_{ab}$ can similarly be derived by varying $\sqrt{-g} \mathcal L_{\rm int}$ with respect to the (inverse) metric; since this term involves curvature quantities, variation by parts results in mixed derivatives on functions of $\vartheta_A$ and the metric, see Sec.~\ref{sec:ScalarExamples} for an example.
From here we restrict to the case $T^{\rm matter}_{ab} = 0$.
Further, for convenience, we take the nonstandard convention of setting $\kappa_0 = 1$. 
These factors can be restored in the equations we derive below by multiplying each instance of a stress energy tensor $T^{\vartheta}_{ab}$ or interaction term $V^{\rm int}_{ab}$ by $\kappa_0$. 

\subsection{Notation for expanding operators} 
\label{ssec:Notation}

It is useful at this point to define a notation for expanding operators when evaluated on perturbative series expansions of the fields and the metric.
We define a two-index notation for perturbations around the background values $\vartheta_A = \vartheta^{(0)}_A= 0$ and $g_{ab} = g^{(0)}_{ab}$, and use single parenthetical superscripts ${}^{(j)}$ to indicate orders in $\epsilon$.
For a given, generally nonlinear, operator $\mathcal F(\vartheta,g)$ we define
\begin{align}
&\mathcal F^{(j,k)}[\varphi_1, \dots, \varphi_j, h_1, \dots h_k]  \coloneqq
\notag \\
& \frac{1}{j! k!}
\left. \frac{\partial^j \partial^k \mathcal F(\vartheta^{(0)} + \sum_{i=1}^j \epsilon_i \varphi_i , \, 
g^{(0)} _{ab} + \sum_{i=1}^k \kappa_i h_i)}
{\partial \epsilon_1\dots \partial \epsilon_j \partial \kappa_1 \dots \partial \kappa_k} 
\right|_{\substack{\epsilon_1, \dots \to 0 \\ \kappa_1, \dots \to 0}}
\,,
\end{align}
The operators $\mathcal F^{(j,k)}$ are multilinear in their arguments, with $j$ slots for perturbations to the fields and $k$ slots for perturbations to the metric.
They are separately totally symmetric in each slot type.
To formally define our operator expansions, we have used $\epsilon_j$ and $\kappa_k$ as a set of independent parameters, with the limit of all such parameters taken to zero at the end, and a set of independent fields and metric perturbations $\varphi_i $ and $h_i$, indexed by $i$.
We have also assumed that the operators we use admit series expansions around the background values of the fields and metric.
In some cases, we expand quantities that depend only on the metric, for example when expanding the Einstein tensor.
In those cases, we use only a single index in the superscript, for example $G_{\mu\nu}^{(1)}[h]$ for the leading expansion of the Einstein tensor around a perturbed background.

The notation is a bit ungainly, but we only need the expansions to low orders in $j,k$, so it is useful to look at specific examples.
Consider $\mathcal F(\vartheta) = \vartheta^2$, using a single scalar field for the $\vartheta_A$.
Then
\begin{align}
\mathcal F^{(2,0)}[\varphi_1, \varphi_2] & = \varphi_1 \varphi_2 \,.
\end{align} 
Meanwhile, if $\mathcal F(\vartheta) = \vartheta \, \partial_a \vartheta$, we have
\begin{align}
\mathcal F^{(2,0)}[\varphi_1, \varphi_2] & = \frac{1}{2}\left(\varphi_1 \partial_a \varphi_2 + \varphi_2 \partial_a \varphi_1 \right) \,.
\end{align}
Some care needs to be taken with the prefactors when we expand operators using this notation.
Consider again the example $\mathcal F(\vartheta) = \vartheta \, \partial_a \vartheta$, then
\begin{align}
\mathcal F(\epsilon_1 \varphi_1 + \epsilon_2 \varphi_2) 
 = & \epsilon_1^{2}  \mathcal F^{(2,0)}[\varphi_1, \varphi_1] 
 + 2 \epsilon_1 \epsilon_2  \mathcal F^{(2,0)}[\varphi_1, \varphi_2] 
\notag \\ & + \epsilon_2^{2}  \mathcal F^{(2,0)}[\varphi_2, \varphi_2] \,.
\end{align}
Note the factor of two on the mixed term from the combinatorics of the expansion, arising from summing over both orderings and recalling the total symmetry of $\mathcal F^{(j,k)}$.

In terms of this notation, the assumption that $\mathcal L_\vartheta$ is at least quadratic in the fields and $\vartheta^{(0)}_A = 0$ means that
\begin{align}
\mathcal W_A^{(0,k)} & = 0 \,, \\
\label{eq:Tconditions}
T^{\vartheta (0,k)}_{ab} & = T^{\vartheta (1,k)}_{ab} = 0 \,.
\end{align}
Our assumption that $\mathcal L_{\rm int}$ is at least linear in the fields means that
\begin{align}
\label{eq:Vconditions}
V^{{\rm int}(0,k)}_{ab} & = 0 \,,
\end{align}
but, for example, $\rho_A^{(0,0)}$ need not be zero.
In fact, we are interested in the case where $\rho_A^{(0,0)}$ is nonzero, requiring terms linear in the fields in $\mathcal L_{\rm int}$. 
In the case of quadratic gravity theories, we have a single scalar field $\vartheta$ and the interaction term separates as $\mathcal L_{\rm int} = f(\vartheta) \mathcal R[g]$ for some curvature operator $\mathcal R$.
Then $\rho = f'(\vartheta) \mathcal R$, and sources the fields at $O(\epsilon)$ when $f'(0) \neq 0$.
The power counting provided below only holds if  $f'(0) \neq 0$, in which case any $f$ is essentially equivalent at leading order; only the Taylor expansion of $f$ around $\vartheta = 0$ matters in the perturbative expansion~\cite{Witek:2018dmd}.

Pursuing this example further, for the quadratic gravity models of interest we have
\begin{align}
\label{eq:ScalarWaveOp}
\mathcal W(\vartheta) = \Box_{g} \vartheta \,,
\end{align}
which is the scalar wave equation for the metric $g_{ab}$. 
It is linear in the field, so $ \mathcal{W}^{(j,k)}[\vartheta,g]  = 0$ for $j\geq 2$. 
Expanding around $\vartheta = 0 + \epsilon \varphi$ and $g_{ab} = g_{ab}^{(0)} + \epsilon h_{ab}$, we have
\begin{align}
\mathcal W^{(1,0)}[\varphi]  = & \Box_{g^{(0)}} \varphi\,, \\
\mathcal W^{(1,1)}[\varphi,h]  = & -\frac{1}{\sqrt{-\det g^{(0)}_{cd}}}\partial_a\left(\sqrt{-\det g^{(0)}_{cd}} \, h^{ab}  \partial_b \varphi \right)
\notag \\
& + \frac{1}{2} g_0^{ab} (\partial_a h^{c}{}_{c}) \partial_b \varphi
 \,,
\end{align}
with indices raised and lowered using the background metric.
Meanwhile,
\begin{align}
\rho^{(0,0)} & = f'(0) \mathcal R \,, \\
\rho^{(j,0)}[\varphi, \varphi, \dots] & = \frac{1}{j!} \left. \frac{d^j f}{d\vartheta^j}\right|_{\vartheta = 0} \varphi^j \mathcal R  \,.
\end{align}

These models, and many others of interest, have standard scalar field stress-energy tensors,
\begin{align}
T^\vartheta_{ab} & = \partial_a \vartheta \partial_b \vartheta - \frac{1}{2}g_{ab} g^{cd} \partial_c \vartheta \partial_d \vartheta \,.
\end{align}
In this case we have
\begin{align}
\label{eq:ScalarStressEnergy20}
T^{\vartheta(2,0)}_{ab}[\varphi_1, \varphi_2] & =  \partial_{(a} \varphi_1 \partial_{b)} \varphi_2 - \frac{1}{2}g^{(0)}_{ab} g_{(0)}^{cd}  \partial_c \varphi_1 \partial_d \varphi_2 \,,
\end{align}
and 
\begin{align}
\label{eq:ScalarStressEnergy21}
T^{\vartheta(2,1)}_{ab}[\varphi_1, \varphi_2 , h]  = &  \partial_{(a} \varphi_1 \partial_{b)} \varphi_2 
\notag \\ & 
+ \frac{1}{2} (g^{(0)}_{ab} h^{cd} - h_{ab} g_{(0)}^{cd}) \partial_c \varphi_1 \partial_d \varphi_2 \,.
\end{align}
The interactions terms $V^{\rm int}_{ab}$ can be similarly expanded, but conventionally they are somewhat complicated in structure.
We discuss particular cases in Sec.~\ref{sec:Examples} below.

\subsection{Equilibrium solutions}

With the notation settled, our goal is to expand Eqs.~\eqref{eq:WaveEqs} and~\eqref{eq:EFE} around a Kerr background, order by order in $\epsilon$, while also incorporating perturbations representing propagating degrees of freedom in the metric (gravitational perturbations $h_{ab}$), and in the scalar fields.
Before adding in these waves, we consider the how the field equations are solved in equilibrium.

We are interested in cases where as $\epsilon \to 0$, we recover the Kerr solution, which means that we require that $\vartheta_A = 0$ should solve $\mathcal W_A[\vartheta] = 0$ on a black hole background.
This means there should be no potentials $V$ which support nonzero configurations of scalar fields in the limit $\epsilon \to 0$.
With this in mind, we expand our fields in powers of $\epsilon$,
\begin{align}
g_{ab} & = g^{(0)}_{ab} + \epsilon g^{(1)}_{ab} + \epsilon^2 g^{(2)}_{ab} + O(\epsilon^3)  \,, \\
\vartheta_A & = 0 + \epsilon \vartheta^{(1)}_A + O(\epsilon^2) \,,
\end{align}
where we know that $\vartheta_A$ enters first at $O(\epsilon)$, consistent with our requirement that $\vartheta_A = 0$ in the limit $\epsilon \to 0$.
At leading order we find that for the metric
\begin{align}
G_{ab}(g^{(0)}_{cd}) = 0 \,,
\end{align}
which is solved for by the vacuum Kerr solution.
 
Now looking at the Euler Lagrange equation of the fields, at the next order we find
\begin{align}
\label{eq:EqField}
\mathcal W^{(1,0)}_A[\vartheta^{(1)}] = \rho^{(0,0)}_A \,,
\end{align}
which are sourced wave equations that we must solve for $\vartheta_A^{(1)}$.
Similarly at $O(\epsilon)$ we have
\begin{align}
G^{(1)}_{ab}[g^{(1)}] = \mathcal E_{ab}[g^{(1)}] = 0 \,,
\end{align}
where we have noted that our $G^{(1)}$ is just the standard linearized Einstein operator on the background, 
\begin{align}
\label{eq:LinearEFE}
\mathcal E_{ab}[h] \coloneqq \frac{1}{2}& \left[ 2 \nabla^c \nabla_{(a}h_{b)c} - \nabla^c \nabla_c h_{ab}  - \nabla_a \nabla_b h^c{}_c   \right.  
\notag \\
& + \left. g^{(0)}_{ab}(\nabla^c\nabla_c h^d{}_d - \nabla^c \nabla^d h_{cd}) \right]\,.
\end{align}
Here all covariant derivatives are with respect to $g^{(0)}_{ab}$.
There are no source terms for $g^{(1)}_{ab}$ at this order, recalling our requirements from Eqs.~\eqref{eq:Tconditions} and~\eqref{eq:Vconditions}.
As we are interested in equilibrium solutions about a black hole background, we see that $g^{(1)}_{ab} = 0$ and the metric is only deformed away from Kerr at $O(\epsilon^2)$.

At $O(\epsilon^2)$ we have
\begin{align}
\label{eq:EqMetric}
\mathcal E_{ab}[g^{(2)}] = T^{\vartheta (2,0)}_{ab}[\vartheta^{(1)},\vartheta^{(1)}] + V^{\rm{int}(1,0)}_{ab}[\vartheta^{(1)}]\,.
\end{align}
The equilibrium solution to this sourced wave equation gives the deformation $g^{(2)}_{ab}$ to the Kerr metric.

\subsection{Coupled field equations with propagating degrees of freedom}

Now we consider the case where in addition to the equilibrium deviations to the spacetime, we allow for gravitational wave perturbations $h_{ab}$. 
To do this, we further perturb $g^{(0)}_{ab}$ by $h_{ab}$, and we introduce a second small parameter $\eta$ to track these perturbations.
We consider our solutions only up to the leading corrections in $\epsilon$, in order to derive the leading corrections to the ringdown spectrum.

One complication to the usual treatment of gravitational perturbations to Kerr is now the perturbations couple to the additional fields $\vartheta_A$, requiring in general a simultaneous treatment of further, $O(\eta)$ perturbations to both.
Physically, this is because perturbations to the spacetime can ``shake'' the background fields and effectively generate propagating degrees of freedom in them, and {\it vice versa}.
Practically it means that the corrections to the ringdown spectrum arise both due to the deformation of the metric $g^{(2)}_{ab}$ and due to the coupling of the equilibrium fields $\vartheta^{(1)}_A$ to these waves.

With this in mind we write our field expansions as
\begin{align}
g_{ab} & = g^{(0)}_{ab} + \epsilon^2 g^{(2)}_{ab} + \eta h_{ab} + \dots \,, \\
\vartheta_A & = \epsilon \vartheta^{(1)}_A +  \epsilon^2 \vartheta_A^{(2)} + \eta \varphi_A + \dots\,.
\end{align}
Here $\varphi_A$ represent wave degrees of freedom in the fields.
Inserting these expressions into our field equations and expanding, we recover the same $O(\eta^0)$ expressions used to derive $\vartheta^{(1)}_A$ and $g^{(2)}_{ab}$ as before, Eqs.~\eqref{eq:EqField} and \eqref{eq:EqMetric}. 
At $O(\eta)$, we find up to $O(\epsilon^2)$
\begin{align}
\label{eq:GWEqs}
 \mathcal E{}_{ab}[h] & + 2 \epsilon^2 G^{(2)}[h,g^{(2)}] 
= 
\notag \\ & 
\epsilon \left[
2 T^{\vartheta(2,0)}_{ab}[\vartheta^{(1)},\varphi]  
+  V^{\rm{int}(1,0)}_{ab}[\varphi] 
\right]
\notag \\ 
& 
+ \epsilon^2 \left[  2 T^{\vartheta(2,0)}_{ab}[\vartheta^{(2)},\varphi]  
+ T^{\vartheta(2,1)}_{ab}[\vartheta^{(1)},\vartheta^{(1)},h] \right.
 \notag \\
&
\left.
+ 3 T^{\vartheta(3,0)}_{ab}[\vartheta^{(1)},\vartheta^{(1)},\varphi]
+  V^{{\rm int}(1,1)}_{ab}[\vartheta^{(1)}, h]  \right.
\notag \\ &
\left. + 2V^{\rm{int}(2,0)}[\vartheta^{(1)},\varphi] \right]\,.
\end{align}
For the field degrees of freedom, we find to $O(\epsilon)$
\begin{align} 
\label{eq:ExpandWaveEqs}
\mathcal W^{(1,0)}_A [\varphi]  + & 2 \epsilon \mathcal W^{(2,0)}_A[\vartheta^{(1)}, \varphi] + \epsilon \mathcal W^{(1,1)}_A [\vartheta^{(1)},h]  =  
\notag \\ &
 \epsilon \rho^{(1,0)}_A [\varphi]  + \epsilon \rho_A^{(0,1)}[h] \,.
\end{align}

Since our focus is on ringdown, in Eq.~\eqref{eq:GWEqs} we have assumed that there are no $O(\eta)$ matter sources for the gravitational waves, and similarly no $O(\eta)$ sources for the fields in Eq.~\eqref{eq:ExpandWaveEqs}, but these can be added as appropriate.
We see that Eqs.~\eqref{eq:GWEqs} and~\eqref{eq:ExpandWaveEqs} are coupled, due to the nonzero background fields $\vartheta^{(1)}_A$ and the presence of the interaction term in the Lagrangian $\mathcal L_{\rm int}$ which is responsible for $V^{\rm int}_{ab}$ and $\rho_A$. 
To proceed, we ideally decouple this linear system of equations for $h_{ab}$ and $\varphi_A$.

\subsection{Decoupling and partial decoupling of the field equations}
\label{sec:Decoupling}

We know that in the limit $\epsilon \to 0$, Eqs.~\eqref{eq:GWEqs} and~\eqref{eq:ExpandWaveEqs} decouple, meaning that we can find solutions where $\varphi_A = 0$ and $h_{ab}$ obeys the linearized Einstein equations, or where $h_{ab} = 0$ and $\varphi_A$ satisfies the generalized wave equation on the background.
We seek consistent solutions perturbing around each of these cases.
In other situations such an ansatz results in a complete decoupling of the field equations, such as occurs for the electromagnetic (EM) and gravitational QNMs of weakly charged Kerr-Newman black holes~\cite{Mark:2014aja}. 
In the class of field equations treated here the problem is more complicated.

We start with the simpler case, where we seek a solution perturbing around the scalar QNMs,
\begin{align}
\varphi_A & = \varphi^{(0)}_A + \epsilon \varphi^{(1)}_A + O(\epsilon^2) \,, \\
 h_{ab} & = 0 + \epsilon h^{(1)}_{ab} + O(\epsilon^2). 
\end{align}
In this case, we find 
\begin{align}
\label{eq:FieldsDecoupled}
\mathcal W^{(1,0)}_A [\varphi^{(0)}] & + 2\epsilon \mathcal W^{(2,0)}_A[\vartheta^{(1)},\varphi^{(0)}] + \epsilon \mathcal W^{(1,0)}_A [\varphi^{(1)}]  =  
\notag \\ &
 \epsilon \rho^{(1,0)}_A [\varphi^{(0)}] \,,
\end{align}
neglecting terms of $O(\epsilon^2)$.
Meanwhile, assuming $\varphi \sim O(1)$ and neglecting terms of $O(\epsilon^2)$, we see that Eq.~\eqref{eq:GWEqs} admits solutions $h_{ab} = \epsilon h^{(1)}_{ab}$, consistent with our ansatz.
This means that the equations for the fields $\varphi_A$ have decoupled from $h_{ab}$ at leading order. 
We discuss how to solve Eq.~\eqref{eq:FieldsDecoupled} for $O(\epsilon)$ shifts to the QNM frequencies associated with the fields $\varphi_A$ in Sec.~\ref{sec:ScalarShifts} below.
Physically, this is the case where the beyond-GR effects modify the free QNM ringing of the fields $\vartheta_A$ at $O(\epsilon)$, while at the same time the ringdown of $\vartheta_A$ sources gravitational modes at $O(\epsilon)$.

The gravitational case is of greater interest but unfortunately is technically more involved. 
Here only a partial decoupling can be achieved, which still provides a practical route for computing the QNM shifts.
We take as our ansatz
\begin{align}
h_{ab} & = h^{(0)}_{ab} + \epsilon^2 h^{(2)}_{ab}   + O(\epsilon^2) \,, \\
\varphi & = 0 + \epsilon \varphi^{(1)}_A + O(\epsilon^2) \,.
\end{align}
First we apply this ansatz to Eq.~\eqref{eq:ExpandWaveEqs}, giving
\begin{align} 
\label{eq:FieldsCoupled}
\mathcal W^{(1,0)}_A [\varphi^{(1)}] + \mathcal W^{(1,1)}_A [\vartheta^{(1)},h^{(0)}]  
- \rho^{(0,1)}_A[h^{(0)}] 
=   0 
\end{align}
when neglecting terms of $O(\epsilon^2)$.
We see that in this case, we consistently source a solution $\varphi_A \sim O(\epsilon)$ from a gravitational ringdown starting at $O(1)$ in $\epsilon$-counting.
Meanwhile, Eq.~\eqref{eq:GWEqs} becomes
\begin{align}
\label{eq:GWCoupled}
&\mathcal E_{ab}[h^{(0)}] 
+ \epsilon^2 \left[2 G^{(2)}_{ab} [h^{(0)},g^{(2)}] 
- T^{\vartheta (2,1)}_{ab}[\vartheta^{(1)},\vartheta^{(1)},h^{(0)}] 
\right.
\notag \\
&\left.
- V^{\rm{int}(1,1)}_{ab}[\vartheta^{(1)}, h^{(0)}]
-2T^{\vartheta (2,0)}_{ab}[\vartheta^{(1)},\varphi^{(1)}]  -  V^{\rm{int}(1,0)}_{ab}[\varphi^{(1)}]  \right]
\notag \\ &
+ \epsilon^2 \mathcal E_{ab}[h^{(2)}] = 0 \,,
\end{align}
neglecting $O(\epsilon^3)$ terms.
We can see that had we included a term $\epsilon h^{(1)}_{ab}$ in our ansatz, we would have had an equation $\mathcal E_{ab}[h^{(1)}] = 0$ which is no different than the equation obeyed by $h^{(0)}_{ab}$, so this correction can be absorbed into the definition of $h^{(0)}_{ab}$. 
The beyond-GR effects only source modifications to the QNMs at $O(\epsilon^2)$.
Together, Eqs.~\eqref{eq:FieldsCoupled} and~\eqref{eq:GWCoupled} are a coupled set of equations, but that can be solved order by order:
First, a particular QNM solution $h^{(0)}_{ab}$ is selected, and input into the source term in Eq.~\eqref{eq:FieldsCoupled}, which is then solved for $\varphi^{(1)}_A$.
With this, the $O(\epsilon^2)$ part of Eq.~\eqref{eq:GWCoupled} can be solved.

In Sec.~\ref{sec:MasterEqn} we describe a practical approach to compute the shifts to the QNM frequencies from our decoupled and partially decoupled equations.
Before this, we give some explicit examples of the various operators described for particular theories of interest.

\section{Example applications}
\label{sec:Examples}

In Sec.~\ref{sec:FieldEquations} we provide general expressions for how the QNM wave equations are modified in a class of beyond-GR theories.
Here we discuss particular cases in greater detail, focusing on dCS and sGB gravity.
We also discuss how the known approach to computing the QNMs of weakly charged black holes~\cite{Mark:2014aja} fits into our formalism.
This final case is an important example, both for how to treat black hole deformations which are due to nontrivial boundary conditions, and as an example of how a different $\epsilon$-scaling of the fields can be treated in our formalism.

\subsection{Scalar fields coupled to curvature}
\label{sec:ScalarExamples}

Consider the case of a single scalar field $\vartheta$ coupled to curvature quantities, and with a standard kinetic term in the Lagrangian,
\begin{align} 
\mathcal L_\vartheta & = -\frac{1}{2} g^{ab} (\partial_a \vartheta) (\partial_b \vartheta)\,.
\end{align}
The form of the scalar wave equations for this situation has been discussed using our notation below Eq.~\eqref{eq:ScalarWaveOp}.
We can consider two cases of particular interest: dCS and shift-symmetric  sGB gravity.

In the first case, the dCS scalar couples to the Pontryagin-Chern density ${}^*RR$~\cite{Jackiw:2003pm},
\begin{align}
\mathcal L_{\rm int} & = \vartheta \mathcal R_{\rm dCS} \,, \\
\mathcal R_{\rm dCS} & = -\frac{1}{8}{}^*R R \coloneqq - \frac{1}{8} {}^*R^{abcd} R_{abcd} \,,\\
{}^*R^{abcd} & \coloneqq \frac{1}{2} \epsilon^{abef}R_{ef}{}^{cd} \,.
\end{align}
The static field $\vartheta^{(1)}$ solves to leading order
\begin{align}
\Box_{g^{(0)}} \vartheta^{(1)} & = \frac{1}{8} ({}^*RR)^{(0,0)} \,.
\end{align}
and $  ({}^*RR)^{(0,0)}$ is the Pontryagin-Chern density evaluated on the background Kerr metric.
The static deformation to the metric $g^{(2)}_{ab}$ solves Eq.~\eqref{eq:EqMetric} with the interaction term given in terms of the $C$-tensor,
\begin{align}
& V^{\rm{int}(1,0)}_{ab}[\vartheta^{(1)}] = - C^{(0)}_{ab}[\vartheta^{(1)}] \,, \\
& C_{ab}[\vartheta^{(1)}] \coloneqq  (\epsilon_{(a}{}^{cde}\nabla_{|d|} R_{b)c})\nabla_e \vartheta^{(1)}
+ {}^* R_{(a}{}^c{}_{b)}{}^d \nabla_c \nabla_d \vartheta^{(1)} \,.
\end{align}
In the expression for the $C$-tensor, the Riemann tensor, Ricci tensor, and covariant derivative are with respect to the full metric, but for $C^{(0)}_{ab}$ all these are evaluated on the Kerr background.
The solutions to these equations have been found to high order in a slow spin expansion~\cite{Yagi:2012ya,Cano:2019ore} and numerically explored in the rapidly rotating case~\cite{Stein:2014xba}.
For the dynamical perturbations to dCS, the above equations together with Eqs.~\eqref{eq:ScalarStressEnergy20} and~\eqref{eq:ScalarStressEnergy21} can be adapted in a straightforward manner to give the terms in Eq.~\eqref{eq:GWCoupled}.
The only element explicitly missing the is the lengthy expansion of $C_{ab}$ around the background to give $V^{\rm{int}(1,1)}_{ab}[\vartheta^{(1)},h^{(0)}] = - C^{(1)}_{ab}[\vartheta^{(1)},h^{(0)}]$, which we omit here for brevity.

The second case of interest is sGB gravity.
We choose our conventions to conform to those of~\cite{Witek:2018dmd}, where $\vartheta$ is made dimensionless by drawing an overall factor of $1/(2\kappa_0)$ out of $\mathcal L_\vartheta$ and $\mathcal L_{\rm int}$, so that the action is
\begin{align}
S_{GB} & = \frac{1}{2 \kappa_0} \int d^4 x \sqrt{-g} \left[R +  \mathcal L_\vartheta + \epsilon \mathcal L_{\rm int} \right] \,.
\end{align}
Here the curvature coupling is to the Gauss-Bonnet invariant
\begin{align}
\mathcal L_{\rm int} & = 2f(\vartheta) \mathcal R_{\rm GB} \,, \\
\mathcal R_{GB} & = R^{abcd} R_{abcd} - 4 R^{ab} R_{ab} + R^2 \,.
\end{align}
In addition, we must select a potential $f(\vartheta)$.
As mentioned previously, all choices where $f' \neq 0$ are equivalent to leading order, up to rescaling of $\epsilon$, and so we select the simple shift-symmetric case $f = \vartheta$.
Then the operators appearing in the scalar wave equations mirror those in the dCS case, with $\rho^{(0,0)}$ simply twice the Gauss-Bonnet scalar evaluated on the background.
With these conventions, $\epsilon$ is dimensionful, but can be rendered dimensionless by drawing out factors of the total mass of the system, see e.g.~\cite{Okounkova:2020rqw}.

For sGB, the interaction terms in the equations for the metric deformation $g^{(2)}_{ab}$ are~\cite{Witek:2018dmd}
\begin{align}
V^{\rm{int}(1,0)}_{ab}[\vartheta^{(1)}] &= - \mathcal G^{(0)}_{ab}[\vartheta^{(1)}] \,, \\
\mathcal G_{ab}[\vartheta^{(1)}] & \coloneqq 2 g_{c(a} g_{b)d} \epsilon^{edfg} \nabla_h ( {}^*R^{ch}{}_{fg}\nabla_e \vartheta^{(1)}) \,.
\end{align}
The metric, curvature quantities, and covariant derivatives in the expression for $\mathcal G_{ab}$ are with respect to the full metric, but are evaluated on the Kerr background for $\mathcal G^{(0)}_{ab}$.
The solutions to these equations have been found in a slow-spin expansion~\cite{Pani:2009wy,Antoniou:2017hxj,Cano:2019ore}.
As with dCS, the dynamical field equations~\eqref{eq:GWCoupled} directly follow from $V^{\rm int}_{ab}$ at this order and Eqs.~\eqref{eq:ScalarStressEnergy20} and~\eqref{eq:ScalarStressEnergy21}, together with the expansion of $\mathcal G_{ab}$ around the background to give $V^{{\rm int}(1,1)}_{ab} [\vartheta^{(1)},h^{(0)}]$. 
Again, we omit this lengthy expression.

\subsection{Weakly charged black holes}
\label{sec:KNExample}

Consider next the perturbations of weakly charged black holes, where now $\epsilon = Q/M$ is the small dimensionless charge of the black hole.
In this case, the deformation to the spacetime is simply the linearization of the exact Kerr-Newman solution in $\epsilon^2$.
These deformations to the metric arise because of the EM stress-energy provided by the electric and magnetic fields of the charged black hole.
In this situation, the additional fields $\vartheta_A$ can be taken to be the components of the Maxwell stress tensor $F_{ab}$, or equivalently the Maxwell scalars $\phi_0$, $\phi_1$, and $\phi_2$ which are the projections of $F_{ab}$ onto a null tetrad in the NP formalism~\cite{Newman:1961qr,Stephani:2003tm}.

In our language, there is no interaction Lagrangian $\mathcal L_{\rm int}$, and hence both the source terms $\rho_A$ in the wave equations for the fields and the interaction potential $V^{\rm int}_{ab}$ vanish. 
Instead, the fields $\vartheta_A$ are nonzero because of the boundary conditions at the horizon.
Thus the solutions of the leading field equations,
\begin{align}
\mathcal W^{(1,0)}[\vartheta^{(1)}] = 0 \,,
\end{align}
are nonzero, entering in at order $\epsilon$ due to the $\epsilon$-small charge, $\vartheta_A \approx \epsilon \vartheta_A^{(1)}$.
These fields source stationary metric deformations $g^{(2)}_{ab}$ through $T^{\vartheta(2,0)}_{ab}[\vartheta^{(1)},\vartheta^{(1)}]$.

Turning to perturbations of the stationary solution, we note that the linearity of Maxwell's equations means that $\mathcal W_A^{(j,k)} = 0$ for $j\geq 2$, and so the field equations at $O(\eta)$ expand to $O(\epsilon^2)$ as
\begin{align}
\mathcal W^{(1,0)}[\varphi] & + \epsilon \mathcal W_A^{(1,1)}[\vartheta^{(1)},h] + \epsilon^2 \mathcal W_A^{(1,1)}[\vartheta^{(2)},h] 
\notag \\ &
+  \epsilon^2 \mathcal W_A^{(1,1)}[\varphi,g^{(2)}] = 0 \,.
\end{align}
Here we need to go to a higher order than before, because it turns out the terms $\mathcal{W}_A^{(1,1)}[\vartheta^{(1)},h]$ and $\mathcal{W}_A^{(1,1)}[\vartheta^{(2)},h]$ are pure gauge.

To see this, we write out the source-free Maxwell's equations for $F_{ab}$ using $g_{ab} = g^{(0)}_{ab} + h_{ab}$,
\begin{align}
\label{eq:MaxwellExp}
& \nabla_a F^{ab} + S^a{}_{ac} F^{cb} + S^b{}_{ac} F^{ac} = 0\,, 
 \\
& S^a{}_{bc}  \coloneqq \frac{1}{2} g_{(0)}^{ad}\left(\nabla_c h_{db} + \nabla_b h_{dc} - \nabla_d h_{bc} \right) \,.
 \end{align}
Here $\nabla_a$ is taken to be a covariant derivative with respect to the background Kerr metric, and we have expanded to leading order in the perturbation $h_{ab}$. 
We see that the last term in Eq.~\eqref{eq:MaxwellExp} vanishes by the antisymmetry of $F^{ac}$, and by using the definition of $S^a{}_{bc}$ and the fact that $g^{(0)}_{ab}$ commutes with $\nabla_a$ we can simplify,
\begin{align}
\nabla_a F^{ab} + \frac{1}{2} F^{ac} \nabla_c h^a{}_a = 0\,.
\end{align}
However, the trace of the gravitational perturbations can be set to zero by a choice of gauge, and hence we can set terms like $\mathcal W^{(1,1)}_A[\vartheta ,h]$ to zero for Maxwell's equations.

With this, the coupled field equations for the dynamical perturbations are simply
\begin{align}
\mathcal W{}&^{(1,0)}[\varphi] + \epsilon^2 \mathcal W_A^{(1,1)}[\varphi,g^{(2)}] = 0 \,.
\\
\mathcal E&{}_{ab}[h] + 2 \epsilon^2 G^{(2)}[h,g^{(2)}] 
= 
2 \epsilon T^{\vartheta(2,0)}_{ab}[\vartheta^{(1)},\varphi]
\notag \\ 
& 
+ \epsilon^2 \left(  2 T^{\vartheta(2,0)}_{ab}[\vartheta^{(2)},\varphi]  
+ T^{\vartheta(2,1)}_{ab}[\vartheta^{(1)},\vartheta^{(1)},h] 
\right)\,.
\end{align}
Note that $T^{\vartheta(3,0)}_{ab}[\vartheta^{(1)},\vartheta^{(1)},\varphi] = 0$ since the stress-energy is purely quadratic in the electromagnetic fields.

At this point, we run into an issue. 
When treating the problem of coupled gravito-electromagnetic perturbations of Kerr Newman black holes using the NP formalism~\cite{Chandrasekhar:1984siy,Dias:2015wqa}, charge enters the equations as $Q^2$.
Here though it appears that the gravitational equations are coupled at $O(Q)$, through the term $2 \epsilon T^{\vartheta(2,0)}_{ab}[\vartheta^{(1)},\varphi]$.  
We can cure this issue by realizing that the consistent scaling for the dynamical EM fields must be $\varphi = \epsilon \varphi^{(1)} + \dots$, when computing either the EM or gravitational shifts.

The fact that this scaling is appropriate can be justified from a few perspectives. 
One approach is to recognize that one of the gauge-invariant combinations of EM and gravitational perturbations is~\cite{Chitre:1976bb,Dias:2015wqa}
\begin{align}
\Phi_{\rm EM} = 2 \phi_1^A \Psi_1^B - 3 \Psi^A_2 \phi^B_0 \,,
\end{align}
where $\phi^A_1$ is the Maxwell scalar associated with the background fields $\vartheta^{(1)}_A$, $\phi^A_1 \sim O(\epsilon)$, $\Psi_1^B$ is a Weyl curvature scalar associated with the gravitational perturbations $h_{ab}$, $\Psi_2^A$ is the non-vanishing background Weyl curvature scalar, and $\phi_0^B$ is the Maxwell scalar associated with the EM perturbations.
Taking both contributions to $\Phi_{\rm EM}$ on equal footing indicates that $\phi_0^B\sim O(\epsilon)$, hence, in the language of our formalism, $\varphi_A \sim O(\epsilon)$.

Another argument is essentially physical.
The reason for the coupling between EM and gravitational perturbations of Kerr Newman is that a perturbation to the spacetime naturally ``shakes'' the background electric and magnetic field lines, generating propagating degrees of freedom. 
Meanwhile, dynamical perturbations to the field lines naturally alter the curvature sourced by these matter fields.
By insisting that the dynamical perturbations to the EM fields is of the same order as the stationary EM fields, we assert that the QNM ringing is due to the ringing of these field lines, even in the case where we expand around the background EM QNMs by taking $h_{ab} = \epsilon^2 h^{(2)}_{ab}$, which is used to decouple the equations.

Setting this scaling, we arrive at equations with couplings at the expected orders,
\begin{align}
\mathcal W{}&^{(1,0)}[\varphi^{(1)}] + \epsilon^2 \mathcal W_A^{(1,1)}[\varphi^{(1)},g^{(2)}] = 0 \,.
\\
\label{eq:KNGrav}
\mathcal E&{}_{ab}[h] + 2 \epsilon^2 G^{(2)}[h,g^{(2)}] 
= 
2 \epsilon^2 T^{\vartheta(2,0)}_{ab}[\vartheta^{(1)},\varphi^{(1)}]
\notag \\ 
& 
+ \epsilon^2 T^{\vartheta(2,1)}_{ab}[\vartheta^{(1)},\vartheta^{(1)},h] \,.
\end{align}
The fact that the corrections to both leading order equations is $\epsilon^2$ allows for a complete decoupling when computing the QNM shifts using the EVP method~\cite{Mark:2014aja}.
To confirm that the chosen scalings are appropriate, we can project the coupling term $T^{\vartheta(2,0)}_{ab}[\vartheta^{(1)},\varphi^{(1)}]$ into the NP language and compare to the known NP result.
We show that these expressions agree in Appendix~\ref{sec:KNApp}.

To make use of these equations, we would next project both the gravitational and EM expressions into the NP formalism, the former using the method described below in Sec.~\ref{sec:TeukProject}, and the latter using the projection operator $\mathcal S_E^a$ defined in~\cite{Wald:1978vm}. 

\section{Spectral shifts and a modified Teukolsky equation}
\label{sec:MasterEqn}

With our decoupled and partially decoupled field equations from Sec.~\ref{sec:Decoupling}, we can derive the shifts to the QNM frequencies of Kerr due to the deformations of the black hole, and the additional coupling of the gravitational waves to the dynamics of the extra fields $\vartheta_A$.
Our primary tool is the EVP approach given in~\cite{Mark:2014aja}.
To introduce this approach and provide a simple example of the formalism, we first derive an expression for the QNM shifts for the ringdown of the propagating degrees of freedom of the fields, $\varphi_A$.

\subsection{Shifts for the field QNMs}
\label{sec:ScalarShifts}

We take as our starting point the decoupled Eq.~\eqref{eq:FieldsDecoupled}.
Our QNM solutions can be expanded as
\begin{align}
\label{eq:ScalarExpansion}
\varphi^{(0)}_A = e^{-i \omega t}e^{i m \phi} \tilde \varphi^{(0)}_{A, m \omega}(r,\theta) 
\end{align}
in terms of Boyer-Lindquist coordinates $x^\mu = (t,r,\theta,\phi)$.
The symmetries of the background guarantee separation of frequencies and angular modes, so the leading equation is
\begin{align}
\label{eq:LeadingFieldFreq}
\mathcal{W}^{(1,0)}_A [\varphi^{(0)}] \rightarrow \tilde {\mathcal W}^{(1,0)}_{A,m\omega} [\tilde \varphi^{(0)}_{m\omega} (r,\theta)] = 0 \,,
\end{align}
where $\tilde {\mathcal W}^{(1,0)}_A$ depends on $\omega, m, r$ and $\theta$ since azimuthal and time derivatives in the linear operator bring down factors of $-i\omega$ and $i m$.
This is solved for the QNM wavefunctions $\varphi^{(0)}_{A, m\omega}(r,\theta)$ and discrete frequencies $\omega^{(0)}$ by setting outgoing boundary conditions at asymptotic infinity and ingoing boundary conditions at the horizon.
We leave implicit the indexing of these modes and the indexing of their frequencies.
For the separable case of scalar fields, the QNMs are indexed azimuthal quantum number $\ell$ and an overtone number $n$ in addition to the magnetic quantum number $m$.

The $O(\epsilon)$ corrections to the $\varphi_A$ QNMs come in two flavors.
There are those corrections that leave the wavefunctions intact but shift the QNM frequencies,
\begin{align}
\omega = \omega^{(0)} + \epsilon \omega^{(1)} \,,
\end{align}
and those that shift the wavefunctions.
Specifically, we expand Eq.~\eqref{eq:ScalarExpansion} as
\begin{align}
\varphi_A & = e^{-i[\omega^{(0)} + \epsilon \omega^{(1)}] t}e^{i m \phi} (\tilde \varphi^{(0)}_{A,m\omega} + \epsilon \tilde \varphi^{(1)}_{A,m\omega} )
\end{align}
so that, viewing $\tilde{\mathcal W}_A$ as a frequency-dependent linear operator,
\begin{align}
\mathcal W_A(\varphi) \approx &  \left[ \tilde {\mathcal W}^{(1,0)}_{A,m\omega}[\tilde \varphi^{(0)}_{m\omega}] 
+ \epsilon \omega^{(1)} (\partial_\omega \tilde {\mathcal W}^{(1,0)}_A)_{m\omega}[\tilde \varphi^{(0)}_{m\omega}] 
\right.
\notag \\ & 
\left.
+ \epsilon \tilde {\mathcal W}^{(1,0)}_{A,m\omega}[\tilde \varphi^{(1)}_{m\omega}]
\right] e^{-i\omega^{(0)}t}e^{-im\phi} \,.
\end{align}
The parts $\omega^{(1)}$ and $\tilde \varphi^{(1)}_{A,m\omega}$ are in direct analogy to the shifts of the quantum mechanical eigenvalues and wavefunctions in time-independent perturbation theory.
The leading order term vanishes by definition of the unperturbed modes.

At the next order then we find
\begin{align}
\label{eq:FieldPerts}
\omega^{(1)} & (\partial_\omega \tilde {\mathcal W}^{(1,0)}_A)_{m\omega}[\tilde \varphi^{(0)}_{m\omega}] + \tilde{\mathcal U}_{A,m\omega}[\tilde \varphi^{(0)}_{m\omega}] 
\notag \\ 
& + \tilde {\mathcal W}^{(1,0)}_{A,m\omega}[\tilde \varphi^{(1)}_{m\omega}] = 0 \,,
\end{align}
where we have defined the operator $\tilde{\mathcal U}_{A,m\omega}$ via
\begin{align}
\tilde{\mathcal U}_{A,m\omega}[\tilde \varphi^{(0)}_{m\omega}] & = 
2 \tilde{\mathcal W}^{(2,0)}_{A,m\omega}[ \vartheta^{(1)}, \tilde \varphi^{(0)}_{m\omega}]  -  \tilde \rho^{(1,0)}_{A,m\omega} [\tilde \varphi^{(0)}_{m\omega}]  \,
\end{align}
and all quantities are evaluated at the unperturbed QNM frequency $\omega^{(0)}$.
By the symmetry of the background metric, $\vartheta^{(1)}_A$ must be independent of $t$ and $\phi$ and so do not mix frequencies or azimuthal modes.

\begin{figure}
\includegraphics[width=1.0\columnwidth]{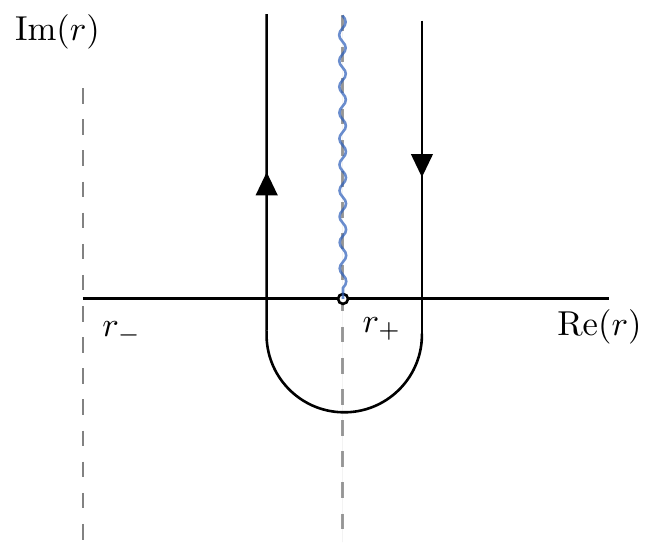} 
\caption{Depiction of the radial integration contour in the complex plane. The QNM wavefunctions have a branch point at the outer horizon $r_+$, which the contour wraps.}
\label{fig:Contour}
\end{figure}

In order to isolate the frequency shifts $\omega^{(1)}$ we apply the same technique used in quantum mechanics: we define a product on QNM wavefunctions with respect to which the leading-order wave operator is self-adjoint,
\begin{align}
\langle \tilde \psi^A | \tilde{\mathcal W}^{(1,0)}_{A,m\omega}[\tilde \xi] \rangle = \langle \tilde{\mathcal W}^{(1,0)}_{A,m\omega}[\tilde \psi] | \tilde \xi^A \rangle \,.
\end{align}
This product must also be finite.
The first requirement is accomplished by integrating the wavefunctions in $r$ and $\theta$ with respect to a weight $w(r,\theta)$ that can be chosen to enforce that $\tilde{\mathcal W}^{(1,0)}_{A,m\omega}$ is self-adjoint.
Ensuring the product is finite is not completely straightforward, since the QNM wavefunctions blow up at the horizon and spatial infinity on slices of constant Boyer-Lindquist time $t$.
However, a trick introduced by Leaver works~\cite{Leaver:1986gd}: we promote $r$ to a complex variable and deform the radial integration contour into the complex plane, wrapping around the outer horizon, where the QNM wavefunctions have a branch point.
By placing both ends of the contour in the upper half plane where the QNM wavefunctions decay exponentially, the integral can be regulated.
The contour is illustrated in Fig.~\ref{fig:Contour}.

We take the product of Eq.~\eqref{eq:FieldPerts} with the leading solution $\tilde \varphi^{(0)}_m$ and use the self-adjoint property of the product to eliminate the term that depends on $\tilde \varphi^{(1)}_{A,m\omega}$. 
The resulting shift is
\begin{align}
\omega^{(1)} & = - \frac{\langle \tilde \varphi^{(0)A}_{m \omega} | \tilde{\mathcal U}_{A,m\omega}[\tilde \varphi^{(0)}_{m \omega}]\rangle}{\langle \tilde \varphi^{(0)A}_{m \omega} |  (\partial_\omega \tilde {\mathcal W}^{(1,0)}_{A,m\omega})[\tilde \varphi^{(0)}_{m \omega}]\rangle} \,.
\end{align}
Given the equilibrium field $\vartheta^{(1)}_A$, the operator $\tilde{\mathcal U}_{A,m\omega}$, and the susceptibility of the leading wave equations $\partial_\omega \tilde{\mathcal W}_{A,m\omega}^{(1,0)}$, the leading order QNMs can thus be used to compute the frequency shifts.

It should be noted that if Eq.~\eqref{eq:LeadingFieldFreq} does not separate, solving for the mode wavefunctions may be computationally challenging, requiring a two-dimensional elliptical solve for each frequency and angular mode, while seeking the particular frequencies that satisfy the boundary conditions. 
Methods to solve for QNMs in similar circumstances have been implemented in many studies, see e.g.~\cite{Dias:2015wqa} and the references therein. 
If numerical, rather than series solutions are employed, an alternative regularization technique is likely needed for the radial integrals.
Finally, we note that we have been fairly careless in specifying how the abstract indices $A$ should be treated in the inner products, as this depends on the particular problem at hand.
For a single scalar field, as with dCS and sGB gravity, the situation is trivial, while for vector fields the Maxwell equations can be projected into the NP language, and once again the problem reduces to the treatment of scalar fields with common modes.
We turn next to the treatment of gravitational perturbations.

\subsection{Modified Teukolsky equation}
\label{sec:TeukProject}

With the method outlined, the next step is to recast the partially coupled gravitational equations \eqref{eq:FieldsCoupled} and~\eqref{eq:GWCoupled} into a form amenable to the EVP method.
The challenge is that there is no known gauge where the metric perturbations of Kerr separate, allowing for convenient computation.
Instead, the Teukolsky formalism provides a separable master equation for spin-weighted scalars $\psi_s$, which are directly proportional to the Weyl curvature scalars $\Psi_0$ and $\Psi_4$~\cite{Teukolsky:1973ha}.
The master equation is derived using the NP formulation of the field equations and Bianchi identities~\cite{Newman:1961qr,Stephani:2003tm}.
Separation of the scalars allows for easy computation of the QNM frequencies $\omega_{\ell m n}$ and wavefunctions, e.g.~using Leaver's method~\cite{Leaver:1985ax}.
In turn, the technique of metric reconstruction allows for the recovery of metric perturbations $h_{ab}$ corresponding to a given curvature perturbation $\psi_s$ in a radiation gauge~\cite{Chrzanowski:1975wv,Wald:1978vm,Ori:2002uv,Keidl:2010pm}. 
The simplest approach works provided there are no sources where the metric is reconstructed, which is the case of interest for us, since $h^{(0)}_{ab}$ are the usual QNMs in vacuum (the impact of the extended fields $\vartheta_A$ enters at higher orders in $\epsilon$).
A method for metric reconstruction in the Lorenz gauge has also been recently derived~\cite{Dolan:2021ijg}.

These conveniences motivate us to recast Eq.~\eqref{eq:GWCoupled} in terms of the Teukolsky scalars $\psi_s$.
For this we make use of the operator formalism introduced by Wald~\cite{Wald:1978vm}.
Let $\mathcal T$ be the linear differential operator that takes a metric perturbation of Kerr $h_{ab}$ into the spin-weighted scalar $\psi_s$.
Then $\psi_s$ obeys the Teukolsky master equation
\begin{align}
\mathcal O \mathcal T[h_{cd}] = \mathcal O[\psi_s] = \mathcal S^{ab}  \mathcal E_{ab}[h_{cd}]\,,
\end{align}
where $\mathcal S^{ab}$ is linear differential operator that can be read off the right hand side of Teukolsky's equation.
The derivation of the Teukolsky equation can be seen as an operator identity which applies for any rank two tensor field $h_{ab}$.
Just as $\mathcal O$ depends on which spin-weighted scalar is considered ($\psi_2$ corresponding to the Weyl scalar $\Psi_0$ and $\psi_{-2}$ corresponding to $\Psi_4$), the projection operator $S^{ab}$ depends on which spin-weight is considered.
Either choice $s = \pm 2$ can be used since, up to gauge and shifts in the mass and spin of the Kerr black hole, all information about the perturbations are present in either quantity~\cite{Wald:1973a}.
The expression for $\mathcal S^{ab}$ is succinctly provided by using the Geroch-Held-Penrose (GHP) formalism~\cite{Geroch:1973am} in Eq.~(B5.a) of \cite{Toomani:2021jlo}.
When using that expression, note that the usual factor of $\kappa_0 = 8\pi$ appearing in front of the stress-energy tensor in the Einstein field equations has been absorbed into $T_{ab}$.

To see that we can apply the operator formalism to our problem, consider the ansatz that $g_{ab} = g^{(0)}_{ab} + \eta h_{ab}$ and linearize the Einstein tensor in $h_{ab}$ about the Kerr background $g^{(0)}_{ab}$.
Similarly, linearize the Bianchi identities about the background.
Neither the definition of the Einstein tensor nor the Bianchi identities rely on the Einstein field equations, and their projection into the NP equations similarly just relies on choosing a null tetrad.
In addition, the commutation relations used by Teukolsky in deriving the master equation hold provided the directional derivatives and spin coefficients are defined on the Kerr background.
Thus, none of the steps in the derivation of $\mathcal O \mathcal T = \mathcal S^{ab} \mathcal E_{ab}$ change, provided all of these operators are defined on the Kerr background.

With the operator identity in hand, we apply $S^{ab}$ to Eq.~\eqref{eq:GWCoupled} and use the identity to write
\begin{align}
\label{eq:ModifiedMaster}
\mathcal O[\psi_s^{(0)}] + & \epsilon^2  \mathcal O[\psi_s^{(2)}] + \epsilon^2 \mathcal V[h^{(0)}] + \epsilon^2 \mathcal C[\varphi^{(1)}] = 0\,,
\\
\label{eq:CorrectionMaster}
\mathcal V[h^{(0)}] = & \mathcal S^{ab}(2G^{(2)}_{ab}[h^{(0)},g^{(2)}] - T^{\vartheta(2,1)}_{ab}[\vartheta^{(1)},\vartheta^{(1)},h^{(0)}]
\notag \\ & 
- V^{\rm{int}(1,1)}_{ab}[\vartheta^{(1)}, h^{(0)}]
) \,,
\\
\label{eq:CouplingOp}
\mathcal C[\varphi^{(1)}] = & - \mathcal S^{ab} 
(2T^{\vartheta(2,0)}_{ab}[\vartheta^{(1)},\varphi^{(1)}]  
+  V^{\rm{int}(1,0)}_{ab}[\varphi^{(1)}]
) \,,
\end{align}
Equations~\eqref{eq:ModifiedMaster}--\eqref{eq:CouplingOp} are a modified Teukolsky equation.
They incorporate both corrections to the leading-order expression due to the deformation to the background $g^{(2)}_{ab}$ and through the modified dynamics, which couple the gravitational waves to the field degrees of freedom.
It is not fully decoupled from~\eqref{eq:FieldsCoupled}, but as discussed previously Eq.~\eqref{eq:ModifiedMaster} can be solved at $O(1)$ in the usual manner for the separable QNM wavefunctions and frequencies.
These in turn can be used to reconstruct the leading order QNM metric perturbations $h^{(0)}_{ab}$, for example using the methods of~\cite{Keidl:2010pm,Nichols:2012jn}.
Those can be used to solve for the fields $\varphi^{(1)}_A$, which therefore can be viewed as being given by complicated linear operators on $h^{(0)}_{ab}$ (i.e. through convolution with a Green's function).
Finally, the solutions $\varphi^{(1)}_A$ can be fed back into Eq.~\eqref{eq:ModifiedMaster} to solve for the $O(\epsilon^2)$ corrections to the waves.

\subsection{Shifts for the gravitational wave QNMs}

Now that we have a modified Teukolsky equation, we can repeat our expansions in terms of frequency and angular harmonics and apply EVP theory.
We encounter a new conceptual issue as compared to the scalar case in Sec.~\ref{sec:ScalarShifts} and to previous applications of the EVP formalism to the shifts of QNM frequencies in Kerr.
The complication is that the modified Teukolsky equation for the complex scalar $\psi_s$ depends on the real quantity $h^{(0)}_{ab}$.
Schematically, given a QNM $\psi_s$, the reconstructed $h^{(0)}_{ab}$ involves a linear combination of both $\psi^{(0)}_s$ and its complex conjugate $\psi^{(0)*}_s$,
\begin{align}
h^{(0)}_{ab} & = \mathcal K_{ab}[\psi^{(0)}_s] + \mathcal K_{ab}^*[\psi^{(0)*}_s] \,,
\end{align}
where $\mathcal K_{ab}$ is implicitly defined through this expression, and is a linear operator used to carry out metric reconstruction.
This combination of operators inevitably mixes together two closely related families of QNM solutions in the harmonic expansion of Eqs.~\eqref{eq:ModifiedMaster}--\eqref{eq:CouplingOp}.

To proceed we divide the QNM frequencies into two sets according to the sign of their real part, $\omega^+_{\ell m n}$ and $\omega^-_{\ell m n}$. 
In Kerr, each QNM frequency with positive real part is paired with a corresponding mode with the same imaginary part but negative real part, and with opposite magnetic quantum number $m$ (see e.g.~\cite{Cook:2014cta}).
This means these modes obey
\begin{align}
\omega^+_{\ell m n} & = \Omega_{\ell m n} - i \gamma_{\ell m n} \,, \\
\omega^-_{\ell -m n} & = - \Omega_{\ell m n} - i \gamma_{\ell m n} \,,
\end{align}
where we have denoted the real part of the frequency as $\Omega_{\ell m n}$ and the decay rate of the mode as $\gamma_{\ell m n}$.
Hence $\omega^+_{\ell m n}  = - (\omega^-_{\ell -m n} )^*$.
These frequency pairs can be viewed as degenerate eigenvalues for perturbations of Kerr, and to apply the EVP method on the deformed Kerr spacetime we must consider combinations of both modes.

Consider the situation where the Weyl scalars are made up of a single pair of positive and negative frequency harmonics, of the form
\begin{align}
\label{eq:PsiCombo}
\psi_s & = \psi^{+}_s + \alpha^* \psi^{-}_s \,, \\
\psi^{+}_s & = \tilde \psi^+_{s \ell m n}(r,\theta) e^{-i \omega^+_{\ell m n}t + i m \phi}  \,, \\
\psi^{-}_s & = \tilde \psi^-_{s \ell -m n}(r,\theta) e^{-i \omega^{-}_{\ell -m n} t - i m \phi}  \,.
\end{align}
Here $\alpha$ is a complex constant, and we have absorbed an overall amplitude and phase into the definition of the wavefunctions, so that all that matters is their relative amplitude and phase.
We use the complex conjugate $\alpha^*$ for later convenience.
As we did when computing the frequency shifts for $\varphi_A$, we divide the perturbations to the QNMs into frequency shifts and perturbations to the wavefunctions, so that
\begin{align}
\label{eq:PsiExpand}
\psi^{+}_s 
&\approx \psi^{+(0)}_s + \epsilon^2 \psi^{+(2)}_s \notag \\
& \approx \exp[-i(\omega^{+(0)}_{\ell m n} + \epsilon^2 \omega^{+(2)}_{\ell m n})t  + i m \phi](\tilde \psi^{+(0)}_{s\ell mn} + \epsilon^2 \tilde \psi^{+(2)}_{s\ell mn}) 
\end{align} 
and similarly for $\psi^-_s$.
The leading order wavefunctions are given by
\begin{align}
\tilde \psi^{+(0)}_{s\ell m n}& =  {}_sR_{\ell m \omega}(r) {}_sS_{\ell m \omega}(\theta) \,, \\
\tilde \psi^{-(0)}_{s\ell -m n}& =  {}_sR_{s \ell -m -\omega^*}(r) {}_sS_{\ell -m -\omega^*}(\theta) \,,
\end{align}
where ${}_sR_{\ell m \omega} $ solves the radial Teukolsky equation for a QNM frequency $\omega^{+(0)}_{\ell m n}$, and ${}_sS_{\ell m \omega}$ is the corresponding spin-weighted spheroidal harmonic~\cite{Teukolsky:1973ha}.
Similarly, ${}_sR_{s\ell -m -\omega^*}$ is the paired wavefunction with $-m$ for the magnetic quantum number, and $-(\omega^{+(0)}_{\ell m n})^* =  \omega^{-(0)}_{\ell -m n}$ inserted for the frequency, with $S_{s\ell -m -\omega^*}$ the corresponding spin-weighted spheroidal harmonic.

Viewing the additional fields $\varphi^{(1)}_{A}$ as linear functionals of $h^{(0)}_{ab}$, so that $\varphi^{(1)}_A = \varphi_A[h^{(0)}]$, Eq.~\eqref{eq:ModifiedMaster} becomes
\begin{align}
\label{eq:PairedSystem}
& \mathcal O[\psi^{+(0)}_s + \epsilon^2 \psi^{+(2)}_s] + \epsilon^2 \left( \mathcal F[\psi^{+(0)}_s] + \alpha \mathcal G[\psi^{-(0)}_s{}^*] \right)
\notag \\ & 
+ \alpha^* \mathcal O[\psi^{-(0)}_s + \epsilon^2 \psi^{-(2)}_s] + \epsilon^2 \left(\alpha^* \mathcal F [\psi^{-(0)}_s] + \mathcal G[\psi^{+(0)}_s{}^*]\right) 
\nonumber \\
& = 0 \,.
\end{align}
This expression is organized so that the terms on the first line and the second line must vanish independently once expanded in time and angular harmonics.
We have defined
\begin{align}
\mathcal F & = (\mathcal V + \mathcal C \varphi)\mathcal K \,, \\
\mathcal G & = (\mathcal V + \mathcal C \varphi)\mathcal K^* \,. 
\end{align}
Next we expand Eq.~\eqref{eq:PairedSystem} in harmonics.

By now the rapidly multiplying decorations on each quantity have become unmanageable, so from here we leave the $s$, $\ell$  and $n$ indices implicit.
As before, we write the action of a linear operator on a positive-frequency harmonic expansion as
\begin{align}
\mathcal O[\tilde \psi^+_m e^{-i\omega^+ t + i m \phi }] & = e^{-i\omega^+ t + i m \phi} \tilde{\mathcal O}_{m\omega}[\tilde \psi^+_m] \,.
\end{align}
Importantly, we define frequency-domain operators $\tilde {\mathcal O}_{m \omega}$ as evaluated on the positive frequency modes $\omega^+_m$ and their corresponding $m$, so that we do not need to further specify which set of frequencies they are evaluated on. 
Then, when expanding operators on the negative frequency modes, we exploit the relationship between the positive and negative frequency modes to write, for example,
\begin{align}
\mathcal O[\tilde \psi^{-}_{-m} e^{-i\omega^- t - i m \phi }] & = e^{-i\omega^- t - i m \phi} \tilde{\mathcal O}_{-m-\omega^*}[\tilde \psi^-_{-m}] \,,
\end{align}
Since at our level of approximation, the operators are always evaluated on the leading order frequencies, there is no ambiguity in relating positive and negative frequency modes.

With this, the action of the Teukolsky operator on $\psi^\pm$, when expanded in $\epsilon$, are
\begin{align}
\mathcal O[\psi^+] \approx 
& \epsilon^2 (\omega^{+(2)}_{m} (\partial_\omega \tilde{\mathcal O})_{m \omega}[\psi^{+ (0)}_{m}] 
+ \tilde{\mathcal O}_{m \omega}[\psi^{+ (2)}_{m}] )\,, \\
\mathcal O[\psi^-] \approx 
& \epsilon^2 ( \omega^{-(2)}_{-m} (\partial_\omega \tilde{\mathcal O})_{-m -\omega^*}[\psi^{-(0)}_{-m}] 
\nonumber \\ &
+ \tilde{\mathcal O}_{-m -\omega^*}[\psi^{-(2)}_{-m}]) \,,
\end{align}
where
\begin{align}
(\partial_\omega \tilde{\mathcal O})_{-m -\omega^*} 
= [\partial_\omega(\tilde{\mathcal O}_{m \omega})]_{m \to -m, \omega \to - \omega^*} \,.
 \end{align}
With this, we expand the modified Teukolsky equation in harmonics and equate each independent harmonic to zero. 
This gives the following equations:
\begin{align}
\label{eq:Eigen1}
&  \omega^{+(2)}_{m} (\partial_\omega \tilde{\mathcal O})_{m \omega}[\psi^{+ (0)}_{m}]  
+ \tilde {\mathcal F}_{m\omega} [\tilde \psi^{+(0)}_{m}] 
+ \alpha \tilde{\mathcal G}_{m\omega}[(\psi^{-(0)}_{-m})^*] 
\notag \\ &
+ \tilde{\mathcal O}_{m\omega}[\tilde \psi^{+(2)}_{m}] = 0 \,,
\\
\label{eq:Eigen2}
& \alpha^* \omega^{-(2)}_{-m} (\partial_\omega \tilde{\mathcal O})_{-m -\omega^*}[\psi^{-(0)}_{-m}]  
+ \alpha^* \tilde {\mathcal F}_{-m -\omega^*} [\tilde \psi^{-(0)}_{-m}] 
\notag \\ 
&
+ \tilde{\mathcal G}_{-m -\omega^*} [(\tilde \psi^{+(0)}_{m})^*] 
+ \alpha^* \tilde{\mathcal O}_{-m-\omega^*}[\tilde \psi^{-(2)}_{-m}] = 0 \,.
\end{align}
In addition, we have the complex conjugates of these equations, which forms an eigensystem to solve for the frequency shifts. 

In fact, the system to use is Eq.~\eqref{eq:Eigen1} and the complex conjugate of Eq.~\eqref{eq:Eigen2}.
The zeroth order frequencies from those expressions are $\omega^{+(0)}_{\ell m n}$ and $-(\omega^{-(0)}_{\ell -m n})^*$, which are equal as noted. 
Hence the problem is one of degenerate perturbation theory: we seek two independent linear combinations of $\psi^{+}_s$ and $(\psi^{-}_s)^*$, which can be viewed as the correct modes whose frequencies are shifted.
This is the reason for making the prefactor $\alpha^*$ explicit in Eq.~\eqref{eq:PsiCombo}.
To proceed, we set 
\begin{align}
\omega^{+(2)}_{m} = -(\omega^{-(2)}_{-m})^* \coloneqq \omega^{(2)}_m
\end{align}
and solve for the ratio of amplitudes $\alpha$ that allows for a consistent solution for $\omega^{(2)}_m$.

First, we must eliminate the perturbed wavefunctions $\tilde \psi^{+(2)}_s$ and $(\tilde \psi^{+(2)}_s)^*$, which we accomplish as in Sec.~\ref{sec:ScalarShifts}, 
by left multiplication by the appropriate zeroth-order wavefunctions, and using contour integration with a weight to make $\tilde{\mathcal O}_{m\omega}$ and $(\tilde{\mathcal O}_{-m-\omega^*})^*$ self-adjoint.
After doing so, we make the convenient definitions
\begin{align}
\langle \delta \mathcal O_+ \rangle & = 
\langle \psi^{+(0)}_m | (\partial_\omega \tilde{\mathcal O})_{m\omega}[\psi^{+(0)}_m]\rangle \,, \\
\langle \mathcal F_+ \rangle & = 
\langle \tilde \psi^{+(0)}_{m} |\tilde {\mathcal F}_{m\omega} [\tilde \psi^{+(0)}_{m}]\rangle \,,\\
\langle \mathcal G_+ \rangle & =  \langle \psi^{+(0)}_{m} | \tilde{\mathcal G}_{m\omega}[(\psi^{-(0)}_{-m})^*] \rangle \,,
\\
\langle \delta \mathcal O_- \rangle & = 
\langle \psi^{-(0)}_{-m} | (\partial_\omega \tilde{\mathcal O})_{-m -\omega^*}[\psi^{-(0)}_{-m}] \rangle^* \,, \\
\langle \mathcal F_- \rangle & = 
\langle \tilde \psi^{-(0)}_{-m} | \tilde {\mathcal F}_{-m -\omega^*} [\tilde \psi^{-(0)}_{-m}] \rangle^*  \\
\langle \mathcal G_- \rangle & = \langle  \psi^{-(0)}_{-m} | \tilde{\mathcal G}_{-m -\omega^*} [(\tilde \psi^{+(0)}_{m})^*] \rangle^* \,.
\end{align}
In terms of these, our system is
\begin{align}
\omega^{(2)} \langle \delta \mathcal O_+ \rangle + \langle \mathcal F_+ \rangle + \alpha \langle \mathcal G_+ \rangle & = 0 \,, \\
- \alpha \omega^{(2)}  \langle \delta \mathcal O_- \rangle + \alpha \langle \mathcal F_- \rangle + \langle \mathcal G_- \rangle & = 0 \,.
\end{align}
We get a consistent solution provided $\alpha$ obeys the quadratic equation
\begin{align}
\label{eq:QuadraticAlpha}
\alpha^2 
+ \left[
\frac{\langle \mathcal F_+\rangle}{\langle \mathcal G_+\rangle}
+ \frac{\langle \mathcal F_-\rangle \langle \delta \mathcal O_+ \rangle}
{\langle \mathcal G_+\rangle \langle \delta \mathcal O_- \rangle}
 \right] \alpha + 
\frac{\langle \mathcal G_- \rangle \langle \delta \mathcal O_+ \rangle}
{\langle \mathcal G_+ \rangle \langle \delta \mathcal O_- \rangle}
= 0 \,.
\end{align}
The solution is
\begin{align}
\label{eq:GravShiftSln}
\omega^{(2)} & = -\frac{\langle \mathcal F_+\rangle + \alpha \langle \mathcal G_+ \rangle }{\langle \delta \mathcal O_+ \rangle}\,.
\end{align}

We see that in general there are two solutions for how the QNM frequency pair is perturbed, splitting the degenerate positive and negative frequency modes into two distinct linear combinations of modes.
The shift of Eq.~\eqref{eq:GravShiftSln} can be further refined using the two solutions to Eq.~\eqref{eq:QuadraticAlpha}.
However, the current form is useful also in the (presumably rare) cases where the positive and negative frequency modes do not couple in the modified Teukolsky equation, as is the case for perturbations of weakly-charged Kerr Newman black holes.

In Kerr, the presence of the modes with $\omega^-_{\ell m n}$ with their relation to the $\omega^+_{\ell m n}$ modes guarantees that pairs of QNMs with opposite parity, even and odd, share the same frequency, see Appendix~C of \cite{Nichols:2012jn}. 
Thus the pairing is a manifestation of the famous isospectrality of axial and polar perturbations of Schwarzschild (see e.g.~\cite{Berti:2009kk}) in the Kerr spacetime. 
The splitting of these modes under a generic perturbation appears to be related to a breaking of isospectrality under generic deformations of Kerr.

This completes our derivation of the QNM shifts.
Much remains to be desired, including the selection of a specific metric reconstruction approach and the simplifications of the various operators defined implicitly in this subsection.
We leave this and a further discussion of issues of QNM parity and isospectrality breaking to future work.

\section{Newman Penrose approach to a generalized Teukolsky equation}
\label{sec:NPApproach}

An alternative way to get the corrections to the Teukolsky operator of Eq.~\eqref{eq:ModifiedMaster} is to redo the derivation in the original paper by Teukolsky~\cite{Teukolsky:1973ha}, keeping all the terms that were set to zero based on assumptions about the background spacetime and chosen tetrad. 
We give the basic character of the results here, but defer the derivation and full result to Appendix~\ref{sec:NPApp}.
For completeness, we state the ingredients that go into deriving the Teukolsky equation. 
For this derivation, we consider a splitting into ``background'' perturbations which are $O(\eta^0)$ and ``dynamical'' perturbations which are $O(\eta)$.
Thus, the background in principle includes stationary deformations to Kerr, such as the $O(\epsilon)$ corrections $g^{(2)}_{ab}$ discussed previously, and all background NP quantities are defined with respect to this background metric and an appropriate null tetrad.
The dynamical degrees of freedom then propagate on this background, and are due to the $O(\eta)$ perturbations $h_{ab}$ to the metric (i.e.~due to the QNMs, including $O(\epsilon^2)$ corrections to them) and to the dynamical fields $\varphi_A$.

First one needs two Bianchi identities in the NP formalism, 
\begin{align}
	\label{Bianchi1}
	(\delta^{*}-4 \alpha+& \pi)\Psi_{0} -(D-4 \rho-2 \epsilon) \Psi_{1}-3 \kappa \Psi_{2} 
	= \notag \\  
	& \left(\delta+\pi^{*}-2 \alpha^{*}-2 \beta\right) \Phi_{00} 
	-\left(D-2 \epsilon-2 \rho^{*}\right) \Phi_{01} 
	\notag \\  
	&+2 \sigma \Phi_{10}-2 \kappa \Phi_{11}-\kappa^{*} \Phi_{02}\,,
\end{align}
and
\begin{align}
	\label{Bianchi2}
	(\Delta-4 \gamma+&\mu) \Psi_{0}-(\delta-4 \tau-2 \beta) \Psi_{1}-3 \sigma \Psi_{2} = 
	\notag\\ 
	& \left(\delta+2 \pi^{*}-2 \beta\right) \Phi_{01}
	-\left(D-2 \epsilon+2 \epsilon^{*}-\rho^{*}\right) \Phi_{02} 
	\notag\\  
	& -\lambda^{*} \Phi_{00}+2 \sigma \Phi_{11}-2 \kappa \Phi_{12}\,,
\end{align}
as well as one spin coefficient equation,
\begin{align}
	\label{DdeltaCommutator}
	(D&-\rho-\rho^{*}-3 \epsilon+\epsilon^{*}) \sigma 
	\notag \\ 
	& -\left(\delta-\tau+\pi^{*}-\alpha^{*}-3 \beta\right) \kappa-\Psi_{0} =0\,.
\end{align}
Lastly, we require a modification to the commutator identity used by Teukolksy,
\begin{align}
	\label{TypeDCommutatorFull}
	\left[D-(p+1) \epsilon+\epsilon^{*}\right.\left.+q \rho-\rho^{*}\right](\delta-p \beta+q \tau) - \nonumber \\
	\left[\delta-(p+1) \beta-\alpha^{*}+\pi^{*}+q \tau\right](D-p \epsilon+q \rho)=E_{p,q}\,,
\end{align}
where
\begin{align}
	E_{p,q} = & \sigma \delta^*-\kappa \Delta 
	+ q[
	(\tau^*+\pi-\bar{\beta}+3 \alpha) \sigma 
	\notag\\
	& +(\mu^*-\mu-\gamma^*-3 \gamma) \kappa
	+2 \Psi_{1}] -  
	\notag\\
	&p[
	(\alpha+\pi) \sigma 
	+(-\gamma-\mu) \kappa 
	+\Psi_{1}
	] \,,
\end{align}
for any constants $p$ and $q$, as derived in Appendix~\ref{sec:NPApp}.
For backgrounds where $\kappa = \sigma = 0 $ and $ \Psi_1 = 0 $, the corrections $E_{p,q}$ to the original identity vanish.
In our derivation of the modified Teukolsky equation, no Ricci Identities are used, so any change to the equations of motion coming from beyond-GR effects can be absorbed into the Ricci scalars $\Phi_{ij}$, and so do not modify our derivation.
 
Similar to the derivation in~\cite{Teukolsky:1973ha}, we expand all the tetrads, NP scalars and derivatives into background and dynamical parts. 
Schematically, we write them as $\psi^A + \eta\psi^B$ for any NP quantity or derivative,
where the superscript $B$ denotes the wavelike perturbation of the quantity and the superscript $A$ denotes the background value of the quantity.
Expanding, we collect the $O(\eta)$ terms, since as before, the $O(\eta^0)$ equations must be satisfied by the background solution.

Simply stating the results we derived in Appendix~\ref{sec:NPApp} here, we find that schematically the modified Teukolsky equation takes the form
\begin{align}
	\label{TermsThatVanishOnKerr}
	\mathcal{O}^A[\psi^B_0]= T^A_{0}[\Phi^B_{ij}] + K
\end{align}
This notation follows that of~\cite{Teukolsky:1973ha}.
In Eq.~\eqref{TermsThatVanishOnKerr}, $\mathcal{O}^A$ and $T^A_0$ are made up of the same NP quantities as the Teukolsky master equation, except that they incorporate the $O(\epsilon^2)$ deformations to the metric and the corresponding corrections to the tetrad, specifically
\begin{align}
\mathcal{O}^A & = (D-3 \epsilon+\epsilon^{*}-4 \rho-\rho^{*})^A(\Delta-4 \gamma+\mu)^A \notag\\&- (\delta+\pi^{*}-\alpha^{*}-3 \beta-4 \tau)^A(\delta^{*}+\pi-4 \alpha)^A -3 \Psi^A_{2}\,,
\end{align}
and
\begin{align}
	T^A_0&[\Phi^B_{ij}] = \nonumber \\
	&[D-3 \epsilon+\epsilon^{*}-4 \rho-\rho^{*}]^A(\delta+2 \pi^{*}-2 \beta)^A \Phi^B_{01} \nonumber \\
	- &[D-3 \epsilon+\epsilon^{*}-4 \rho-\rho^{*}]^A(D-2 \epsilon-2 \rho^{*})^A \Phi^B_{02} \nonumber \\ 
	- &[\delta-3 \beta-\alpha^{*}+\pi^{*}-4 \tau]^A(\delta+\pi^{*}-2 \alpha^{*}-2 \beta)^A \Phi^B_{00}  \nonumber \\
	+ &[\delta-3 \beta-\alpha^{*}+\pi^{*}-4 \tau]^A\left(D-2 \epsilon+2 \epsilon^{*}-\rho^{*}\right)^A \Phi^B_{01}\,,
\end{align}
where $A$ denotes that the quantity is evaluated on the beyond Kerr background.
Meanwhile, the term $K$ [provided in~\eqref{eq:AllExtraTermsNP}] includes any additional modifications which cannot be captured in this way.
This equation is supplemented by an equation governing $\Psi_4^B$, which is the GHP dual of Eq.~\eqref{TermsThatVanishOnKerr}.

While it is true, and is shown in the appendix, that $K = 0$ on vacuum Type D backgrounds in relativity, 
for beyond-GR theories we do not expect that the Goldberg-Sachs theorem~\cite{Newman:1961qr} enforces the additional simplifications $\kappa = \sigma = 0$ (and their GHP dual relations) that usually arise for Type D spacetimes.
To avoid any ambiguity we spell out the conditions required for $K$ to vanish.
We must firstly have that 
\begin{align}
	\label{eq:TypeD-like-condition}
	\sigma^A = \kappa^A = \lambda^A= \Psi^A_0 = \Psi^A_1 = 0\,.
\end{align}
Next we must have a background that is a vacuum, specifically all the background Ricci scalars vanish, 
\begin{align}
	\label{eq:vacuum-condition}
	\Phi^A_{ij} = 0\,,
\end{align}
and lastly one needs that the background $\Psi_2$ follows the two equations
\begin{align}
\label{eq:vacuum-type-d-condition}
	D\Psi^A_2 = 3\rho^A\Psi^A_2 \quad \text{and} \quad \delta\Psi^A_2 = 3\tau^A\Psi^A_2 \,.
\end{align}
Since we are interested in beyond-GR theories, the condition of Eq.~\eqref{eq:vacuum-condition} does not hold in our case, but some or all of the conditions in Eq.~\eqref{eq:TypeD-like-condition} may hold, particularly when the background metric remains Type D, in which case $\Psi_0^A = \Psi_1^A = 0$ (as well as the dual relations $\Psi_3^A = \Psi_4^A = 0$).
The full equation we have derived does not assume any of these conditions, but the above conditions achieve the reduction of Eq.~\eqref{TermsThatVanishOnKerr} to the Teukolsky equation. 
When only a portion of the conditions hold, various contributions to $K$ vanish.

The simplicity of Eq.~\eqref{TermsThatVanishOnKerr} belies the underlying complexity of the terms in $K$, and the resultant equation couples to the wave perturbations of many other NP scalars, necessitating some tetrad-reconstruction procedure in general.
As with our alternate derivation in Secs.~\ref{sec:FieldEquations} and~\ref{sec:MasterEqn}, this equation also couples to the additional fields, which appear in the Ricci scalars, and must be supplemented with equations of motion for those fields.
For work that focuses on this approach to computing the QNM shifts, and exploits gauge and tetrad choices to further simplify the above expressions, we refer to~\cite{Li2022InPrep}.

\section{Conclusions}
\label{sec:Conclusions}

In this work we have derived a modified Teukolsky equation for gravitational perturbations in a broad class of beyond-GR theories, with the goal of computing the shifts to the QNM spectrum in such theories.
Our approach is primarily adapted to quadratic gravity models when the modifications to gravity are perturbative, such as dCS and sGB gravity in the decoupling limit.
However it can be modified to capture other cases, which we have illustrated by considering the case of weakly charged black holes.
The modified Teukolsky equation is coupled to additional fields, those which are non-minimally coupled to the curvature and source deformations to the background Kerr solution.
Our equation incorporates corrections from both the deformation to the Kerr background and the changes to the dynamics of the fields arising from the modified equations of motion.

By using as an ansatz that we seek solutions which perturb around the QNMs of Kerr black holes, we can partially decouple the additional fields from the gravitational QNMs. 
This allows for a hierarchical approach to computing the shifts to the QNM spectra as follows.
First one computes the unperturbed QNM wavefunction for a given mode on Kerr, including the reconstructed metric perturbation $h^{(0)}_{ab}$ for that mode.
This mode serves to source the additional fields $\vartheta_A^{(1)}$, usually a non-minimally coupled scalar field.
With the solution to this sourced scalar, and the unperturbed QNM wavefunction, the correction to the gravitational QNM frequency can be computed.
Finally, we illustrate how these equations can be used in a concrete expression for the gravitational QNM shifts, using EVP theory.
Along the way, we have shown that in general deviations from Kerr lift a degeneracy between positive and negative frequency modes, requiring degenerate perturbation theory to resolve the spectral shifts.
The connection between these degeneracies, parity breaking, and the loss of isospectrality will be the subject of future studies.

In this work we do not compute the QNM shifts for any particular theory.
Practical application of our formalism requires a number of nontrivial steps, which are the target of future work.
To apply our approach, we first must choose a beyond-GR theory, such as dCS or sGB, compute the non-minimally coupled fields which are sourced by the background Kerr curvature, and use these as a source for solving for the stationary metric deformation, which we denote $g^{(2)}_{ab}$ in this work.
Next, we require the solutions for the dynamical field degrees of freedom $\varphi^{(1)}_A$, generically sourced by the dynamical gravitational QNM $h^{(0)}_{ab}$.
With these elements in place it is straightforward to compute the QNM shifts.
However, our EVP approach requires the ability to evaluate these quantities for complex $r$, both inside and outside of the outer horizon $r_+$.
Series solutions for these quantities would be ideal for this purpose, particularly solutions which are nonperturbative in the black hole spin parameter $\chi$. 
If particular cases require direct numerical solutions for any of these quantities, our approach can be adapted by regularizing the required integrals by some other means.
For example, hyperboloidal slicing could provide a promising alternative approach~\cite{Zenginoglu:2011jz,PanossoMacedo:2019npm,Ripley:2022ypi}.
It would also be valuable to compare specific QNM shifts to those observed in numerical simulations of dCS and sGB binary black hole mergers~\cite{Okounkova:2019zjf,Okounkova:2020rqw}.

Our approach may also prove valuable in extending approaches that predict QNMs from generic, parametrized deviations from the Regge-Wheeler and Zerilli potentials around non-spinning and slowly spinning black holes~\cite{Cardoso:2019mqo,McManus:2019ulj}, and those that seek to reconstruct the deformations of the effective potential from the QNM shifts~\cite{Volkel:2022aca}.
The method presented here would allow for the mapping of specific, stationary deformations from the Kerr spacetime, such as those arising from ``bumpy'' black holes~\cite{Vigeland:2011ji}, onto QNM shifts.
However, without an underlying theory for how such deviations are supported, our derivation shows that contributions to the frequency shifts from modifications to the equations of motion and coupling to additional degrees of freedom would be missed.

The formalism presented here represents a first step towards a concrete prediction of the full QNM spectrum in specific theories beyond-GR, for Kerr black holes with arbitrary spin.
With such predictions, direct constraints on the coupling parameters can be derived by applying Bayesian inference on past and future gravitational wave detections.
Unlike the case of parametrized null tests, by using specific theories it is straightforward to combine a large number of detections in precision searches for beyond-GR effects in black hole ringdown.
In addition, if parametrized ringdown tests uncover a violation of the predictions of relativity, the ability to predict the shifts to the QNM spectra in particular theories is critical to identifying the physics underlying such deviations.
As we move into the era of precision gravitational wave physics, we can hope that such subtle deviations will point the way to a new paradigm for gravitation.

\acknowledgements
We thank Dongjun Li, Pratik Wagle, Yanbei Chen and Nico Yunes for valuable discussions. We especially thank Dongjun Li for detailed discussions of some of the results presented in~\cite{Li2022InPrep}, as well as discussions around degenerate EVP and definite-parity modes. 
AZ~thanks Yanbei Chen, Zachary Mark and Huan Yang for past discussions on the EVP method and its application to charged black holes.
AH and AZ are supported by NSF Grant Number PHY-1912578.
This research was supported in part by the National Science Foundation under Grant No.~NSF PHY-1748958. 

\appendix

\section{Operator $G^{(2)}$}
\label{sec:G2}

In order to compute the QNM shifts, the expression $G^{(2)}_{ab}[h^{(0)},g^{(2)}]$ is required, as in Eq.~\eqref{eq:GWCoupled}.
To derive it, we first consider the general problem of expanding the Einstein tensor $G_{ab}$ around a generic perturbation to a background metric, $g_{ab} = g^{(0)}_{ab} + h_{ab}$.
In what follows, all covariant derivatives are with respect to the background metric $g^{(0)}_{ab}$ and all indices are raised and lowered with the background metric.

First we expand the Riemann tensor in powers of $h_{ab}$,
\begin{align}
R_{ab}(g^{(0)}+ h) = R^{(0)}_{ab} + R^{(1)}_{ab}[h] + R^{(2)}_{ab}[h,h] + \dots \,,
\end{align}
where $R^{(0)}_{ab} = R_{ab}(g^{(0)})$ is the full Ricci tensor evaluated on the background metric, $R^{(1)}_{ab}$ is linear in $h_{ab}$ and $R^{(2)}_{ab}$ is quadratic in $h_{ab}$.
We have (e.g.~\cite{Misner:1973prb})
\begin{align}
R^{(1)}_{ab}[h]  = & \nabla^c \nabla_{(a} h_{b)c} - \frac 12 (\nabla^c \nabla_c h_{ab} + \nabla_a \nabla_b h^c{}_c) \,, \\
R^{(2)}_{ab}[h,h] = & \frac{1}{2}\left[ 
h^{cd}(\nabla_a \nabla_b h_{cd} + \nabla_c \nabla_d h_{ab} - 2\nabla_d \nabla_{(a}h_{b)c})
\right. \notag \\
& 
-(\nabla_c \bar h^{cd})(2\nabla_{(a} h_{b)d} - \nabla_d h_{ab}) \notag \\
& 
+ \frac{1}{2}(\nabla_a h_{cd})(\nabla_b h^{cd})
\notag \\
&
\left.
+ (\nabla^c h_{b}{}^d)(\nabla_c h_{ad} - \nabla_d h_{ac})
\right] \,,
\end{align}
where we have defined the trace-reversed perturbation $\bar h_{ab} = h_{ab} - (1/2) g^{(0)}_{ab}g_{(0)}^{cd} h_{cd}$.
The Einstein tensor is then
\begin{align}
G_{ab}(g) & = G^{(0)}_{ab} + G^{(1)}_{ab}[h] + G^{(2)}_{ab}[h,h] + \dots \,,
\end{align}
where $G^{(0)}_{ab} = G_{ab}(g^{(0)})$ is the full Einstein tensor on the background, and for example
\begin{align}
G^{(1)}_{ab}[h] & = R^{(1)}_{ab}[h] - \frac 12 \left(h_{ab} R^{(0)} +g^{(0)}_{ab} R^{(1)}[h] \right) \,,
\end{align}
where the Ricci scalars $R^{(0)}$, $R^{(1)}$ are defined as the trace of the Ricci tensors at each order.

Before discussing the next order, we specialize to the case where the background is vacuum, so that $R^{(0)}_{ab} = 0$.
Then $G^{(1)}_{ab}[h] = \mathcal E_{ab}[h]$ as given in Eq.~\eqref{eq:LinearEFE}.
If the background is not vacuum, $G^{(1)}_{ab}$ still gives the linearized Einstein equation for $h_{ab}$, but with additional terms present in Eq.~\eqref{eq:LinearEFE}.
With this specialization, the expression for $G^{(2)}_{ab}$ simplifies to
\begin{align}
G^{(2)}_{ab}[h,h] = R^{(2)}_{ab}[h,h] + \frac{1}{2}(g^{(0)}_{ab} h^{cd} R^{(1)}_{cd}[h] - \notag\\ h_{ab} R^{(1)}[h] - g^{(0)}_{ab}R^{(2)}[h,h]) \,.
\end{align}
Further, we are interested in cases where the perturbation $h_{ab}$ is a solution to the linearized equations, with or without source,
\begin{align}
\mathcal E_{ab}[h] = \tau_{ab} \,,
\end{align}
so that $R^{(1)}_{ab}[h]$ and $R^{(1)}[h]$ can be further reduced.
For example, when $h_{ab}$ is a QNM perturbation $h^{(0)}_{ab}$ as in Eq.~\eqref{eq:GWCoupled}, $R^{(1)}_{ab}[h^{(0)}] = 0$.
Meanwhile, when $h_{ab}$ is the static deformation $g^{(2)}_{ab}$, if convenient we can substitute Ricci terms for source terms,
\begin{align}
R^{(1)}_{ab}[g^{(2)}] & = \tau_{ab} - \frac{1}{2} g^{(0)}_{ab} \tau\,, \\
\tau_{ab} & = T^{\vartheta (2,0)}_{ab}[\vartheta^{(1)},\vartheta^{(1)}] + V^{\rm{int}(1,0)}_{ab}[\vartheta^{(1)}]\,.
\end{align}

With this we express $G^{(2)}_{ab}[h^{(0)},g^{(2)}]$ by taking $G^{(2)}_{ab}[h,h]$ and, for each quadratic term in $h_{ab}$, ensuring that one copy is replaced by $h^{(0)}_{ab}$ and one by $g^{(2)}_{ab}$, summing over both possible substitutions.
Further simplifications are made for the $R^{(1)}_{ab}$ terms in each case. 
We have
\begin{widetext}
\begin{align}
G^{(2)}_{ab}[h^{(0)},g^{(2)}]  = &  
\frac{1}{4}\left[ 
g_{(2)}^{cd}(\nabla_a \nabla_b h^{(0)}_{cd} + \nabla_c \nabla_d h^{(0)}_{ab} - 2\nabla_d \nabla_{(a}h^{(0)}_{b)c})
-(\nabla_c \bar g_{(2)}^{cd})(2\nabla_{(a} h^{(0)}_{b)d} - \nabla_d h^{(0)}_{ab})
+ \frac{1}{2}(\nabla_a g^{(2)}_{cd})(\nabla_b h_{(0)}^{cd})
\right.
\notag \\
&
\left.
+ (\nabla^c g^{(2)}_{b}{}^d)(\nabla_c h^{(0)}_{ad} - \nabla_d h^{(0)}_{ac})
\right] 
-\frac{1}{8}g^{(0)}_{ab}\left[ 
g_{(2)}^{cd}(\nabla_e \nabla^e h^{(0)}_{cd} + \nabla_c \nabla_d h^{(0)} - 2\nabla_d \nabla^e h^{(0)}_{ec})
\right.
\notag \\
&
\left.
-(\nabla_c \bar g_{(2)}^{cd})(2\nabla^e h^{(0)}_{ed} - \nabla_d h^{(0)})
+ \frac{1}{2}(\nabla_e g^{(2)}_{cd})(\nabla^e h_{(0)}^{cd})
+ (\nabla^c g_{(2)}^{ed})(\nabla_c h^{(0)}_{ed} - \nabla_d h^{(0)}_{ec})
\right] 
+ (g^{(2)}_{ab} \leftrightarrow h^{(0)}_{ab})
\notag \\ &
-\frac{1}{4} g^{(0)}_{ab} h_{(0)}^{cd}\tau_{cd} + \frac{1}{4} \bar h^{(0)}_{ab} \tau \,,
\end{align}
\end{widetext}
where the term $(g^{(2)}_{ab} \leftrightarrow h^{(0)}_{ab})$ indicates all the previous terms in the expression with the two types of perturbations exchanged.
The final contribution is asymmetric in the perturbation types because each is sourced differently, as noted above.

\section{Comparison to the Newman Penrose approach for weakly charged black holes}
\label{sec:KNApp}

Here we confirm that the $\epsilon$-scaling selected in Sec.~\ref{sec:KNExample} matches known results on the perturbation of weakly charged black holes.
We focus on matching a single term between our modified Teukolsky equation for $s = +2$ and the full perturbation equations derived in the NP formalism for the corresponding scalars, $\Psi_0$ and $\phi_0$.
We use the expressions given in~\cite{Dias:2015wqa}, hereafter DGS, and we cite the NP equations as given in~\cite{Stephani:2003tm}, hereafter SKMHH, whose definitions of the NP scalars are appropriate for our metric signature.
We denote the $O(\eta)$ dynamical perturbations to the geometry and EM fields with the superscript $B$, and all other quantities are assumed to be background quantities unless noted.
When necessary for clarity, these $O(\eta^0)$ quantities are decorated with the superscript $A$. 

Our point of comparison is the coupling term present in the GHP dual of DGS Eqs.~(3) and~(6), keeping in mind that we set $Q=0$ in these equations because we only need the leading order expressions.
Using the background NP relations, SKMHH Eqs.~(7.32e) and~(7.32h), $D \Psi_2 = 3 \rho \Psi_2 $ and $\delta \Psi_2 = 3 \tau \Psi_2 $, we cast the coupling term of DGS in the form
\begin{align}
\label{eq:DGSCoupling}
\Phi_{11} \mathcal Q_{2} \varphi_1  = 2\kappa_0 \phi_1^{*A}&\left[ (D - 2 \rho)(\delta - 2 \beta - 3 \tau) 
\right. 
\notag \\ &
\left.
+ (\tau - \pi^*)(D- 3 \rho)\right] \phi_0^B \,,
\end{align}
by commuting factors of $\Psi_2^{-1}$ through the directional derivatives.
Here we have chosen to set the NP spin coefficient $\epsilon = 0$ at leading order using our background tetrad, and
we have also selected a perturbation to tetrad such that $\Psi^B_1 = 0$.
We have also restored the factor $\kappa_0 = 8 \pi G$.
From the background EM fields, the Ricci scalars are $\Phi_{ij} = \kappa_0 \phi_i \phi_j^*$, so that only $\Phi_{11}$ and its complex conjugate are nonvanishing at leading order.\footnote{
The GHP dual of $\Phi_{11}$ is itself, and so it would appear that there should be no complex conjugate on the $\phi_1$ term in the denominator of the $\mathcal Q_{-2}$ operator in DGS Eq.~(6).}

In our approach, the coupling between the gravitational QNMs and the EM QNMs is given by the term
\begin{align}
2 \kappa_0 S^{ab} T^{(2,0)}_{ab}[\vartheta^{(1)},\varphi^{(1)}]
\end{align}
when projecting Eq.~\eqref{eq:KNGrav} using $S^{ab}$ and restoring $\kappa_0$.
To match the coupling term to Eq.~\eqref{eq:DGSCoupling}, we expand the projection operator in terms of NP quantities.
The only nonzero projection of $T^{(2,0)}_{ab}[\vartheta^{(1)}, \varphi^{(1)}]$ onto the tetrad is the one that involves a single copy of the background Maxwell scalar $\phi^A_1$ and the dynamical perturbation $\phi^B_0$, which is 
\begin{align}
T^{(2,0)}_{lm} = T^{(2,0)}_{ab}[\phi^A_1,\phi^B_0]l^a m^b  = 2 \phi_1^{*A}\phi_0^B \,.
\end{align}
With this, reading $S^{ab}$ off of the source term in~\cite{Teukolsky:1973ha}, we have
\begin{align}
\label{eq:ProjT1}
2 \kappa_0 S^{ab} T^{(2,0)}_{ab}&
= 2\kappa_0 \left[ 
(\delta + \pi^* - \alpha^* - 3 \beta - 4 \tau)(D-2 \rho^*)
\right.
\notag \\ & 
\left . 
+(D-4 \rho -\rho^*)(\delta + 2 \pi^* -2 \beta)
\right](\phi^{*A}_1 \phi^B_0) \,.
\end{align}

To make progress, we use Maxwell's equations on the background EM field,
\begin{align}
D\phi^{*A}_1 & = 2 \rho^* \phi^{*A}_1\,, & \delta \phi^{*A}_1 & = - 2\pi^* \phi^{*A}_1 \,,
\end{align}
to bring factors of $\phi^{*A}_1$ through the derivative operators in Eq.~\eqref{eq:ProjT1}.
Next we use the commutator relation SKMHH Eq.~(7.6b),
\begin{align}
\delta D \phi^B_0 & = D \delta \phi^B_0 + (\alpha^* + \beta - \pi^*)D \phi^B_0 - \rho^* \delta \phi^B_0 \,.
\end{align}
to match the ordering of derivatives in Eq.~\eqref{eq:DGSCoupling}.
The result is
\begin{align}
\label{eq:ProjT2}
2 \kappa_0 S^{ab} T^{(2,0)}_{ab}&
=2 \kappa_0 \phi^{*A}_1 \left[(D-2 \rho)\delta + (\tau - \pi^*)D \right.
\notag \\
& 
\left.
 - (\beta + 3 \tau) D - \beta \rho^* -  D \beta +4 \beta \rho
\right] \phi^B_0 \,. 
\end{align}
The remaining differences can be removed using the background NP equations,
\begin{align}
D\beta &=  \beta \rho^* \,, & D\tau & = (\tau + \pi^*)\rho\,,
\end{align}
bringing Eqs.~\eqref{eq:DGSCoupling} and~\eqref{eq:ProjT2} into agreement. 
This shows that the coupling terms are in agreement between the formalisms, and similar manipulations are expected to demonstrate agreement between the remaining terms.

\section{Details of the Newman Penrose approach to a generalized Teukolsky equation}
\label{sec:NPApp}

In this section we derive the modified Teukolsky equation directly from the NP formalism, and connect it to the discussion in Sec.~\ref{sec:NPApproach}. 
We expand the tetrads and all the resultant NP quantities and derivatives into background and wave parts. Background quantities are $O(\eta^0)$ and have the superscript $A$, while the wave perturbation of the quantities are $O(\eta)$ have the superscript $B$.

\subsection{Notation}
Before we move on we state some convenient notation that makes the resulting modified Teukolsky equation more succinct. In particular we create a new notation for some derivative quantities, since they frequently reappear in expressions.
We define 
\begin{align}
	\label{eq:OperatorsPQ}
	\delta_{p,q} &= (\delta-p \beta+q \tau)\,, \\
	D_{p,q} &= (D-p \epsilon+q \rho) \,,\\
	P_{p,q} &= \left[D_{p,q}-\epsilon+\epsilon^{*}-\rho^{*}\right] \,,\\
	Q_{p,q} &= \left[\delta_{p,q}- \beta-\alpha^{*}+\pi^{*}\right]\,.
\end{align}
We define further operators
\begin{align}
	\label{eq:RemainingOperatorsNP}
	\delta^*_{00}  &= \delta^* - 4\alpha + \pi\,, \\
	\Delta_{10}  &= \Delta - 4\gamma + \mu\,,
\end{align}
and operators acting on Ricci scalars,
\begin{align}
	\label{eq:C0Definition}
	C_{0}^{ij} [\Phi_{ij}] = & \left(\delta+\pi^{*}-2 \alpha^{*}-2 \beta\right) \Phi_{00}
	\nonumber \\ 
	&- \left(D-2 \epsilon-2 \rho^{*}\right) \Phi_{01}+2 \sigma \Phi_{10} 
	\nonumber \\ 
	&-2 \kappa \Phi_{11}-\kappa^{*} \Phi_{02}\,,
\end{align}
and
\begin{align}
	\label{eq:C1Definition}
	C_{1}^{ij} [\Phi_{ij}] = & \left(\delta+2 \pi^{*}-2 \beta\right) \Phi_{01} 
	\nonumber \\
	&-\left(D-2 \epsilon+2 \epsilon^{*}-\rho^{*}\right) \Phi_{02}-\lambda^{*} \Phi_{00} 
	\nonumber \\ 
	& +2 \sigma \Phi_{11}-2 \kappa \Phi_{12}\,.
\end{align}
This converts the Eqs.~\eqref{Bianchi1}--\eqref{TypeDCommutatorFull} into
\begin{align}
	\delta^*_{00} [\Psi_0] - \hat{D}_{2,-4}[\Psi_1] - 3\kappa\Psi_2 =  C_{0}^{ij} [\Phi_{ij}]\,, \label{Bianchi1Compact} \\
	\Delta_{10} [\Psi_0] - \hat{\delta}_{2,-4}[\Psi_1 ]- 3\sigma\Psi_2 =  C_{1}^{ij}[\Phi_{ij}]\,, \label{Bianchi2Compact}\\
	P_{2,-1}[\sigma] - Q_{2,-1}[\kappa] =  \Psi_0 \,, \label{SpinCoefficientCompact}\\
	P_{p,q} \delta_{p,q} - Q_{p,q}D_{p,q} = 0 + E_{p,q} \,. \label{OperatorIdentityCompact}
\end{align}
Note that Eq.~\eqref{OperatorIdentityCompact} is an operator identity. 

\subsection{Expanding around Kerr Background}

Expanding all the NP quantities and derivatives we get four new equations which are later combined to form the Teukolsky equation.

\begin{widetext}

\subsubsection{First Bianchi Identity}
Using the first Bianchi identity we get
\begin{align}
	\delta^{*\ A}_{00} [\Psi^B_0] - \hat{D}^A_{2,-4}[\Psi^B_1] - 3\kappa^B\Psi^A_2
	& =
	C_{0}^{ij A} [\Phi^B_{ij}]
	+C_{0}^{ij B} [\Phi^A_{ij}]
	-\delta^{*\ B}_{00} [\Psi^A_0] + \hat{D}^B_{2,-4}[\Psi^A_1] + 3\kappa^A\Psi^B_2 \,.
\end{align}
We reorder terms and put all the extra terms that are absent in the original derivation into the expression $S_0^{AB}$,
\begin{align}
	\label{Bianchi1New}
	\delta^{*\ A}_{00} [\Psi^B_0] - \hat{D}^A_{2,-4}[\Psi^B_1] - 3\kappa^B\Psi^A_2
	=
	C_{0}^{ij A} [\Phi^B_{ij}] + S_0^{AB}\,,
\end{align}
where
\begin{align}
	\label{eq:S0Definition}
	S^{AB}_{0}  &= C_{0}^{ij B} [\Phi^A_{ij}]
	-\delta^{*\ B}_{00} [\Psi^A_0] + \hat{D}^B_{2,-4}[\Psi^A_1] + 3\kappa^A\Psi^B_2 \,.
\end{align}

\subsubsection{Second Bianchi Identity}
With the second Bianchi identity we get another equation,
\begin{align}
	\Delta^{*\ A}_{10} [\Psi^B_0] - \hat{\delta}^{A}_{2,-4}[\Psi^B_1 ]- 3\sigma^B\Psi^A_2
	& =
	C_{0}^{ij A} [\Phi^B_{ij}]
	+C_{0}^{ij B} [\Phi^A_{ij}]
	 -\Delta^{*\ B}_{10} [\Psi^A_0] + \hat{\delta}^B_{2,-4}[\Psi^A_1 ]+ 3\sigma^A\Psi^B_2 \,.
\end{align}
We again reorder terms and put all the extra terms that are absent in the original derivation into the expression $S_1^{AB}$,
\begin{align}
	\label{Bianchi2New}
	\Delta^{*\ A}_{10} [\Psi^B_0] - \hat{\delta}^{A}_{2,-4}[\Psi^B_1 ]- 3\sigma^B\Psi^A_2
	= 
	C_{1}^{ij A} [\Phi^B_{ij}] + S_1^{AB}\,,
\end{align}
where
\begin{align}
	\label{eq:S1Definition}
S^{AB}_1 =  C_{1}^{ij B} [\Phi^A_{ij}]
-\Delta^{*\ B}_{10} [\Psi^A_0] + \delta^B_{2,-4}[\Psi^A_1 ]+ 3\sigma^A\Psi^B_2 \,.
\end{align}
Notice so far that for a Kerr background, $\Phi^A_{ij} = 0, \sigma^A = \kappa^A = 0$ and $\Psi^A_0 = \Psi^A_1 = 0$, which makes $S^{AB}_1 = 0$ and $S^{AB}_0 = 0$.
We keep these terms for our analysis. 

\subsubsection{The spin coefficient equation}

Expanding the spin coefficient equations we get
\begin{align}
	P^A_{2,-1}[\sigma^B] - Q^A_{2,-1}[\kappa^B] 
	+P^B_{2,-1}[\sigma^A] - Q^B_{2,-1}[\kappa^A]  
	=
	\Psi^B_0 \,.
\end{align}
Multiplying both sides by $\Psi^A_2$,
\begin{align}
	P^A_{2,-1}[\sigma^B]\Psi^A_2 - Q^A_{2,-1}[\kappa^B] \Psi^A_2
	+P^B_{2,-1}[\sigma^A]\Psi^A_2 - Q^B_{2,-1}[\kappa^A]  \Psi^A_2
	=
	\Psi^B_0 \Psi^A_2\,,
\end{align}
and using the product rule,
\begin{align}
P_{p,q}[fg] = P_{p,q}[f]g + D[g] f \,,\notag\\
Q_{p,q}[fg] = Q_{p,q}[f]g + \delta[g] f\,,
\end{align}
we can rearrange and get
\begin{align}
	\label{halfwaySpinCoefficient}
	P^A_{2,-1}[\sigma^B\Psi^A_2] - \sigma^B D[\Psi^A_2] 
	-Q^A_{2,-1}[\kappa^B\Psi^A_2] +  \kappa^B\delta[\Psi^A_2]
	&=
	\Psi^B_0\Psi^A_2 - P^B_{2,-1}[\sigma^A]\Psi^A_2 + Q^B_{2,-1}[\kappa^A] \Psi^A_2 \,.
\end{align}
Now by looking at the definitions of $P^A_{p,q}$ and $Q^A_{p,q}$, we can show that
\begin{align}
	P^A_{p,q-n} &= P^A_{p,q} - n \rho^A \,,\\ 
	Q^A_{p,q-n} &= Q^A_{p,q} - n \tau^A \,.
\end{align}
So now if we want to convert our $P^A_{2,-1}$ into $P^A_{2,-4}$ and $Q^A_{2,-1}$ into $Q^A_{2,-4}$ so that they match the terms in the Bianchi Identities, we need to only add and subtract factors of $3\rho^A$ and $3\tau^A$, respectively.

Continuing with Eq.~\eqref{halfwaySpinCoefficient}, we add and subtract $3\rho^A$ and $3\tau^A$ to get
\begin{align}
	(P^A_{2,-1}-3\rho^A)[\sigma^B\Psi^A_2] & - \sigma^B (D-3\rho^A)[\Psi^A_2] 
	-(Q^A_{2,-1}-3\tau^A)[\kappa^B\Psi^A_2] +  \kappa^B(\delta-3\tau^A)[\Psi^A_2]
	\notag \\
	& =
	\Psi^B_0\Psi^A_2 - P^B_{2,-1}[\sigma^A]\Psi^A_2 + Q^B_{2,-1}[\kappa^A] \Psi^A_2 \,,
\end{align}
which can be made more compact by writing
\begin{align}
	\label{eq:SpinCoefficientOpenedUp}
	P^A_{2,-4}[\sigma^B\Psi^A_2] & -Q^A_{2,-4}[\kappa^B\Psi^A_2]
	=
	\Psi^B_0\Psi^A_2 - S^{AB}_2 \,,
	\end{align}
where
\begin{align}
	\label{eq:S2Definition}
	 S^{AB}_2 &= P^B_{2,-1}[\sigma^A]\Psi^A_2 + Q^B_{2,-1}[\kappa^A] \Psi^A_2 + \sigma^B (D-3\rho^A)[\Psi^A_2] - \kappa^B(\delta-3\tau^A)[\Psi^A_2] \,.
\end{align}
We see that $S^{AB}_2$ only vanishes when $\sigma^A = \kappa^A = 0$, and the background quantities obey
\begin{align}
	D\Psi^A_2 = 3\rho^A\Psi^A_2 \quad \text{and} \quad \delta\Psi^A_2 = 3\tau^A\Psi^A_2 \,.
\end{align}
The former are satisfied for any vacuum Type D spacetime in general relativity. 

\subsubsection{Operator Identity}
In this section we prove the operator identity~\eqref{TypeDCommutatorFull}. 
We can expand the expression as
\begin{align}
	P_{p,q}\delta_{p,q}-Q_{p,q}D_{p,q} = [D_{p,q},\delta_{p,q}]  - (\epsilon - \epsilon^* + \rho^*)\delta_{p,q} 
	+ (\beta + \alpha^* - \pi^*)\Delta_{p,q}\,,
\end{align}
where
\begin{align}
	\label{eq:expanded_PQ}
	 [D_{p,q},\delta_{p,q}] = [D,\delta]  - [(p\epsilon - q\rho),\delta] - [D,(p\beta - q\tau)]\,.
\end{align}
In the above operator expression, any NP scalar is essentially an operator that multiplies any function with itself (e.g $\rho[\Psi] = \rho\Psi$).
This means that the commutator of the derivative operator with an NP scalar can be defined, and simplified as
\begin{align}
	[D,f](\Psi) &= D(f\Psi) - f D(\Psi) 
	= D(f) \Psi + f D(\Psi) - f D(\Psi) 
	= D(f) \Psi\,,
\end{align}
implying that the commutator of a derivative with a scalar is just the derivative acting on the scalar.
This simplifies what we have above and gives us
\begin{align}
	[D_{p,q},\delta_{p,q}] = [D,\delta]  + \delta[(p\epsilon - q\rho)] - D[(p\beta - q\tau)]\,.
\end{align}
We define the result of the operator expression,
\begin{align}
	E_{p,q} &:= P_{p,q}\delta_{p,q}-Q_{p,q}D_{p,q}\,.
\end{align}
Upon further simplifications we get
\begin{align}
	E_{p,q} &= [D,\delta]  - p(D\beta - \delta\epsilon) - q(\delta\rho - D\tau) 
	- (\epsilon - \epsilon^* + \rho^*)\delta_{p,q}+ (\beta + \alpha^* - \pi^*)D_{p,q}\,,
\end{align}
which then follows to,
\begin{align}
	E_{p,q} = & [D,\delta]  - (\epsilon - \epsilon^* + \rho^*)\delta+ (\beta + \alpha^* - \pi^*)D
	- p(D\beta - \delta\epsilon - \beta(\epsilon - \epsilon^* + \rho^*) + \epsilon(\beta + \alpha^* - \pi^*)) 
	\notag\\ & 
	- q(\delta\rho - D\tau + \tau(\epsilon - \epsilon^* + \rho^*) - \rho(\beta + \alpha^* - \pi^*))\,.
\end{align}
Using the commutation relation
\begin{align}
	[D,\delta] = \sigma\bar{\delta} - \kappa\Delta + (\rho^* + \epsilon - \epsilon^*)\delta - (\alpha^* + \beta - \pi^*)D\,,
\end{align}
and three spin coefficient equations,
\begin{align}
	D \beta-\delta \varepsilon&=(\alpha+\pi) \sigma+(\bar{\rho}-\bar{\varepsilon}) \beta-(\mu+\gamma) \kappa-(\bar{\alpha}-\bar{\pi}) \varepsilon
	+\Psi_{1}\,,
	\\
	\delta \rho-\bar{\delta} \sigma& = \rho(\bar{\alpha}+\beta)-\sigma(3 \alpha-\bar{\beta})+(\rho-\bar{\rho}) \tau+(\mu-\bar{\mu}) \kappa
	-\Psi_{1}+\Phi_{01}\,,		
	\\ 
	D \tau-\Delta \kappa& =(\tau+\bar{\pi}) \rho+(\bar{\tau}+\pi) \sigma+(\varepsilon-\bar{\varepsilon}) \tau-(3 \gamma+\bar{\gamma}) \kappa 
	+\Psi_{1}+\Phi_{01}\,,
\end{align}
we find that the commutator becomes,
\begin{align}
	\hat{E}_{p,q} = \sigma \bar{\delta}-\kappa \Delta +
		q[
			(\bar{\tau}+\pi-\bar{\beta}+3 \alpha) \sigma 
			+(\bar{\mu}-\mu-\bar{\gamma}-3 \gamma) \kappa
			+2 \Psi_{1}] -  
	p[
			(\alpha+\pi) \sigma 
			+(-\gamma-\mu) \kappa 
			+\Psi_{1}
		]\,.
\end{align}
This operator is the right hand side of the operator identity used by Teukolsky in~\cite{Teukolsky:1973ha}, where it is set to zero by the fact that the background is Type D. 
This is manifest in the above, where we can see that any background where $\sigma = \kappa = \Psi_1 = 0$ makes this operator vanish. 

\subsubsection{Combining the two Bianchi Identities and the operator equation}

To get the Teukolsky equation we compute
\begin{equation}
	P^A_{2,-4}[\text{Equation \eqref{Bianchi2New}}] - Q^A_{2,-4}[\text{Equation \eqref{Bianchi1New}}]\,.
\end{equation}
This gives us
\begin{align}
		P^A_{2,-4}\left[\Delta^{*\ A}_{10} [\Psi^B_0] - \hat{\delta}^{A}_{2,-4}[\Psi^B_1 ]\right] 
		& - Q^A_{2,-4}\left[\delta^{*\ A}_{00} [\Psi^B_0] - \hat{D}^A_{2,-4}[\Psi^B_1]\right]
	-3 \left(
		P^A_{2,-4}[\sigma^B\Psi^A_2] 
		- Q^A_{2,-4}[\kappa^B\Psi^A_2]
	\right)
	\notag \\ &
		= P^A_{2,-4}[ C_{1}^{ij A} [\Phi^B_{ij}] + S_1^{AB}] 
		- Q^A_{2,-4}[C_{0}^{ij A} [\Phi^B_{ij}] + S_0^{AB}]\,.
\end{align}
Using~\eqref{eq:SpinCoefficientOpenedUp} we get,
\begin{align}
		P^A_{2,-4}\left[\Delta^{*\ A}_{10} [\Psi^B_0] - \hat{\delta}^{A}_{2,-4}[\Psi^B_1 ]\right]
		- Q^A_{2,-4}\left[\delta^{*\ A}_{00} [\Psi^B_0] - \hat{D}^A_{2,-4}[\Psi^B_1]\right]
	-3 \left(
		\Psi^B_0\Psi^A_2 
		- S^{AB}_2
	\right)
	\notag \\
	= P^A_{2,-4}[ C_{1}^{ij A} [\Phi^B_{ij}] + S_1^{AB}] 
	- Q^A_{2,-4}[C_{0}^{ij A} [\Phi^B_{ij}] + S_0^{AB}]
	\,.
\end{align}
Now rearranging the first terms gives us,
\begin{align}
		(P^A_{2,-4}\Delta^{*\ A}_{10}  - Q^A_{2,-4}\delta^{*\ A}_{00} )\Psi^B_0 
		- (P^A_{2,-4}\hat{\delta}^{A}_{2,-4} - 
		Q^A_{2,-4}\hat{D}^A_{2,-4})\Psi^B_1 
	-3 \left(
		\Psi^B_0\Psi^A_2
		- S^{AB}_2
	\right)
	 \notag \\
		= P^A_{2,-4}[ C_{1}^{ij A} [\Phi^B_{ij}] + S_1^{AB}] 
		- Q^A_{2,-4}[C_{0}^{ij A} [\Phi^B_{ij}] + S_0^{AB}]
		\,.
\end{align}
The operator acting on $\Psi^B_1$ is equivalent to our commutator operator relation from Eq.~\eqref{TypeDCommutatorFull} on the background, so we can substitute that in, yielding
\begin{align}
		(P^A_{2,-4}\Delta^{*\ A}_{10}  - Q^A_{2,-4}\delta^{*\ A}_{00} )\Psi^B_0
		- E^A_{2,-4}[\Psi^B_1] 
	-3\left( 
		\Psi^B_0\Psi^A_2
		- S^{AB}_2
	\right)
	=
		P^A_{2,-4}[ C_{1}^{ij A} [\Phi^B_{ij}] + S_1^{AB}] 
		- Q^A_{2,-4}[C_{0}^{ij A} [\Phi^B_{ij}] + S_0^{AB}]
		\,.
\end{align}
Now we can get it into a form suggestive of Teukoslky's equation for $\Psi^B_0$,
\begin{align}
	(P^A_{2,-4}\Delta^{*\ A}_{10}  - Q^A_{2,-4}\delta^{*\ A}_{00} -3\Psi^A_2  )\Psi^B_0 
		= P^A_{2,-4}[ C_{1}^{ij A} [\Phi^B_{ij}] + S_1^{AB}] + E^A_{2,-4}[\Psi^B_1]  
		- Q^A_{2,-4}[C_{0}^{ij A} [\Phi^B_{ij}] + S_0^{AB}] - 3 S^{AB}_2\,,
\end{align}
which is the Teukolsky equation for $\Psi_0$. 
Note that the $T_0$ source terms that normally exist on the right hand side of the Teukolsky equation are a subset of the 
$P^A_{2,-4}[ C_{1}^{ij A} [\Phi^B_{ij}]]- Q^A_{2,-4}[C_{0}^{ij A} [\Phi^B_{ij}]]$ 
terms above.
We can break up the $C^{ij A}_{a}[\Phi_{ij}]$ expressions by defining
\begin{align}
	C_{0}^{ij} [\Phi_{ij}] & = \left(\delta+\pi^{*}-2 \alpha^{*}-2 \beta\right) \Phi_{00}
	- \left(D-2 \epsilon-2 \rho^{*}\right) \Phi_{01}+F_{0}^{ij}[\Phi_{ij}]\,, \\
	\label{eq:F0Definition}
	F_{0}^{ij}[\Phi_{ij}] & = 2 \sigma \Phi_{10}-2 \kappa \Phi_{11}-\kappa^{*} \Phi_{02} \,,
\end{align}
and
\begin{align}
	C_{1}^{ij} [\Phi_{ij}] &= \left(\delta+2 \pi^{*}-2 \beta\right) \Phi_{01}
	-\left(D-2 \epsilon+2 \epsilon^{*}-\rho^{*}\right) \Phi_{02} + F_{1}^{ij}[\Phi_{ij}] \,, \\
	\label{eq:F1Definition}
	F_{1}^{ij}[\Phi_{ij}] & = -\lambda^{*} \Phi_{00} +2 \sigma \Phi_{11}-2 \kappa \Phi_{12}\,.
\end{align}
This splitting up gives us the usual form of the Teukolsky equation,
\begin{align}
	\mathcal{O}^A & [\Psi^B_0] = T^A_{0}[\Phi^B_{ij}] + K  \,, \\
	\label{eq:AllExtraTermsNP}
		K & \coloneqq P^A_{2,-4}[ F_{1}^{ij A} [\Phi^B_{ij}] + S_1^{AB}] + E^A_{2,-4}[\Psi^B_1]  
		- Q^A_{2,-4}[F_{0}^{ij A} [\Phi^B_{ij}] + S_0^{AB}] - 3 S^{AB}_2\,.
\end{align}
The above can be expanded using the Eqs. \eqref{eq:OperatorsPQ}--\eqref{eq:RemainingOperatorsNP}, \eqref{eq:C0Definition}, \eqref{eq:C1Definition}, \eqref{eq:S0Definition}, \eqref{eq:S1Definition} and \eqref{eq:S2Definition}.
One can then get the equation for $\Psi_4$ using the GHP dual, interchanging $l^a\leftrightarrow n^a$ and  $m^a \leftrightarrow m^{a*}$.

To reiterate the conditions under which the above is zero, we have shown that,
\begin{align}
	\sigma^A  = \kappa^A = \lambda^A = \Psi^A_0 = \Psi^A_1 = 0 
	 \implies F_{0}^{ij A} [\Phi^B_{ij}]   = F_{1}^{ij A} [\Phi^B_{ij}] = E^A_{p,q} = 0\,.
\end{align}
Additional assumptions on top of the ones above yield further simplifications, such as
\begin{align}
	\Phi^A_{ij}= 0 \implies S_0^{AB} = S_1^{AB} = 0\,,
\end{align}
and
\begin{align}
	(D -3\rho)^A\Psi^A_2 = (\delta-3\tau)^A\Psi^A_2= 0 \implies S_2^{AB} = 0\,,
\end{align}
which would then completely set $K = 0$.

\end{widetext}

\bibliography{main.bbl}

\begin{thebibliography}{108}%
\makeatletter
\providecommand \@ifxundefined [1]{%
 \@ifx{#1\undefined}
}%
\providecommand \@ifnum [1]{%
 \ifnum #1\expandafter \@firstoftwo
 \else \expandafter \@secondoftwo
 \fi
}%
\providecommand \@ifx [1]{%
 \ifx #1\expandafter \@firstoftwo
 \else \expandafter \@secondoftwo
 \fi
}%
\providecommand \natexlab [1]{#1}%
\providecommand \enquote  [1]{``#1''}%
\providecommand \bibnamefont  [1]{#1}%
\providecommand \bibfnamefont [1]{#1}%
\providecommand \citenamefont [1]{#1}%
\providecommand \href@noop [0]{\@secondoftwo}%
\providecommand \href [0]{\begingroup \@sanitize@url \@href}%
\providecommand \@href[1]{\@@startlink{#1}\@@href}%
\providecommand \@@href[1]{\endgroup#1\@@endlink}%
\providecommand \@sanitize@url [0]{\catcode `\\12\catcode `\$12\catcode
  `\&12\catcode `\#12\catcode `\^12\catcode `\_12\catcode `\%12\relax}%
\providecommand \@@startlink[1]{}%
\providecommand \@@endlink[0]{}%
\providecommand \url  [0]{\begingroup\@sanitize@url \@url }%
\providecommand \@url [1]{\endgroup\@href {#1}{\urlprefix }}%
\providecommand \urlprefix  [0]{URL }%
\providecommand \Eprint [0]{\href }%
\providecommand \doibase [0]{https://doi.org/}%
\providecommand \selectlanguage [0]{\@gobble}%
\providecommand \bibinfo  [0]{\@secondoftwo}%
\providecommand \bibfield  [0]{\@secondoftwo}%
\providecommand \translation [1]{[#1]}%
\providecommand \BibitemOpen [0]{}%
\providecommand \bibitemStop [0]{}%
\providecommand \bibitemNoStop [0]{.\EOS\space}%
\providecommand \EOS [0]{\spacefactor3000\relax}%
\providecommand \BibitemShut  [1]{\csname bibitem#1\endcsname}%
\let\auto@bib@innerbib\@empty
\bibitem [{\citenamefont {Abbott}\ \emph
  {et~al.}(2016{\natexlab{a}})\citenamefont {Abbott} \emph
  {et~al.}}]{LIGOScientific:2016aoc}%
  \BibitemOpen
  \bibfield  {author} {\bibinfo {author} {\bibfnamefont {B.~P.}\ \bibnamefont
  {Abbott}} \emph {et~al.} (\bibinfo {collaboration} {LIGO Scientific,
  Virgo}),\ }\bibfield  {title} {\bibinfo {title} {{Observation of
  Gravitational Waves from a Binary Black Hole Merger}},\ }\href
  {https://doi.org/10.1103/PhysRevLett.116.061102} {\bibfield  {journal}
  {\bibinfo  {journal} {Phys. Rev. Lett.}\ }\textbf {\bibinfo {volume} {116}},\
  \bibinfo {pages} {061102} (\bibinfo {year} {2016}{\natexlab{a}})},\ \Eprint
  {https://arxiv.org/abs/1602.03837} {arXiv:1602.03837 [gr-qc]} \BibitemShut
  {NoStop}%
\bibitem [{\citenamefont {Abbott}\ \emph
  {et~al.}(2019{\natexlab{a}})\citenamefont {Abbott} \emph
  {et~al.}}]{LIGOScientific:2018mvr}%
  \BibitemOpen
  \bibfield  {author} {\bibinfo {author} {\bibfnamefont {B.~P.}\ \bibnamefont
  {Abbott}} \emph {et~al.} (\bibinfo {collaboration} {LIGO Scientific,
  Virgo}),\ }\bibfield  {title} {\bibinfo {title} {{GWTC-1: A
  Gravitational-Wave Transient Catalog of Compact Binary Mergers Observed by
  LIGO and Virgo during the First and Second Observing Runs}},\ }\href
  {https://doi.org/10.1103/PhysRevX.9.031040} {\bibfield  {journal} {\bibinfo
  {journal} {Phys. Rev. X}\ }\textbf {\bibinfo {volume} {9}},\ \bibinfo {pages}
  {031040} (\bibinfo {year} {2019}{\natexlab{a}})},\ \Eprint
  {https://arxiv.org/abs/1811.12907} {arXiv:1811.12907 [astro-ph.HE]}
  \BibitemShut {NoStop}%
\bibitem [{\citenamefont {Abbott}\ \emph
  {et~al.}(2021{\natexlab{a}})\citenamefont {Abbott} \emph
  {et~al.}}]{LIGOScientific:2020ibl}%
  \BibitemOpen
  \bibfield  {author} {\bibinfo {author} {\bibfnamefont {R.}~\bibnamefont
  {Abbott}} \emph {et~al.} (\bibinfo {collaboration} {LIGO Scientific,
  Virgo}),\ }\bibfield  {title} {\bibinfo {title} {{GWTC-2: Compact Binary
  Coalescences Observed by LIGO and Virgo During the First Half of the Third
  Observing Run}},\ }\href {https://doi.org/10.1103/PhysRevX.11.021053}
  {\bibfield  {journal} {\bibinfo  {journal} {Phys. Rev. X}\ }\textbf {\bibinfo
  {volume} {11}},\ \bibinfo {pages} {021053} (\bibinfo {year}
  {2021}{\natexlab{a}})},\ \Eprint {https://arxiv.org/abs/2010.14527}
  {arXiv:2010.14527 [gr-qc]} \BibitemShut {NoStop}%
\bibitem [{\citenamefont {Abbott}\ \emph
  {et~al.}(2021{\natexlab{b}})\citenamefont {Abbott} \emph
  {et~al.}}]{LIGOScientific:2021usb}%
  \BibitemOpen
  \bibfield  {author} {\bibinfo {author} {\bibfnamefont {R.}~\bibnamefont
  {Abbott}} \emph {et~al.} (\bibinfo {collaboration} {LIGO Scientific,
  VIRGO}),\ }\bibfield  {title} {\bibinfo {title} {{GWTC-2.1: Deep Extended
  Catalog of Compact Binary Coalescences Observed by LIGO and Virgo During the
  First Half of the Third Observing Run}},\ }\href@noop {} {\  (\bibinfo {year}
  {2021}{\natexlab{b}})},\ \Eprint {https://arxiv.org/abs/2108.01045}
  {arXiv:2108.01045 [gr-qc]} \BibitemShut {NoStop}%
\bibitem [{\citenamefont {Abbott}\ \emph
  {et~al.}(2021{\natexlab{c}})\citenamefont {Abbott} \emph
  {et~al.}}]{LIGOScientific:2021djp}%
  \BibitemOpen
  \bibfield  {author} {\bibinfo {author} {\bibfnamefont {R.}~\bibnamefont
  {Abbott}} \emph {et~al.} (\bibinfo {collaboration} {LIGO Scientific, VIRGO,
  KAGRA}),\ }\bibfield  {title} {\bibinfo {title} {{GWTC-3: Compact Binary
  Coalescences Observed by LIGO and Virgo During the Second Part of the Third
  Observing Run}},\ }\href@noop {} {\  (\bibinfo {year}
  {2021}{\natexlab{c}})},\ \Eprint {https://arxiv.org/abs/2111.03606}
  {arXiv:2111.03606 [gr-qc]} \BibitemShut {NoStop}%
\bibitem [{\citenamefont {Nitz}\ \emph {et~al.}(2019)\citenamefont {Nitz},
  \citenamefont {Capano}, \citenamefont {Nielsen}, \citenamefont {Reyes},
  \citenamefont {White}, \citenamefont {Brown},\ and\ \citenamefont
  {Krishnan}}]{Nitz:2018imz}%
  \BibitemOpen
  \bibfield  {author} {\bibinfo {author} {\bibfnamefont {A.~H.}\ \bibnamefont
  {Nitz}}, \bibinfo {author} {\bibfnamefont {C.}~\bibnamefont {Capano}},
  \bibinfo {author} {\bibfnamefont {A.~B.}\ \bibnamefont {Nielsen}}, \bibinfo
  {author} {\bibfnamefont {S.}~\bibnamefont {Reyes}}, \bibinfo {author}
  {\bibfnamefont {R.}~\bibnamefont {White}}, \bibinfo {author} {\bibfnamefont
  {D.~A.}\ \bibnamefont {Brown}},\ and\ \bibinfo {author} {\bibfnamefont
  {B.}~\bibnamefont {Krishnan}},\ }\bibfield  {title} {\bibinfo {title}
  {{1-OGC: The first open gravitational-wave catalog of binary mergers from
  analysis of public Advanced LIGO data}},\ }\href
  {https://doi.org/10.3847/1538-4357/ab0108} {\bibfield  {journal} {\bibinfo
  {journal} {Astrophys. J.}\ }\textbf {\bibinfo {volume} {872}},\ \bibinfo
  {pages} {195} (\bibinfo {year} {2019})},\ \Eprint
  {https://arxiv.org/abs/1811.01921} {arXiv:1811.01921 [gr-qc]} \BibitemShut
  {NoStop}%
\bibitem [{\citenamefont {Nitz}\ \emph {et~al.}(2020)\citenamefont {Nitz},
  \citenamefont {Dent}, \citenamefont {Davies}, \citenamefont {Kumar},
  \citenamefont {Capano}, \citenamefont {Harry}, \citenamefont {Mozzon},
  \citenamefont {Nuttall}, \citenamefont {Lundgren},\ and\ \citenamefont
  {T\'apai}}]{Nitz:2020oeq}%
  \BibitemOpen
  \bibfield  {author} {\bibinfo {author} {\bibfnamefont {A.~H.}\ \bibnamefont
  {Nitz}}, \bibinfo {author} {\bibfnamefont {T.}~\bibnamefont {Dent}}, \bibinfo
  {author} {\bibfnamefont {G.~S.}\ \bibnamefont {Davies}}, \bibinfo {author}
  {\bibfnamefont {S.}~\bibnamefont {Kumar}}, \bibinfo {author} {\bibfnamefont
  {C.~D.}\ \bibnamefont {Capano}}, \bibinfo {author} {\bibfnamefont
  {I.}~\bibnamefont {Harry}}, \bibinfo {author} {\bibfnamefont
  {S.}~\bibnamefont {Mozzon}}, \bibinfo {author} {\bibfnamefont
  {L.}~\bibnamefont {Nuttall}}, \bibinfo {author} {\bibfnamefont
  {A.}~\bibnamefont {Lundgren}},\ and\ \bibinfo {author} {\bibfnamefont
  {M.}~\bibnamefont {T\'apai}},\ }\bibfield  {title} {\bibinfo {title} {{2-OGC:
  Open Gravitational-wave Catalog of binary mergers from analysis of public
  Advanced LIGO and Virgo data}},\ }\href
  {https://doi.org/10.3847/1538-4357/ab733f} {\bibfield  {journal} {\bibinfo
  {journal} {Astrophys. J.}\ }\textbf {\bibinfo {volume} {891}},\ \bibinfo
  {pages} {123} (\bibinfo {year} {2020})},\ \Eprint
  {https://arxiv.org/abs/1910.05331} {arXiv:1910.05331 [astro-ph.HE]}
  \BibitemShut {NoStop}%
\bibitem [{\citenamefont {Nitz}\ \emph
  {et~al.}(2021{\natexlab{a}})\citenamefont {Nitz}, \citenamefont {Capano},
  \citenamefont {Kumar}, \citenamefont {Wang}, \citenamefont {Kastha},
  \citenamefont {Sch\"afer}, \citenamefont {Dhurkunde},\ and\ \citenamefont
  {Cabero}}]{Nitz:2021uxj}%
  \BibitemOpen
  \bibfield  {author} {\bibinfo {author} {\bibfnamefont {A.~H.}\ \bibnamefont
  {Nitz}}, \bibinfo {author} {\bibfnamefont {C.~D.}\ \bibnamefont {Capano}},
  \bibinfo {author} {\bibfnamefont {S.}~\bibnamefont {Kumar}}, \bibinfo
  {author} {\bibfnamefont {Y.-F.}\ \bibnamefont {Wang}}, \bibinfo {author}
  {\bibfnamefont {S.}~\bibnamefont {Kastha}}, \bibinfo {author} {\bibfnamefont
  {M.}~\bibnamefont {Sch\"afer}}, \bibinfo {author} {\bibfnamefont
  {R.}~\bibnamefont {Dhurkunde}},\ and\ \bibinfo {author} {\bibfnamefont
  {M.}~\bibnamefont {Cabero}},\ }\bibfield  {title} {\bibinfo {title} {{3-OGC:
  Catalog of Gravitational Waves from Compact-binary Mergers}},\ }\href
  {https://doi.org/10.3847/1538-4357/ac1c03} {\bibfield  {journal} {\bibinfo
  {journal} {Astrophys. J.}\ }\textbf {\bibinfo {volume} {922}},\ \bibinfo
  {pages} {76} (\bibinfo {year} {2021}{\natexlab{a}})},\ \Eprint
  {https://arxiv.org/abs/2105.09151} {arXiv:2105.09151 [astro-ph.HE]}
  \BibitemShut {NoStop}%
\bibitem [{\citenamefont {Nitz}\ \emph
  {et~al.}(2021{\natexlab{b}})\citenamefont {Nitz}, \citenamefont {Kumar},
  \citenamefont {Wang}, \citenamefont {Kastha}, \citenamefont {Wu},
  \citenamefont {Sch\"afer}, \citenamefont {Dhurkunde},\ and\ \citenamefont
  {Capano}}]{Nitz:2021zwj}%
  \BibitemOpen
  \bibfield  {author} {\bibinfo {author} {\bibfnamefont {A.~H.}\ \bibnamefont
  {Nitz}}, \bibinfo {author} {\bibfnamefont {S.}~\bibnamefont {Kumar}},
  \bibinfo {author} {\bibfnamefont {Y.-F.}\ \bibnamefont {Wang}}, \bibinfo
  {author} {\bibfnamefont {S.}~\bibnamefont {Kastha}}, \bibinfo {author}
  {\bibfnamefont {S.}~\bibnamefont {Wu}}, \bibinfo {author} {\bibfnamefont
  {M.}~\bibnamefont {Sch\"afer}}, \bibinfo {author} {\bibfnamefont
  {R.}~\bibnamefont {Dhurkunde}},\ and\ \bibinfo {author} {\bibfnamefont
  {C.~D.}\ \bibnamefont {Capano}},\ }\bibfield  {title} {\bibinfo {title}
  {{4-OGC: Catalog of gravitational waves from compact-binary mergers}},\
  }\href@noop {} {\  (\bibinfo {year} {2021}{\natexlab{b}})},\ \Eprint
  {https://arxiv.org/abs/2112.06878} {arXiv:2112.06878 [astro-ph.HE]}
  \BibitemShut {NoStop}%
\bibitem [{\citenamefont {Zackay}\ \emph {et~al.}(2019)\citenamefont {Zackay},
  \citenamefont {Venumadhav}, \citenamefont {Dai}, \citenamefont {Roulet},\
  and\ \citenamefont {Zaldarriaga}}]{Zackay:2019tzo}%
  \BibitemOpen
  \bibfield  {author} {\bibinfo {author} {\bibfnamefont {B.}~\bibnamefont
  {Zackay}}, \bibinfo {author} {\bibfnamefont {T.}~\bibnamefont {Venumadhav}},
  \bibinfo {author} {\bibfnamefont {L.}~\bibnamefont {Dai}}, \bibinfo {author}
  {\bibfnamefont {J.}~\bibnamefont {Roulet}},\ and\ \bibinfo {author}
  {\bibfnamefont {M.}~\bibnamefont {Zaldarriaga}},\ }\bibfield  {title}
  {\bibinfo {title} {{Highly spinning and aligned binary black hole merger in
  the Advanced LIGO first observing run}},\ }\href
  {https://doi.org/10.1103/PhysRevD.100.023007} {\bibfield  {journal} {\bibinfo
   {journal} {Phys. Rev. D}\ }\textbf {\bibinfo {volume} {100}},\ \bibinfo
  {pages} {023007} (\bibinfo {year} {2019})},\ \Eprint
  {https://arxiv.org/abs/1902.10331} {arXiv:1902.10331 [astro-ph.HE]}
  \BibitemShut {NoStop}%
\bibitem [{\citenamefont {Venumadhav}\ \emph {et~al.}(2019)\citenamefont
  {Venumadhav}, \citenamefont {Zackay}, \citenamefont {Roulet}, \citenamefont
  {Dai},\ and\ \citenamefont {Zaldarriaga}}]{Venumadhav:2019tad}%
  \BibitemOpen
  \bibfield  {author} {\bibinfo {author} {\bibfnamefont {T.}~\bibnamefont
  {Venumadhav}}, \bibinfo {author} {\bibfnamefont {B.}~\bibnamefont {Zackay}},
  \bibinfo {author} {\bibfnamefont {J.}~\bibnamefont {Roulet}}, \bibinfo
  {author} {\bibfnamefont {L.}~\bibnamefont {Dai}},\ and\ \bibinfo {author}
  {\bibfnamefont {M.}~\bibnamefont {Zaldarriaga}},\ }\bibfield  {title}
  {\bibinfo {title} {{New search pipeline for compact binary mergers: Results
  for binary black holes in the first observing run of Advanced LIGO}},\ }\href
  {https://doi.org/10.1103/PhysRevD.100.023011} {\bibfield  {journal} {\bibinfo
   {journal} {Phys. Rev. D}\ }\textbf {\bibinfo {volume} {100}},\ \bibinfo
  {pages} {023011} (\bibinfo {year} {2019})},\ \Eprint
  {https://arxiv.org/abs/1902.10341} {arXiv:1902.10341 [astro-ph.IM]}
  \BibitemShut {NoStop}%
\bibitem [{\citenamefont {Venumadhav}\ \emph {et~al.}(2020)\citenamefont
  {Venumadhav}, \citenamefont {Zackay}, \citenamefont {Roulet}, \citenamefont
  {Dai},\ and\ \citenamefont {Zaldarriaga}}]{Venumadhav:2019lyq}%
  \BibitemOpen
  \bibfield  {author} {\bibinfo {author} {\bibfnamefont {T.}~\bibnamefont
  {Venumadhav}}, \bibinfo {author} {\bibfnamefont {B.}~\bibnamefont {Zackay}},
  \bibinfo {author} {\bibfnamefont {J.}~\bibnamefont {Roulet}}, \bibinfo
  {author} {\bibfnamefont {L.}~\bibnamefont {Dai}},\ and\ \bibinfo {author}
  {\bibfnamefont {M.}~\bibnamefont {Zaldarriaga}},\ }\bibfield  {title}
  {\bibinfo {title} {{New binary black hole mergers in the second observing run
  of Advanced LIGO and Advanced Virgo}},\ }\href
  {https://doi.org/10.1103/PhysRevD.101.083030} {\bibfield  {journal} {\bibinfo
   {journal} {Phys. Rev. D}\ }\textbf {\bibinfo {volume} {101}},\ \bibinfo
  {pages} {083030} (\bibinfo {year} {2020})},\ \Eprint
  {https://arxiv.org/abs/1904.07214} {arXiv:1904.07214 [astro-ph.HE]}
  \BibitemShut {NoStop}%
\bibitem [{\citenamefont {Zackay}\ \emph {et~al.}(2021)\citenamefont {Zackay},
  \citenamefont {Dai}, \citenamefont {Venumadhav}, \citenamefont {Roulet},\
  and\ \citenamefont {Zaldarriaga}}]{Zackay:2019btq}%
  \BibitemOpen
  \bibfield  {author} {\bibinfo {author} {\bibfnamefont {B.}~\bibnamefont
  {Zackay}}, \bibinfo {author} {\bibfnamefont {L.}~\bibnamefont {Dai}},
  \bibinfo {author} {\bibfnamefont {T.}~\bibnamefont {Venumadhav}}, \bibinfo
  {author} {\bibfnamefont {J.}~\bibnamefont {Roulet}},\ and\ \bibinfo {author}
  {\bibfnamefont {M.}~\bibnamefont {Zaldarriaga}},\ }\bibfield  {title}
  {\bibinfo {title} {{Detecting gravitational waves with disparate detector
  responses: Two new binary black hole mergers}},\ }\href
  {https://doi.org/10.1103/PhysRevD.104.063030} {\bibfield  {journal} {\bibinfo
   {journal} {Phys. Rev. D}\ }\textbf {\bibinfo {volume} {104}},\ \bibinfo
  {pages} {063030} (\bibinfo {year} {2021})},\ \Eprint
  {https://arxiv.org/abs/1910.09528} {arXiv:1910.09528 [astro-ph.HE]}
  \BibitemShut {NoStop}%
\bibitem [{\citenamefont {Olsen}\ \emph {et~al.}(2022)\citenamefont {Olsen},
  \citenamefont {Venumadhav}, \citenamefont {Mushkin}, \citenamefont {Roulet},
  \citenamefont {Zackay},\ and\ \citenamefont {Zaldarriaga}}]{Olsen:2022pin}%
  \BibitemOpen
  \bibfield  {author} {\bibinfo {author} {\bibfnamefont {S.}~\bibnamefont
  {Olsen}}, \bibinfo {author} {\bibfnamefont {T.}~\bibnamefont {Venumadhav}},
  \bibinfo {author} {\bibfnamefont {J.}~\bibnamefont {Mushkin}}, \bibinfo
  {author} {\bibfnamefont {J.}~\bibnamefont {Roulet}}, \bibinfo {author}
  {\bibfnamefont {B.}~\bibnamefont {Zackay}},\ and\ \bibinfo {author}
  {\bibfnamefont {M.}~\bibnamefont {Zaldarriaga}},\ }\bibfield  {title}
  {\bibinfo {title} {{New binary black hole mergers in the LIGO--Virgo O3a
  data}},\ }\href@noop {} {\  (\bibinfo {year} {2022})},\ \Eprint
  {https://arxiv.org/abs/2201.02252} {arXiv:2201.02252 [astro-ph.HE]}
  \BibitemShut {NoStop}%
\bibitem [{\citenamefont {Aasi}\ \emph {et~al.}(2015)\citenamefont {Aasi} \emph
  {et~al.}}]{LIGOScientific:2014pky}%
  \BibitemOpen
  \bibfield  {author} {\bibinfo {author} {\bibfnamefont {J.}~\bibnamefont
  {Aasi}} \emph {et~al.} (\bibinfo {collaboration} {LIGO Scientific}),\
  }\bibfield  {title} {\bibinfo {title} {{Advanced LIGO}},\ }\href
  {https://doi.org/10.1088/0264-9381/32/7/074001} {\bibfield  {journal}
  {\bibinfo  {journal} {Class. Quant. Grav.}\ }\textbf {\bibinfo {volume}
  {32}},\ \bibinfo {pages} {074001} (\bibinfo {year} {2015})},\ \Eprint
  {https://arxiv.org/abs/1411.4547} {arXiv:1411.4547 [gr-qc]} \BibitemShut
  {NoStop}%
\bibitem [{\citenamefont {Acernese}\ \emph {et~al.}(2015)\citenamefont
  {Acernese} \emph {et~al.}}]{VIRGO:2014yos}%
  \BibitemOpen
  \bibfield  {author} {\bibinfo {author} {\bibfnamefont {F.}~\bibnamefont
  {Acernese}} \emph {et~al.} (\bibinfo {collaboration} {VIRGO}),\ }\bibfield
  {title} {\bibinfo {title} {{Advanced Virgo: a second-generation
  interferometric gravitational wave detector}},\ }\href
  {https://doi.org/10.1088/0264-9381/32/2/024001} {\bibfield  {journal}
  {\bibinfo  {journal} {Class. Quant. Grav.}\ }\textbf {\bibinfo {volume}
  {32}},\ \bibinfo {pages} {024001} (\bibinfo {year} {2015})},\ \Eprint
  {https://arxiv.org/abs/1408.3978} {arXiv:1408.3978 [gr-qc]} \BibitemShut
  {NoStop}%
\bibitem [{\citenamefont {Abbott}\ \emph
  {et~al.}(2016{\natexlab{b}})\citenamefont {Abbott} \emph
  {et~al.}}]{LIGOScientific:2016lio}%
  \BibitemOpen
  \bibfield  {author} {\bibinfo {author} {\bibfnamefont {B.~P.}\ \bibnamefont
  {Abbott}} \emph {et~al.} (\bibinfo {collaboration} {LIGO Scientific,
  Virgo}),\ }\bibfield  {title} {\bibinfo {title} {{Tests of general relativity
  with GW150914}},\ }\href {https://doi.org/10.1103/PhysRevLett.116.221101}
  {\bibfield  {journal} {\bibinfo  {journal} {Phys. Rev. Lett.}\ }\textbf
  {\bibinfo {volume} {116}},\ \bibinfo {pages} {221101} (\bibinfo {year}
  {2016}{\natexlab{b}})},\ \bibinfo {note} {[Erratum: Phys.Rev.Lett. 121,
  129902 (2018)]},\ \Eprint {https://arxiv.org/abs/1602.03841}
  {arXiv:1602.03841 [gr-qc]} \BibitemShut {NoStop}%
\bibitem [{\citenamefont {Yunes}\ \emph {et~al.}(2016)\citenamefont {Yunes},
  \citenamefont {Yagi},\ and\ \citenamefont {Pretorius}}]{Yunes:2016jcc}%
  \BibitemOpen
  \bibfield  {author} {\bibinfo {author} {\bibfnamefont {N.}~\bibnamefont
  {Yunes}}, \bibinfo {author} {\bibfnamefont {K.}~\bibnamefont {Yagi}},\ and\
  \bibinfo {author} {\bibfnamefont {F.}~\bibnamefont {Pretorius}},\ }\bibfield
  {title} {\bibinfo {title} {{Theoretical Physics Implications of the Binary
  Black-Hole Mergers GW150914 and GW151226}},\ }\href
  {https://doi.org/10.1103/PhysRevD.94.084002} {\bibfield  {journal} {\bibinfo
  {journal} {Phys. Rev. D}\ }\textbf {\bibinfo {volume} {94}},\ \bibinfo
  {pages} {084002} (\bibinfo {year} {2016})},\ \Eprint
  {https://arxiv.org/abs/1603.08955} {arXiv:1603.08955 [gr-qc]} \BibitemShut
  {NoStop}%
\bibitem [{\citenamefont {Abbott}\ \emph
  {et~al.}(2016{\natexlab{c}})\citenamefont {Abbott} \emph
  {et~al.}}]{LIGOScientific:2016dsl}%
  \BibitemOpen
  \bibfield  {author} {\bibinfo {author} {\bibfnamefont {B.~P.}\ \bibnamefont
  {Abbott}} \emph {et~al.} (\bibinfo {collaboration} {LIGO Scientific,
  Virgo}),\ }\bibfield  {title} {\bibinfo {title} {{Binary Black Hole Mergers
  in the first Advanced LIGO Observing Run}},\ }\href
  {https://doi.org/10.1103/PhysRevX.6.041015} {\bibfield  {journal} {\bibinfo
  {journal} {Phys. Rev. X}\ }\textbf {\bibinfo {volume} {6}},\ \bibinfo {pages}
  {041015} (\bibinfo {year} {2016}{\natexlab{c}})},\ \bibinfo {note} {[Erratum:
  Phys.Rev.X 8, 039903 (2018)]},\ \Eprint {https://arxiv.org/abs/1606.04856}
  {arXiv:1606.04856 [gr-qc]} \BibitemShut {NoStop}%
\bibitem [{\citenamefont {Abbott}\ \emph
  {et~al.}(2019{\natexlab{b}})\citenamefont {Abbott} \emph
  {et~al.}}]{LIGOScientific:2018dkp}%
  \BibitemOpen
  \bibfield  {author} {\bibinfo {author} {\bibfnamefont {B.~P.}\ \bibnamefont
  {Abbott}} \emph {et~al.} (\bibinfo {collaboration} {LIGO Scientific,
  Virgo}),\ }\bibfield  {title} {\bibinfo {title} {{Tests of General Relativity
  with GW170817}},\ }\href {https://doi.org/10.1103/PhysRevLett.123.011102}
  {\bibfield  {journal} {\bibinfo  {journal} {Phys. Rev. Lett.}\ }\textbf
  {\bibinfo {volume} {123}},\ \bibinfo {pages} {011102} (\bibinfo {year}
  {2019}{\natexlab{b}})},\ \Eprint {https://arxiv.org/abs/1811.00364}
  {arXiv:1811.00364 [gr-qc]} \BibitemShut {NoStop}%
\bibitem [{\citenamefont {Abbott}\ \emph
  {et~al.}(2019{\natexlab{c}})\citenamefont {Abbott} \emph
  {et~al.}}]{LIGOScientific:2019fpa}%
  \BibitemOpen
  \bibfield  {author} {\bibinfo {author} {\bibfnamefont {B.~P.}\ \bibnamefont
  {Abbott}} \emph {et~al.} (\bibinfo {collaboration} {LIGO Scientific,
  Virgo}),\ }\bibfield  {title} {\bibinfo {title} {{Tests of General Relativity
  with the Binary Black Hole Signals from the LIGO-Virgo Catalog GWTC-1}},\
  }\href {https://doi.org/10.1103/PhysRevD.100.104036} {\bibfield  {journal}
  {\bibinfo  {journal} {Phys. Rev. D}\ }\textbf {\bibinfo {volume} {100}},\
  \bibinfo {pages} {104036} (\bibinfo {year} {2019}{\natexlab{c}})},\ \Eprint
  {https://arxiv.org/abs/1903.04467} {arXiv:1903.04467 [gr-qc]} \BibitemShut
  {NoStop}%
\bibitem [{\citenamefont {Abbott}\ \emph
  {et~al.}(2021{\natexlab{d}})\citenamefont {Abbott} \emph
  {et~al.}}]{LIGOScientific:2020tif}%
  \BibitemOpen
  \bibfield  {author} {\bibinfo {author} {\bibfnamefont {R.}~\bibnamefont
  {Abbott}} \emph {et~al.} (\bibinfo {collaboration} {LIGO Scientific,
  Virgo}),\ }\bibfield  {title} {\bibinfo {title} {{Tests of general relativity
  with binary black holes from the second LIGO-Virgo gravitational-wave
  transient catalog}},\ }\href {https://doi.org/10.1103/PhysRevD.103.122002}
  {\bibfield  {journal} {\bibinfo  {journal} {Phys. Rev. D}\ }\textbf {\bibinfo
  {volume} {103}},\ \bibinfo {pages} {122002} (\bibinfo {year}
  {2021}{\natexlab{d}})},\ \Eprint {https://arxiv.org/abs/2010.14529}
  {arXiv:2010.14529 [gr-qc]} \BibitemShut {NoStop}%
\bibitem [{\citenamefont {Abbott}\ \emph
  {et~al.}(2021{\natexlab{e}})\citenamefont {Abbott} \emph
  {et~al.}}]{LIGOScientific:2021sio}%
  \BibitemOpen
  \bibfield  {author} {\bibinfo {author} {\bibfnamefont {R.}~\bibnamefont
  {Abbott}} \emph {et~al.} (\bibinfo {collaboration} {LIGO Scientific, VIRGO,
  KAGRA}),\ }\bibfield  {title} {\bibinfo {title} {{Tests of General Relativity
  with GWTC-3}},\ }\href@noop {} {\  (\bibinfo {year} {2021}{\natexlab{e}})},\
  \Eprint {https://arxiv.org/abs/2112.06861} {arXiv:2112.06861 [gr-qc]}
  \BibitemShut {NoStop}%
\bibitem [{\citenamefont {Zimmerman}\ \emph {et~al.}(2019)\citenamefont
  {Zimmerman}, \citenamefont {Haster},\ and\ \citenamefont
  {Chatziioannou}}]{Zimmerman:2019wzo}%
  \BibitemOpen
  \bibfield  {author} {\bibinfo {author} {\bibfnamefont {A.}~\bibnamefont
  {Zimmerman}}, \bibinfo {author} {\bibfnamefont {C.-J.}\ \bibnamefont
  {Haster}},\ and\ \bibinfo {author} {\bibfnamefont {K.}~\bibnamefont
  {Chatziioannou}},\ }\bibfield  {title} {\bibinfo {title} {{On combining
  information from multiple gravitational wave sources}},\ }\href
  {https://doi.org/10.1103/PhysRevD.99.124044} {\bibfield  {journal} {\bibinfo
  {journal} {Phys. Rev. D}\ }\textbf {\bibinfo {volume} {99}},\ \bibinfo
  {pages} {124044} (\bibinfo {year} {2019})},\ \Eprint
  {https://arxiv.org/abs/1903.11008} {arXiv:1903.11008 [astro-ph.IM]}
  \BibitemShut {NoStop}%
\bibitem [{\citenamefont {Isi}\ \emph {et~al.}(2019{\natexlab{a}})\citenamefont
  {Isi}, \citenamefont {Chatziioannou},\ and\ \citenamefont
  {Farr}}]{Isi:2019asy}%
  \BibitemOpen
  \bibfield  {author} {\bibinfo {author} {\bibfnamefont {M.}~\bibnamefont
  {Isi}}, \bibinfo {author} {\bibfnamefont {K.}~\bibnamefont {Chatziioannou}},\
  and\ \bibinfo {author} {\bibfnamefont {W.~M.}\ \bibnamefont {Farr}},\
  }\bibfield  {title} {\bibinfo {title} {{Hierarchical test of general
  relativity with gravitational waves}},\ }\href
  {https://doi.org/10.1103/PhysRevLett.123.121101} {\bibfield  {journal}
  {\bibinfo  {journal} {Phys. Rev. Lett.}\ }\textbf {\bibinfo {volume} {123}},\
  \bibinfo {pages} {121101} (\bibinfo {year} {2019}{\natexlab{a}})},\ \Eprint
  {https://arxiv.org/abs/1904.08011} {arXiv:1904.08011 [gr-qc]} \BibitemShut
  {NoStop}%
\bibitem [{\citenamefont {Berti}\ \emph {et~al.}(2009)\citenamefont {Berti},
  \citenamefont {Cardoso},\ and\ \citenamefont {Starinets}}]{Berti:2009kk}%
  \BibitemOpen
  \bibfield  {author} {\bibinfo {author} {\bibfnamefont {E.}~\bibnamefont
  {Berti}}, \bibinfo {author} {\bibfnamefont {V.}~\bibnamefont {Cardoso}},\
  and\ \bibinfo {author} {\bibfnamefont {A.~O.}\ \bibnamefont {Starinets}},\
  }\bibfield  {title} {\bibinfo {title} {{Quasinormal modes of black holes and
  black branes}},\ }\href {https://doi.org/10.1088/0264-9381/26/16/163001}
  {\bibfield  {journal} {\bibinfo  {journal} {Class. Quant. Grav.}\ }\textbf
  {\bibinfo {volume} {26}},\ \bibinfo {pages} {163001} (\bibinfo {year}
  {2009})},\ \Eprint {https://arxiv.org/abs/0905.2975} {arXiv:0905.2975
  [gr-qc]} \BibitemShut {NoStop}%
\bibitem [{\citenamefont {Dreyer}\ \emph {et~al.}(2004)\citenamefont {Dreyer},
  \citenamefont {Kelly}, \citenamefont {Krishnan}, \citenamefont {Finn},
  \citenamefont {Garrison},\ and\ \citenamefont
  {Lopez-Aleman}}]{Dreyer:2003bv}%
  \BibitemOpen
  \bibfield  {author} {\bibinfo {author} {\bibfnamefont {O.}~\bibnamefont
  {Dreyer}}, \bibinfo {author} {\bibfnamefont {B.~J.}\ \bibnamefont {Kelly}},
  \bibinfo {author} {\bibfnamefont {B.}~\bibnamefont {Krishnan}}, \bibinfo
  {author} {\bibfnamefont {L.~S.}\ \bibnamefont {Finn}}, \bibinfo {author}
  {\bibfnamefont {D.}~\bibnamefont {Garrison}},\ and\ \bibinfo {author}
  {\bibfnamefont {R.}~\bibnamefont {Lopez-Aleman}},\ }\bibfield  {title}
  {\bibinfo {title} {{Black hole spectroscopy: Testing general relativity
  through gravitational wave observations}},\ }\href
  {https://doi.org/10.1088/0264-9381/21/4/003} {\bibfield  {journal} {\bibinfo
  {journal} {Class. Quant. Grav.}\ }\textbf {\bibinfo {volume} {21}},\ \bibinfo
  {pages} {787} (\bibinfo {year} {2004})},\ \Eprint
  {https://arxiv.org/abs/gr-qc/0309007} {arXiv:gr-qc/0309007} \BibitemShut
  {NoStop}%
\bibitem [{\citenamefont {Berti}\ \emph {et~al.}(2006)\citenamefont {Berti},
  \citenamefont {Cardoso},\ and\ \citenamefont {Will}}]{Berti:2005ys}%
  \BibitemOpen
  \bibfield  {author} {\bibinfo {author} {\bibfnamefont {E.}~\bibnamefont
  {Berti}}, \bibinfo {author} {\bibfnamefont {V.}~\bibnamefont {Cardoso}},\
  and\ \bibinfo {author} {\bibfnamefont {C.~M.}\ \bibnamefont {Will}},\
  }\bibfield  {title} {\bibinfo {title} {{On gravitational-wave spectroscopy of
  massive black holes with the space interferometer LISA}},\ }\href
  {https://doi.org/10.1103/PhysRevD.73.064030} {\bibfield  {journal} {\bibinfo
  {journal} {Phys. Rev. D}\ }\textbf {\bibinfo {volume} {73}},\ \bibinfo
  {pages} {064030} (\bibinfo {year} {2006})},\ \Eprint
  {https://arxiv.org/abs/gr-qc/0512160} {arXiv:gr-qc/0512160} \BibitemShut
  {NoStop}%
\bibitem [{\citenamefont {Berti}\ \emph {et~al.}(2018)\citenamefont {Berti},
  \citenamefont {Yagi}, \citenamefont {Yang},\ and\ \citenamefont
  {Yunes}}]{Berti:2018vdi}%
  \BibitemOpen
  \bibfield  {author} {\bibinfo {author} {\bibfnamefont {E.}~\bibnamefont
  {Berti}}, \bibinfo {author} {\bibfnamefont {K.}~\bibnamefont {Yagi}},
  \bibinfo {author} {\bibfnamefont {H.}~\bibnamefont {Yang}},\ and\ \bibinfo
  {author} {\bibfnamefont {N.}~\bibnamefont {Yunes}},\ }\bibfield  {title}
  {\bibinfo {title} {{Extreme Gravity Tests with Gravitational Waves from
  Compact Binary Coalescences: (II) Ringdown}},\ }\href
  {https://doi.org/10.1007/s10714-018-2372-6} {\bibfield  {journal} {\bibinfo
  {journal} {Gen. Rel. Grav.}\ }\textbf {\bibinfo {volume} {50}},\ \bibinfo
  {pages} {49} (\bibinfo {year} {2018})},\ \Eprint
  {https://arxiv.org/abs/1801.03587} {arXiv:1801.03587 [gr-qc]} \BibitemShut
  {NoStop}%
\bibitem [{\citenamefont {Giesler}\ \emph {et~al.}(2019)\citenamefont
  {Giesler}, \citenamefont {Isi}, \citenamefont {Scheel},\ and\ \citenamefont
  {Teukolsky}}]{Giesler:2019uxc}%
  \BibitemOpen
  \bibfield  {author} {\bibinfo {author} {\bibfnamefont {M.}~\bibnamefont
  {Giesler}}, \bibinfo {author} {\bibfnamefont {M.}~\bibnamefont {Isi}},
  \bibinfo {author} {\bibfnamefont {M.~A.}\ \bibnamefont {Scheel}},\ and\
  \bibinfo {author} {\bibfnamefont {S.}~\bibnamefont {Teukolsky}},\ }\bibfield
  {title} {\bibinfo {title} {{Black Hole Ringdown: The Importance of
  Overtones}},\ }\href {https://doi.org/10.1103/PhysRevX.9.041060} {\bibfield
  {journal} {\bibinfo  {journal} {Phys. Rev. X}\ }\textbf {\bibinfo {volume}
  {9}},\ \bibinfo {pages} {041060} (\bibinfo {year} {2019})},\ \Eprint
  {https://arxiv.org/abs/1903.08284} {arXiv:1903.08284 [gr-qc]} \BibitemShut
  {NoStop}%
\bibitem [{\citenamefont {Carullo}\ \emph {et~al.}(2019)\citenamefont
  {Carullo}, \citenamefont {Del~Pozzo},\ and\ \citenamefont
  {Veitch}}]{Carullo:2019flw}%
  \BibitemOpen
  \bibfield  {author} {\bibinfo {author} {\bibfnamefont {G.}~\bibnamefont
  {Carullo}}, \bibinfo {author} {\bibfnamefont {W.}~\bibnamefont {Del~Pozzo}},\
  and\ \bibinfo {author} {\bibfnamefont {J.}~\bibnamefont {Veitch}},\
  }\bibfield  {title} {\bibinfo {title} {{Observational Black Hole
  Spectroscopy: A time-domain multimode analysis of GW150914}},\ }\href
  {https://doi.org/10.1103/PhysRevD.99.123029} {\bibfield  {journal} {\bibinfo
  {journal} {Phys. Rev. D}\ }\textbf {\bibinfo {volume} {99}},\ \bibinfo
  {pages} {123029} (\bibinfo {year} {2019})},\ \bibinfo {note} {[Erratum:
  Phys.Rev.D 100, 089903 (2019)]},\ \Eprint {https://arxiv.org/abs/1902.07527}
  {arXiv:1902.07527 [gr-qc]} \BibitemShut {NoStop}%
\bibitem [{\citenamefont {Finch}\ and\ \citenamefont
  {Moore}(2021)}]{Finch:2021qph}%
  \BibitemOpen
  \bibfield  {author} {\bibinfo {author} {\bibfnamefont {E.}~\bibnamefont
  {Finch}}\ and\ \bibinfo {author} {\bibfnamefont {C.~J.}\ \bibnamefont
  {Moore}},\ }\bibfield  {title} {\bibinfo {title} {{Frequency-domain analysis
  of black-hole ringdowns}},\ }\href
  {https://doi.org/10.1103/PhysRevD.104.123034} {\bibfield  {journal} {\bibinfo
   {journal} {Phys. Rev. D}\ }\textbf {\bibinfo {volume} {104}},\ \bibinfo
  {pages} {123034} (\bibinfo {year} {2021})},\ \Eprint
  {https://arxiv.org/abs/2108.09344} {arXiv:2108.09344 [gr-qc]} \BibitemShut
  {NoStop}%
\bibitem [{\citenamefont {Isi}\ and\ \citenamefont {Farr}(2021)}]{Isi:2021iql}%
  \BibitemOpen
  \bibfield  {author} {\bibinfo {author} {\bibfnamefont {M.}~\bibnamefont
  {Isi}}\ and\ \bibinfo {author} {\bibfnamefont {W.~M.}\ \bibnamefont {Farr}},\
  }\bibfield  {title} {\bibinfo {title} {{Analyzing black-hole ringdowns}},\
  }\href@noop {} {\  (\bibinfo {year} {2021})},\ \Eprint
  {https://arxiv.org/abs/2107.05609} {arXiv:2107.05609 [gr-qc]} \BibitemShut
  {NoStop}%
\bibitem [{\citenamefont {Isi}\ \emph {et~al.}(2019{\natexlab{b}})\citenamefont
  {Isi}, \citenamefont {Giesler}, \citenamefont {Farr}, \citenamefont
  {Scheel},\ and\ \citenamefont {Teukolsky}}]{Isi:2019aib}%
  \BibitemOpen
  \bibfield  {author} {\bibinfo {author} {\bibfnamefont {M.}~\bibnamefont
  {Isi}}, \bibinfo {author} {\bibfnamefont {M.}~\bibnamefont {Giesler}},
  \bibinfo {author} {\bibfnamefont {W.~M.}\ \bibnamefont {Farr}}, \bibinfo
  {author} {\bibfnamefont {M.~A.}\ \bibnamefont {Scheel}},\ and\ \bibinfo
  {author} {\bibfnamefont {S.~A.}\ \bibnamefont {Teukolsky}},\ }\bibfield
  {title} {\bibinfo {title} {{Testing the no-hair theorem with GW150914}},\
  }\href {https://doi.org/10.1103/PhysRevLett.123.111102} {\bibfield  {journal}
  {\bibinfo  {journal} {Phys. Rev. Lett.}\ }\textbf {\bibinfo {volume} {123}},\
  \bibinfo {pages} {111102} (\bibinfo {year} {2019}{\natexlab{b}})},\ \Eprint
  {https://arxiv.org/abs/1905.00869} {arXiv:1905.00869 [gr-qc]} \BibitemShut
  {NoStop}%
\bibitem [{\citenamefont {Ghosh}\ \emph {et~al.}(2021)\citenamefont {Ghosh},
  \citenamefont {Brito},\ and\ \citenamefont {Buonanno}}]{Ghosh:2021mrv}%
  \BibitemOpen
  \bibfield  {author} {\bibinfo {author} {\bibfnamefont {A.}~\bibnamefont
  {Ghosh}}, \bibinfo {author} {\bibfnamefont {R.}~\bibnamefont {Brito}},\ and\
  \bibinfo {author} {\bibfnamefont {A.}~\bibnamefont {Buonanno}},\ }\bibfield
  {title} {\bibinfo {title} {{Constraints on quasinormal-mode frequencies with
  LIGO-Virgo binary\textendash{}black-hole observations}},\ }\href
  {https://doi.org/10.1103/PhysRevD.103.124041} {\bibfield  {journal} {\bibinfo
   {journal} {Phys. Rev. D}\ }\textbf {\bibinfo {volume} {103}},\ \bibinfo
  {pages} {124041} (\bibinfo {year} {2021})},\ \Eprint
  {https://arxiv.org/abs/2104.01906} {arXiv:2104.01906 [gr-qc]} \BibitemShut
  {NoStop}%
\bibitem [{\citenamefont {Capano}\ \emph {et~al.}(2021)\citenamefont {Capano},
  \citenamefont {Cabero}, \citenamefont {Westerweck}, \citenamefont {Abedi},
  \citenamefont {Kastha}, \citenamefont {Nitz}, \citenamefont {Nielsen},\ and\
  \citenamefont {Krishnan}}]{Capano:2021etf}%
  \BibitemOpen
  \bibfield  {author} {\bibinfo {author} {\bibfnamefont {C.~D.}\ \bibnamefont
  {Capano}}, \bibinfo {author} {\bibfnamefont {M.}~\bibnamefont {Cabero}},
  \bibinfo {author} {\bibfnamefont {J.}~\bibnamefont {Westerweck}}, \bibinfo
  {author} {\bibfnamefont {J.}~\bibnamefont {Abedi}}, \bibinfo {author}
  {\bibfnamefont {S.}~\bibnamefont {Kastha}}, \bibinfo {author} {\bibfnamefont
  {A.~H.}\ \bibnamefont {Nitz}}, \bibinfo {author} {\bibfnamefont {A.~B.}\
  \bibnamefont {Nielsen}},\ and\ \bibinfo {author} {\bibfnamefont
  {B.}~\bibnamefont {Krishnan}},\ }\bibfield  {title} {\bibinfo {title}
  {{Observation of a multimode quasi-normal spectrum from a perturbed black
  hole}},\ }\href@noop {} {\  (\bibinfo {year} {2021})},\ \Eprint
  {https://arxiv.org/abs/2105.05238} {arXiv:2105.05238 [gr-qc]} \BibitemShut
  {NoStop}%
\bibitem [{\citenamefont {Cotesta}\ \emph {et~al.}(2022)\citenamefont
  {Cotesta}, \citenamefont {Carullo}, \citenamefont {Berti},\ and\
  \citenamefont {Cardoso}}]{Cotesta:2022pci}%
  \BibitemOpen
  \bibfield  {author} {\bibinfo {author} {\bibfnamefont {R.}~\bibnamefont
  {Cotesta}}, \bibinfo {author} {\bibfnamefont {G.}~\bibnamefont {Carullo}},
  \bibinfo {author} {\bibfnamefont {E.}~\bibnamefont {Berti}},\ and\ \bibinfo
  {author} {\bibfnamefont {V.}~\bibnamefont {Cardoso}},\ }\bibfield  {title}
  {\bibinfo {title} {{On the detection of ringdown overtones in GW150914}},\
  }\href@noop {} {\  (\bibinfo {year} {2022})},\ \Eprint
  {https://arxiv.org/abs/2201.00822} {arXiv:2201.00822 [gr-qc]} \BibitemShut
  {NoStop}%
\bibitem [{\citenamefont {Isi}\ and\ \citenamefont {Farr}(2022)}]{Isi:2022mhy}%
  \BibitemOpen
  \bibfield  {author} {\bibinfo {author} {\bibfnamefont {M.}~\bibnamefont
  {Isi}}\ and\ \bibinfo {author} {\bibfnamefont {W.~M.}\ \bibnamefont {Farr}},\
  }\bibfield  {title} {\bibinfo {title} {{Revisiting the ringdown of
  GW150914}},\ }\href@noop {} {\  (\bibinfo {year} {2022})},\ \Eprint
  {https://arxiv.org/abs/2202.02941} {arXiv:2202.02941 [gr-qc]} \BibitemShut
  {NoStop}%
\bibitem [{\citenamefont {Lanczos}(1938)}]{Cornelius:1938}%
  \BibitemOpen
  \bibfield  {author} {\bibinfo {author} {\bibfnamefont {C.}~\bibnamefont
  {Lanczos}},\ }\bibfield  {title} {\bibinfo {title} {A remarkable property of
  the riemann-christoffel tensor in four dimensions},\ }\href
  {http://www.jstor.org/stable/1968467} {\bibfield  {journal} {\bibinfo
  {journal} {Annals of Mathematics}\ }\textbf {\bibinfo {volume} {39}},\
  \bibinfo {pages} {842} (\bibinfo {year} {1938})}\BibitemShut {NoStop}%
\bibitem [{\citenamefont {Lovelock}(1971)}]{Lovelock_1971}%
  \BibitemOpen
  \bibfield  {author} {\bibinfo {author} {\bibfnamefont {D.}~\bibnamefont
  {Lovelock}},\ }\bibfield  {title} {\bibinfo {title} {The einstein tensor and
  its generalizations},\ }\href {https://doi.org/10.1063/1.1665613} {\bibfield
  {journal} {\bibinfo  {journal} {Journal of Mathematical Physics}\ }\textbf
  {\bibinfo {volume} {12}},\ \bibinfo {pages} {498} (\bibinfo {year}
  {1971})}\BibitemShut {NoStop}%
\bibitem [{\citenamefont {Deser}\ \emph {et~al.}(1982)\citenamefont {Deser},
  \citenamefont {Jackiw},\ and\ \citenamefont {Templeton}}]{Deser:1981wh}%
  \BibitemOpen
  \bibfield  {author} {\bibinfo {author} {\bibfnamefont {S.}~\bibnamefont
  {Deser}}, \bibinfo {author} {\bibfnamefont {R.}~\bibnamefont {Jackiw}},\ and\
  \bibinfo {author} {\bibfnamefont {S.}~\bibnamefont {Templeton}},\ }\bibfield
  {title} {\bibinfo {title} {{Topologically Massive Gauge Theories}},\ }\href
  {https://doi.org/10.1016/0003-4916(82)90164-6} {\bibfield  {journal}
  {\bibinfo  {journal} {Annals Phys.}\ }\textbf {\bibinfo {volume} {140}},\
  \bibinfo {pages} {372} (\bibinfo {year} {1982})},\ \bibinfo {note} {[Erratum:
  Annals Phys. 185, 406 (1988)]}\BibitemShut {NoStop}%
\bibitem [{\citenamefont {Kanti}\ \emph {et~al.}(1996)\citenamefont {Kanti},
  \citenamefont {Mavromatos}, \citenamefont {Rizos}, \citenamefont {Tamvakis},\
  and\ \citenamefont {Winstanley}}]{Kanti:1995vq}%
  \BibitemOpen
  \bibfield  {author} {\bibinfo {author} {\bibfnamefont {P.}~\bibnamefont
  {Kanti}}, \bibinfo {author} {\bibfnamefont {N.~E.}\ \bibnamefont
  {Mavromatos}}, \bibinfo {author} {\bibfnamefont {J.}~\bibnamefont {Rizos}},
  \bibinfo {author} {\bibfnamefont {K.}~\bibnamefont {Tamvakis}},\ and\
  \bibinfo {author} {\bibfnamefont {E.}~\bibnamefont {Winstanley}},\ }\bibfield
   {title} {\bibinfo {title} {{Dilatonic black holes in higher curvature string
  gravity}},\ }\href {https://doi.org/10.1103/PhysRevD.54.5049} {\bibfield
  {journal} {\bibinfo  {journal} {Phys. Rev. D}\ }\textbf {\bibinfo {volume}
  {54}},\ \bibinfo {pages} {5049} (\bibinfo {year} {1996})},\ \Eprint
  {https://arxiv.org/abs/hep-th/9511071} {arXiv:hep-th/9511071} \BibitemShut
  {NoStop}%
\bibitem [{\citenamefont {Jackiw}\ and\ \citenamefont
  {Pi}(2003)}]{Jackiw:2003pm}%
  \BibitemOpen
  \bibfield  {author} {\bibinfo {author} {\bibfnamefont {R.}~\bibnamefont
  {Jackiw}}\ and\ \bibinfo {author} {\bibfnamefont {S.~Y.}\ \bibnamefont
  {Pi}},\ }\bibfield  {title} {\bibinfo {title} {{Chern-Simons modification of
  general relativity}},\ }\href {https://doi.org/10.1103/PhysRevD.68.104012}
  {\bibfield  {journal} {\bibinfo  {journal} {Phys. Rev. D}\ }\textbf {\bibinfo
  {volume} {68}},\ \bibinfo {pages} {104012} (\bibinfo {year} {2003})},\
  \Eprint {https://arxiv.org/abs/gr-qc/0308071} {arXiv:gr-qc/0308071}
  \BibitemShut {NoStop}%
\bibitem [{\citenamefont {Alexander}\ and\ \citenamefont
  {Yunes}(2009)}]{Alexander:2009tp}%
  \BibitemOpen
  \bibfield  {author} {\bibinfo {author} {\bibfnamefont {S.}~\bibnamefont
  {Alexander}}\ and\ \bibinfo {author} {\bibfnamefont {N.}~\bibnamefont
  {Yunes}},\ }\bibfield  {title} {\bibinfo {title} {{Chern-Simons Modified
  General Relativity}},\ }\href {https://doi.org/10.1016/j.physrep.2009.07.002}
  {\bibfield  {journal} {\bibinfo  {journal} {Phys. Rept.}\ }\textbf {\bibinfo
  {volume} {480}},\ \bibinfo {pages} {1} (\bibinfo {year} {2009})},\ \Eprint
  {https://arxiv.org/abs/0907.2562} {arXiv:0907.2562 [hep-th]} \BibitemShut
  {NoStop}%
\bibitem [{\citenamefont {Cardoso}\ and\ \citenamefont
  {Gualtieri}(2009)}]{Cardoso:2009pk}%
  \BibitemOpen
  \bibfield  {author} {\bibinfo {author} {\bibfnamefont {V.}~\bibnamefont
  {Cardoso}}\ and\ \bibinfo {author} {\bibfnamefont {L.}~\bibnamefont
  {Gualtieri}},\ }\bibfield  {title} {\bibinfo {title} {{Perturbations of
  Schwarzschild black holes in Dynamical Chern-Simons modified gravity}},\
  }\href {https://doi.org/10.1103/PhysRevD.81.089903} {\bibfield  {journal}
  {\bibinfo  {journal} {Phys. Rev. D}\ }\textbf {\bibinfo {volume} {80}},\
  \bibinfo {pages} {064008} (\bibinfo {year} {2009})},\ \bibinfo {note}
  {[Erratum: Phys.Rev.D 81, 089903 (2010)]},\ \Eprint
  {https://arxiv.org/abs/0907.5008} {arXiv:0907.5008 [gr-qc]} \BibitemShut
  {NoStop}%
\bibitem [{\citenamefont {Molina}\ \emph {et~al.}(2010)\citenamefont {Molina},
  \citenamefont {Pani}, \citenamefont {Cardoso},\ and\ \citenamefont
  {Gualtieri}}]{Molina:2010}%
  \BibitemOpen
  \bibfield  {author} {\bibinfo {author} {\bibfnamefont {C.}~\bibnamefont
  {Molina}}, \bibinfo {author} {\bibfnamefont {P.}~\bibnamefont {Pani}},
  \bibinfo {author} {\bibfnamefont {V.}~\bibnamefont {Cardoso}},\ and\ \bibinfo
  {author} {\bibfnamefont {L.}~\bibnamefont {Gualtieri}},\ }\bibfield  {title}
  {\bibinfo {title} {{Gravitational signature of Schwarzschild black holes in
  dynamical Chern-Simons gravity}},\ }\href
  {https://doi.org/10.1103/PhysRevD.81.124021} {\bibfield  {journal} {\bibinfo
  {journal} {Phys. Rev. D}\ }\textbf {\bibinfo {volume} {81}},\ \bibinfo
  {pages} {124021} (\bibinfo {year} {2010})}\BibitemShut {NoStop}%
\bibitem [{\citenamefont {Cano}\ \emph {et~al.}(2020)\citenamefont {Cano},
  \citenamefont {Fransen},\ and\ \citenamefont {Hertog}}]{Cano:2020cao}%
  \BibitemOpen
  \bibfield  {author} {\bibinfo {author} {\bibfnamefont {P.~A.}\ \bibnamefont
  {Cano}}, \bibinfo {author} {\bibfnamefont {K.}~\bibnamefont {Fransen}},\ and\
  \bibinfo {author} {\bibfnamefont {T.}~\bibnamefont {Hertog}},\ }\bibfield
  {title} {\bibinfo {title} {{Ringing of rotating black holes in
  higher-derivative gravity}},\ }\href
  {https://doi.org/10.1103/PhysRevD.102.044047} {\bibfield  {journal} {\bibinfo
   {journal} {Phys. Rev. D}\ }\textbf {\bibinfo {volume} {102}},\ \bibinfo
  {pages} {044047} (\bibinfo {year} {2020})},\ \Eprint
  {https://arxiv.org/abs/2005.03671} {arXiv:2005.03671 [gr-qc]} \BibitemShut
  {NoStop}%
\bibitem [{\citenamefont {Wagle}\ \emph {et~al.}(2021)\citenamefont {Wagle},
  \citenamefont {Yunes},\ and\ \citenamefont {Silva}}]{Wagle:2021tam}%
  \BibitemOpen
  \bibfield  {author} {\bibinfo {author} {\bibfnamefont {P.}~\bibnamefont
  {Wagle}}, \bibinfo {author} {\bibfnamefont {N.}~\bibnamefont {Yunes}},\ and\
  \bibinfo {author} {\bibfnamefont {H.~O.}\ \bibnamefont {Silva}},\ }\bibfield
  {title} {\bibinfo {title} {{Quasinormal modes of slowly-rotating black holes
  in dynamical Chern-Simons gravity}},\ }\href@noop {} {\  (\bibinfo {year}
  {2021})},\ \Eprint {https://arxiv.org/abs/2103.09913} {arXiv:2103.09913
  [gr-qc]} \BibitemShut {NoStop}%
\bibitem [{\citenamefont {Srivastava}\ \emph {et~al.}(2021)\citenamefont
  {Srivastava}, \citenamefont {Chen},\ and\ \citenamefont
  {Shankaranarayanan}}]{Srivastava:2021imr}%
  \BibitemOpen
  \bibfield  {author} {\bibinfo {author} {\bibfnamefont {M.}~\bibnamefont
  {Srivastava}}, \bibinfo {author} {\bibfnamefont {Y.}~\bibnamefont {Chen}},\
  and\ \bibinfo {author} {\bibfnamefont {S.}~\bibnamefont
  {Shankaranarayanan}},\ }\bibfield  {title} {\bibinfo {title} {{Analytical
  computation of quasinormal modes of slowly rotating black holes in dynamical
  Chern-Simons gravity}},\ }\href {https://doi.org/10.1103/PhysRevD.104.064034}
  {\bibfield  {journal} {\bibinfo  {journal} {Phys. Rev. D}\ }\textbf {\bibinfo
  {volume} {104}},\ \bibinfo {pages} {064034} (\bibinfo {year} {2021})},\
  \Eprint {https://arxiv.org/abs/2106.06209} {arXiv:2106.06209 [gr-qc]}
  \BibitemShut {NoStop}%
\bibitem [{\citenamefont {Bl\'azquez-Salcedo}\ \emph
  {et~al.}(2017)\citenamefont {Bl\'azquez-Salcedo}, \citenamefont {Khoo},\ and\
  \citenamefont {Kunz}}]{Blazquez-Salcedo:2017txk}%
  \BibitemOpen
  \bibfield  {author} {\bibinfo {author} {\bibfnamefont {J.~L.}\ \bibnamefont
  {Bl\'azquez-Salcedo}}, \bibinfo {author} {\bibfnamefont {F.~S.}\ \bibnamefont
  {Khoo}},\ and\ \bibinfo {author} {\bibfnamefont {J.}~\bibnamefont {Kunz}},\
  }\bibfield  {title} {\bibinfo {title} {{Quasinormal modes of
  Einstein-Gauss-Bonnet-dilaton black holes}},\ }\href
  {https://doi.org/10.1103/PhysRevD.96.064008} {\bibfield  {journal} {\bibinfo
  {journal} {Phys. Rev. D}\ }\textbf {\bibinfo {volume} {96}},\ \bibinfo
  {pages} {064008} (\bibinfo {year} {2017})},\ \Eprint
  {https://arxiv.org/abs/1706.03262} {arXiv:1706.03262 [gr-qc]} \BibitemShut
  {NoStop}%
\bibitem [{\citenamefont {Bl\'azquez-Salcedo}\ \emph
  {et~al.}(2016)\citenamefont {Bl\'azquez-Salcedo}, \citenamefont {Macedo},
  \citenamefont {Cardoso}, \citenamefont {Ferrari}, \citenamefont {Gualtieri},
  \citenamefont {Khoo}, \citenamefont {Kunz},\ and\ \citenamefont
  {Pani}}]{Blazquez-Salcedo:2016enn}%
  \BibitemOpen
  \bibfield  {author} {\bibinfo {author} {\bibfnamefont {J.~L.}\ \bibnamefont
  {Bl\'azquez-Salcedo}}, \bibinfo {author} {\bibfnamefont {C.~F.~B.}\
  \bibnamefont {Macedo}}, \bibinfo {author} {\bibfnamefont {V.}~\bibnamefont
  {Cardoso}}, \bibinfo {author} {\bibfnamefont {V.}~\bibnamefont {Ferrari}},
  \bibinfo {author} {\bibfnamefont {L.}~\bibnamefont {Gualtieri}}, \bibinfo
  {author} {\bibfnamefont {F.~S.}\ \bibnamefont {Khoo}}, \bibinfo {author}
  {\bibfnamefont {J.}~\bibnamefont {Kunz}},\ and\ \bibinfo {author}
  {\bibfnamefont {P.}~\bibnamefont {Pani}},\ }\bibfield  {title} {\bibinfo
  {title} {{Perturbed black holes in Einstein-dilaton-Gauss-Bonnet gravity:
  Stability, ringdown, and gravitational-wave emission}},\ }\href
  {https://doi.org/10.1103/PhysRevD.94.104024} {\bibfield  {journal} {\bibinfo
  {journal} {Phys. Rev. D}\ }\textbf {\bibinfo {volume} {94}},\ \bibinfo
  {pages} {104024} (\bibinfo {year} {2016})},\ \Eprint
  {https://arxiv.org/abs/1609.01286} {arXiv:1609.01286 [gr-qc]} \BibitemShut
  {NoStop}%
\bibitem [{\citenamefont {Endlich}\ \emph {et~al.}(2017)\citenamefont
  {Endlich}, \citenamefont {Gorbenko}, \citenamefont {Huang},\ and\
  \citenamefont {Senatore}}]{Endlich:2017tqa}%
  \BibitemOpen
  \bibfield  {author} {\bibinfo {author} {\bibfnamefont {S.}~\bibnamefont
  {Endlich}}, \bibinfo {author} {\bibfnamefont {V.}~\bibnamefont {Gorbenko}},
  \bibinfo {author} {\bibfnamefont {J.}~\bibnamefont {Huang}},\ and\ \bibinfo
  {author} {\bibfnamefont {L.}~\bibnamefont {Senatore}},\ }\bibfield  {title}
  {\bibinfo {title} {{An effective formalism for testing extensions to General
  Relativity with gravitational waves}},\ }\href
  {https://doi.org/10.1007/JHEP09(2017)122} {\bibfield  {journal} {\bibinfo
  {journal} {JHEP}\ }\textbf {\bibinfo {volume} {09}},\ \bibinfo {pages}
  {122}},\ \Eprint {https://arxiv.org/abs/1704.01590} {arXiv:1704.01590
  [gr-qc]} \BibitemShut {NoStop}%
\bibitem [{\citenamefont {Cano}\ \emph {et~al.}(2022)\citenamefont {Cano},
  \citenamefont {Fransen}, \citenamefont {Hertog},\ and\ \citenamefont
  {Maenaut}}]{Cano:2021myl}%
  \BibitemOpen
  \bibfield  {author} {\bibinfo {author} {\bibfnamefont {P.~A.}\ \bibnamefont
  {Cano}}, \bibinfo {author} {\bibfnamefont {K.}~\bibnamefont {Fransen}},
  \bibinfo {author} {\bibfnamefont {T.}~\bibnamefont {Hertog}},\ and\ \bibinfo
  {author} {\bibfnamefont {S.}~\bibnamefont {Maenaut}},\ }\bibfield  {title}
  {\bibinfo {title} {{Gravitational ringing of rotating black holes in
  higher-derivative gravity}},\ }\href
  {https://doi.org/10.1103/PhysRevD.105.024064} {\bibfield  {journal} {\bibinfo
   {journal} {Phys. Rev. D}\ }\textbf {\bibinfo {volume} {105}},\ \bibinfo
  {pages} {024064} (\bibinfo {year} {2022})},\ \Eprint
  {https://arxiv.org/abs/2110.11378} {arXiv:2110.11378 [gr-qc]} \BibitemShut
  {NoStop}%
\bibitem [{\citenamefont {Okounkova}\ \emph {et~al.}(2017)\citenamefont
  {Okounkova}, \citenamefont {Stein}, \citenamefont {Scheel},\ and\
  \citenamefont {Hemberger}}]{Okounkova:2017yby}%
  \BibitemOpen
  \bibfield  {author} {\bibinfo {author} {\bibfnamefont {M.}~\bibnamefont
  {Okounkova}}, \bibinfo {author} {\bibfnamefont {L.~C.}\ \bibnamefont
  {Stein}}, \bibinfo {author} {\bibfnamefont {M.~A.}\ \bibnamefont {Scheel}},\
  and\ \bibinfo {author} {\bibfnamefont {D.~A.}\ \bibnamefont {Hemberger}},\
  }\bibfield  {title} {\bibinfo {title} {{Numerical binary black hole mergers
  in dynamical Chern-Simons gravity: Scalar field}},\ }\href
  {https://doi.org/10.1103/PhysRevD.96.044020} {\bibfield  {journal} {\bibinfo
  {journal} {Phys. Rev. D}\ }\textbf {\bibinfo {volume} {96}},\ \bibinfo
  {pages} {044020} (\bibinfo {year} {2017})},\ \Eprint
  {https://arxiv.org/abs/1705.07924} {arXiv:1705.07924 [gr-qc]} \BibitemShut
  {NoStop}%
\bibitem [{\citenamefont {Okounkova}\ \emph {et~al.}(2019)\citenamefont
  {Okounkova}, \citenamefont {Scheel},\ and\ \citenamefont
  {Teukolsky}}]{Okounkova:2018pql}%
  \BibitemOpen
  \bibfield  {author} {\bibinfo {author} {\bibfnamefont {M.}~\bibnamefont
  {Okounkova}}, \bibinfo {author} {\bibfnamefont {M.~A.}\ \bibnamefont
  {Scheel}},\ and\ \bibinfo {author} {\bibfnamefont {S.~A.}\ \bibnamefont
  {Teukolsky}},\ }\bibfield  {title} {\bibinfo {title} {{Evolving Metric
  Perturbations in dynamical Chern-Simons Gravity}},\ }\href
  {https://doi.org/10.1103/PhysRevD.99.044019} {\bibfield  {journal} {\bibinfo
  {journal} {Phys. Rev. D}\ }\textbf {\bibinfo {volume} {99}},\ \bibinfo
  {pages} {044019} (\bibinfo {year} {2019})},\ \Eprint
  {https://arxiv.org/abs/1811.10713} {arXiv:1811.10713 [gr-qc]} \BibitemShut
  {NoStop}%
\bibitem [{\citenamefont {Okounkova}\ \emph {et~al.}(2020)\citenamefont
  {Okounkova}, \citenamefont {Stein}, \citenamefont {Moxon}, \citenamefont
  {Scheel},\ and\ \citenamefont {Teukolsky}}]{Okounkova:2019zjf}%
  \BibitemOpen
  \bibfield  {author} {\bibinfo {author} {\bibfnamefont {M.}~\bibnamefont
  {Okounkova}}, \bibinfo {author} {\bibfnamefont {L.~C.}\ \bibnamefont
  {Stein}}, \bibinfo {author} {\bibfnamefont {J.}~\bibnamefont {Moxon}},
  \bibinfo {author} {\bibfnamefont {M.~A.}\ \bibnamefont {Scheel}},\ and\
  \bibinfo {author} {\bibfnamefont {S.~A.}\ \bibnamefont {Teukolsky}},\
  }\bibfield  {title} {\bibinfo {title} {{Numerical relativity simulation of
  GW150914 beyond general relativity}},\ }\href
  {https://doi.org/10.1103/PhysRevD.101.104016} {\bibfield  {journal} {\bibinfo
   {journal} {Phys. Rev. D}\ }\textbf {\bibinfo {volume} {101}},\ \bibinfo
  {pages} {104016} (\bibinfo {year} {2020})},\ \Eprint
  {https://arxiv.org/abs/1911.02588} {arXiv:1911.02588 [gr-qc]} \BibitemShut
  {NoStop}%
\bibitem [{\citenamefont {Witek}\ \emph {et~al.}(2019)\citenamefont {Witek},
  \citenamefont {Gualtieri}, \citenamefont {Pani},\ and\ \citenamefont
  {Sotiriou}}]{Witek:2018dmd}%
  \BibitemOpen
  \bibfield  {author} {\bibinfo {author} {\bibfnamefont {H.}~\bibnamefont
  {Witek}}, \bibinfo {author} {\bibfnamefont {L.}~\bibnamefont {Gualtieri}},
  \bibinfo {author} {\bibfnamefont {P.}~\bibnamefont {Pani}},\ and\ \bibinfo
  {author} {\bibfnamefont {T.~P.}\ \bibnamefont {Sotiriou}},\ }\bibfield
  {title} {\bibinfo {title} {{Black holes and binary mergers in scalar
  Gauss-Bonnet gravity: scalar field dynamics}},\ }\href
  {https://doi.org/10.1103/PhysRevD.99.064035} {\bibfield  {journal} {\bibinfo
  {journal} {Phys. Rev. D}\ }\textbf {\bibinfo {volume} {99}},\ \bibinfo
  {pages} {064035} (\bibinfo {year} {2019})},\ \Eprint
  {https://arxiv.org/abs/1810.05177} {arXiv:1810.05177 [gr-qc]} \BibitemShut
  {NoStop}%
\bibitem [{\citenamefont {Okounkova}(2020)}]{Okounkova:2020rqw}%
  \BibitemOpen
  \bibfield  {author} {\bibinfo {author} {\bibfnamefont {M.}~\bibnamefont
  {Okounkova}},\ }\bibfield  {title} {\bibinfo {title} {{Numerical relativity
  simulation of GW150914 in Einstein dilaton Gauss-Bonnet gravity}},\ }\href
  {https://doi.org/10.1103/PhysRevD.102.084046} {\bibfield  {journal} {\bibinfo
   {journal} {Phys. Rev. D}\ }\textbf {\bibinfo {volume} {102}},\ \bibinfo
  {pages} {084046} (\bibinfo {year} {2020})},\ \Eprint
  {https://arxiv.org/abs/2001.03571} {arXiv:2001.03571 [gr-qc]} \BibitemShut
  {NoStop}%
\bibitem [{\citenamefont {Silva}\ \emph {et~al.}(2021)\citenamefont {Silva},
  \citenamefont {Witek}, \citenamefont {Elley},\ and\ \citenamefont
  {Yunes}}]{Silva:2020omi}%
  \BibitemOpen
  \bibfield  {author} {\bibinfo {author} {\bibfnamefont {H.~O.}\ \bibnamefont
  {Silva}}, \bibinfo {author} {\bibfnamefont {H.}~\bibnamefont {Witek}},
  \bibinfo {author} {\bibfnamefont {M.}~\bibnamefont {Elley}},\ and\ \bibinfo
  {author} {\bibfnamefont {N.}~\bibnamefont {Yunes}},\ }\bibfield  {title}
  {\bibinfo {title} {{Dynamical Descalarization in Binary Black Hole
  Mergers}},\ }\href {https://doi.org/10.1103/PhysRevLett.127.031101}
  {\bibfield  {journal} {\bibinfo  {journal} {Phys. Rev. Lett.}\ }\textbf
  {\bibinfo {volume} {127}},\ \bibinfo {pages} {031101} (\bibinfo {year}
  {2021})},\ \Eprint {https://arxiv.org/abs/2012.10436} {arXiv:2012.10436
  [gr-qc]} \BibitemShut {NoStop}%
\bibitem [{\citenamefont {Witek}\ \emph {et~al.}(2020)\citenamefont {Witek},
  \citenamefont {Gualtieri},\ and\ \citenamefont {Pani}}]{Witek:2020uzz}%
  \BibitemOpen
  \bibfield  {author} {\bibinfo {author} {\bibfnamefont {H.}~\bibnamefont
  {Witek}}, \bibinfo {author} {\bibfnamefont {L.}~\bibnamefont {Gualtieri}},\
  and\ \bibinfo {author} {\bibfnamefont {P.}~\bibnamefont {Pani}},\ }\bibfield
  {title} {\bibinfo {title} {{Towards numerical relativity in scalar
  Gauss-Bonnet gravity: $3+1$ decomposition beyond the small-coupling limit}},\
  }\href {https://doi.org/10.1103/PhysRevD.101.124055} {\bibfield  {journal}
  {\bibinfo  {journal} {Phys. Rev. D}\ }\textbf {\bibinfo {volume} {101}},\
  \bibinfo {pages} {124055} (\bibinfo {year} {2020})},\ \Eprint
  {https://arxiv.org/abs/2004.00009} {arXiv:2004.00009 [gr-qc]} \BibitemShut
  {NoStop}%
\bibitem [{\citenamefont {Cayuso}\ \emph {et~al.}(2017)\citenamefont {Cayuso},
  \citenamefont {Ortiz},\ and\ \citenamefont {Lehner}}]{Cayuso:2017iqc}%
  \BibitemOpen
  \bibfield  {author} {\bibinfo {author} {\bibfnamefont {J.}~\bibnamefont
  {Cayuso}}, \bibinfo {author} {\bibfnamefont {N.}~\bibnamefont {Ortiz}},\ and\
  \bibinfo {author} {\bibfnamefont {L.}~\bibnamefont {Lehner}},\ }\bibfield
  {title} {\bibinfo {title} {{Fixing extensions to general relativity in the
  nonlinear regime}},\ }\href {https://doi.org/10.1103/PhysRevD.96.084043}
  {\bibfield  {journal} {\bibinfo  {journal} {Phys. Rev. D}\ }\textbf {\bibinfo
  {volume} {96}},\ \bibinfo {pages} {084043} (\bibinfo {year} {2017})},\
  \Eprint {https://arxiv.org/abs/1706.07421} {arXiv:1706.07421 [gr-qc]}
  \BibitemShut {NoStop}%
\bibitem [{\citenamefont {Cayuso}\ and\ \citenamefont
  {Lehner}(2020)}]{Cayuso:2020lca}%
  \BibitemOpen
  \bibfield  {author} {\bibinfo {author} {\bibfnamefont {R.}~\bibnamefont
  {Cayuso}}\ and\ \bibinfo {author} {\bibfnamefont {L.}~\bibnamefont
  {Lehner}},\ }\bibfield  {title} {\bibinfo {title} {{Nonlinear, noniterative
  treatment of EFT-motivated gravity}},\ }\href
  {https://doi.org/10.1103/PhysRevD.102.084008} {\bibfield  {journal} {\bibinfo
   {journal} {Phys. Rev. D}\ }\textbf {\bibinfo {volume} {102}},\ \bibinfo
  {pages} {084008} (\bibinfo {year} {2020})},\ \Eprint
  {https://arxiv.org/abs/2005.13720} {arXiv:2005.13720 [gr-qc]} \BibitemShut
  {NoStop}%
\bibitem [{\citenamefont {East}\ and\ \citenamefont
  {Ripley}(2021{\natexlab{a}})}]{East:2020hgw}%
  \BibitemOpen
  \bibfield  {author} {\bibinfo {author} {\bibfnamefont {W.~E.}\ \bibnamefont
  {East}}\ and\ \bibinfo {author} {\bibfnamefont {J.~L.}\ \bibnamefont
  {Ripley}},\ }\bibfield  {title} {\bibinfo {title} {{Evolution of
  Einstein-scalar-Gauss-Bonnet gravity using a modified harmonic
  formulation}},\ }\href {https://doi.org/10.1103/PhysRevD.103.044040}
  {\bibfield  {journal} {\bibinfo  {journal} {Phys. Rev. D}\ }\textbf {\bibinfo
  {volume} {103}},\ \bibinfo {pages} {044040} (\bibinfo {year}
  {2021}{\natexlab{a}})},\ \Eprint {https://arxiv.org/abs/2011.03547}
  {arXiv:2011.03547 [gr-qc]} \BibitemShut {NoStop}%
\bibitem [{\citenamefont {East}\ and\ \citenamefont
  {Ripley}(2021{\natexlab{b}})}]{East:2021bqk}%
  \BibitemOpen
  \bibfield  {author} {\bibinfo {author} {\bibfnamefont {W.~E.}\ \bibnamefont
  {East}}\ and\ \bibinfo {author} {\bibfnamefont {J.~L.}\ \bibnamefont
  {Ripley}},\ }\bibfield  {title} {\bibinfo {title} {{Dynamics of Spontaneous
  Black Hole Scalarization and Mergers in Einstein-Scalar-Gauss-Bonnet
  Gravity}},\ }\href {https://doi.org/10.1103/PhysRevLett.127.101102}
  {\bibfield  {journal} {\bibinfo  {journal} {Phys. Rev. Lett.}\ }\textbf
  {\bibinfo {volume} {127}},\ \bibinfo {pages} {101102} (\bibinfo {year}
  {2021}{\natexlab{b}})},\ \Eprint {https://arxiv.org/abs/2105.08571}
  {arXiv:2105.08571 [gr-qc]} \BibitemShut {NoStop}%
\bibitem [{\citenamefont {Hofmann}\ \emph {et~al.}(2016)\citenamefont
  {Hofmann}, \citenamefont {Barausse},\ and\ \citenamefont
  {Rezzolla}}]{Hofmann:2016yih}%
  \BibitemOpen
  \bibfield  {author} {\bibinfo {author} {\bibfnamefont {F.}~\bibnamefont
  {Hofmann}}, \bibinfo {author} {\bibfnamefont {E.}~\bibnamefont {Barausse}},\
  and\ \bibinfo {author} {\bibfnamefont {L.}~\bibnamefont {Rezzolla}},\
  }\bibfield  {title} {\bibinfo {title} {{The final spin from binary black
  holes in quasi-circular orbits}},\ }\href
  {https://doi.org/10.3847/2041-8205/825/2/L19} {\bibfield  {journal} {\bibinfo
   {journal} {Astrophys. J. Lett.}\ }\textbf {\bibinfo {volume} {825}},\
  \bibinfo {pages} {L19} (\bibinfo {year} {2016})},\ \Eprint
  {https://arxiv.org/abs/1605.01938} {arXiv:1605.01938 [gr-qc]} \BibitemShut
  {NoStop}%
\bibitem [{\citenamefont {Fishbach}\ \emph {et~al.}(2017)\citenamefont
  {Fishbach}, \citenamefont {Holz},\ and\ \citenamefont
  {Farr}}]{Fishbach:2017dwv}%
  \BibitemOpen
  \bibfield  {author} {\bibinfo {author} {\bibfnamefont {M.}~\bibnamefont
  {Fishbach}}, \bibinfo {author} {\bibfnamefont {D.~E.}\ \bibnamefont {Holz}},\
  and\ \bibinfo {author} {\bibfnamefont {B.}~\bibnamefont {Farr}},\ }\bibfield
  {title} {\bibinfo {title} {{Are LIGO's Black Holes Made From Smaller Black
  Holes?}},\ }\href {https://doi.org/10.3847/2041-8213/aa7045} {\bibfield
  {journal} {\bibinfo  {journal} {Astrophys. J. Lett.}\ }\textbf {\bibinfo
  {volume} {840}},\ \bibinfo {pages} {L24} (\bibinfo {year} {2017})},\ \Eprint
  {https://arxiv.org/abs/1703.06869} {arXiv:1703.06869 [astro-ph.HE]}
  \BibitemShut {NoStop}%
\bibitem [{\citenamefont {Cano}\ and\ \citenamefont
  {Ruip\'erez}(2019)}]{Cano:2019ore}%
  \BibitemOpen
  \bibfield  {author} {\bibinfo {author} {\bibfnamefont {P.~A.}\ \bibnamefont
  {Cano}}\ and\ \bibinfo {author} {\bibfnamefont {A.}~\bibnamefont
  {Ruip\'erez}},\ }\bibfield  {title} {\bibinfo {title} {{Leading
  higher-derivative corrections to Kerr geometry}},\ }\href
  {https://doi.org/10.1007/JHEP05(2019)189} {\bibfield  {journal} {\bibinfo
  {journal} {JHEP}\ }\textbf {\bibinfo {volume} {05}},\ \bibinfo {pages}
  {189}},\ \bibinfo {note} {[Erratum: JHEP 03, 187 (2020)]},\ \Eprint
  {https://arxiv.org/abs/1901.01315} {arXiv:1901.01315 [gr-qc]} \BibitemShut
  {NoStop}%
\bibitem [{\citenamefont {Andersson}\ and\ \citenamefont
  {Glampedakis}(2000)}]{Andersson:1999wj}%
  \BibitemOpen
  \bibfield  {author} {\bibinfo {author} {\bibfnamefont {N.}~\bibnamefont
  {Andersson}}\ and\ \bibinfo {author} {\bibfnamefont {K.}~\bibnamefont
  {Glampedakis}},\ }\bibfield  {title} {\bibinfo {title} {{A Superradiance
  resonance cavity outside rapidly rotating black holes}},\ }\href
  {https://doi.org/10.1103/PhysRevLett.84.4537} {\bibfield  {journal} {\bibinfo
   {journal} {Phys. Rev. Lett.}\ }\textbf {\bibinfo {volume} {84}},\ \bibinfo
  {pages} {4537} (\bibinfo {year} {2000})},\ \Eprint
  {https://arxiv.org/abs/gr-qc/9909050} {arXiv:gr-qc/9909050} \BibitemShut
  {NoStop}%
\bibitem [{\citenamefont {Yang}\ \emph {et~al.}(2013)\citenamefont {Yang},
  \citenamefont {Zimmerman}, \citenamefont {Zengino\u{g}lu}, \citenamefont
  {Zhang}, \citenamefont {Berti},\ and\ \citenamefont {Chen}}]{Yang:2013uba}%
  \BibitemOpen
  \bibfield  {author} {\bibinfo {author} {\bibfnamefont {H.}~\bibnamefont
  {Yang}}, \bibinfo {author} {\bibfnamefont {A.}~\bibnamefont {Zimmerman}},
  \bibinfo {author} {\bibfnamefont {A.}~\bibnamefont {Zengino\u{g}lu}},
  \bibinfo {author} {\bibfnamefont {F.}~\bibnamefont {Zhang}}, \bibinfo
  {author} {\bibfnamefont {E.}~\bibnamefont {Berti}},\ and\ \bibinfo {author}
  {\bibfnamefont {Y.}~\bibnamefont {Chen}},\ }\bibfield  {title} {\bibinfo
  {title} {{Quasinormal modes of nearly extremal Kerr spacetimes: spectrum
  bifurcation and power-law ringdown}},\ }\href
  {https://doi.org/10.1103/PhysRevD.88.044047} {\bibfield  {journal} {\bibinfo
  {journal} {Phys. Rev. D}\ }\textbf {\bibinfo {volume} {88}},\ \bibinfo
  {pages} {044047} (\bibinfo {year} {2013})},\ \Eprint
  {https://arxiv.org/abs/1307.8086} {arXiv:1307.8086 [gr-qc]} \BibitemShut
  {NoStop}%
\bibitem [{\citenamefont {Gralla}\ \emph {et~al.}(2016)\citenamefont {Gralla},
  \citenamefont {Zimmerman},\ and\ \citenamefont {Zimmerman}}]{Gralla:2016sxp}%
  \BibitemOpen
  \bibfield  {author} {\bibinfo {author} {\bibfnamefont {S.~E.}\ \bibnamefont
  {Gralla}}, \bibinfo {author} {\bibfnamefont {A.}~\bibnamefont {Zimmerman}},\
  and\ \bibinfo {author} {\bibfnamefont {P.}~\bibnamefont {Zimmerman}},\
  }\bibfield  {title} {\bibinfo {title} {{Transient Instability of Rapidly
  Rotating Black Holes}},\ }\href {https://doi.org/10.1103/PhysRevD.94.084017}
  {\bibfield  {journal} {\bibinfo  {journal} {Phys. Rev. D}\ }\textbf {\bibinfo
  {volume} {94}},\ \bibinfo {pages} {084017} (\bibinfo {year} {2016})},\
  \Eprint {https://arxiv.org/abs/1608.04739} {arXiv:1608.04739 [gr-qc]}
  \BibitemShut {NoStop}%
\bibitem [{\citenamefont {Comp\`ere}\ \emph {et~al.}(2018)\citenamefont
  {Comp\`ere}, \citenamefont {Fransen}, \citenamefont {Hertog},\ and\
  \citenamefont {Long}}]{Compere:2017hsi}%
  \BibitemOpen
  \bibfield  {author} {\bibinfo {author} {\bibfnamefont {G.}~\bibnamefont
  {Comp\`ere}}, \bibinfo {author} {\bibfnamefont {K.}~\bibnamefont {Fransen}},
  \bibinfo {author} {\bibfnamefont {T.}~\bibnamefont {Hertog}},\ and\ \bibinfo
  {author} {\bibfnamefont {J.}~\bibnamefont {Long}},\ }\bibfield  {title}
  {\bibinfo {title} {{Gravitational waves from plunges into Gargantua}},\
  }\href {https://doi.org/10.1088/1361-6382/aab99e} {\bibfield  {journal}
  {\bibinfo  {journal} {Class. Quant. Grav.}\ }\textbf {\bibinfo {volume}
  {35}},\ \bibinfo {pages} {104002} (\bibinfo {year} {2018})},\ \Eprint
  {https://arxiv.org/abs/1712.07130} {arXiv:1712.07130 [gr-qc]} \BibitemShut
  {NoStop}%
\bibitem [{\citenamefont {Gralla}\ and\ \citenamefont
  {Zimmerman}(2018)}]{Gralla:2017lto}%
  \BibitemOpen
  \bibfield  {author} {\bibinfo {author} {\bibfnamefont {S.~E.}\ \bibnamefont
  {Gralla}}\ and\ \bibinfo {author} {\bibfnamefont {P.}~\bibnamefont
  {Zimmerman}},\ }\bibfield  {title} {\bibinfo {title} {{Critical Exponents of
  Extremal Kerr Perturbations}},\ }\href
  {https://doi.org/10.1088/1361-6382/aab140} {\bibfield  {journal} {\bibinfo
  {journal} {Class. Quant. Grav.}\ }\textbf {\bibinfo {volume} {35}},\ \bibinfo
  {pages} {095002} (\bibinfo {year} {2018})},\ \Eprint
  {https://arxiv.org/abs/1711.00855} {arXiv:1711.00855 [gr-qc]} \BibitemShut
  {NoStop}%
\bibitem [{\citenamefont {Zimmerman}\ \emph {et~al.}(2015)\citenamefont
  {Zimmerman}, \citenamefont {Yang}, \citenamefont {Mark}, \citenamefont
  {Chen},\ and\ \citenamefont {Lehner}}]{Zimmerman:2014aha}%
  \BibitemOpen
  \bibfield  {author} {\bibinfo {author} {\bibfnamefont {A.}~\bibnamefont
  {Zimmerman}}, \bibinfo {author} {\bibfnamefont {H.}~\bibnamefont {Yang}},
  \bibinfo {author} {\bibfnamefont {Z.}~\bibnamefont {Mark}}, \bibinfo {author}
  {\bibfnamefont {Y.}~\bibnamefont {Chen}},\ and\ \bibinfo {author}
  {\bibfnamefont {L.}~\bibnamefont {Lehner}},\ }\bibfield  {title} {\bibinfo
  {title} {{Quasinormal Modes Beyond Kerr}},\ }\href
  {https://doi.org/10.1007/978-3-319-10488-1_19} {\bibfield  {journal}
  {\bibinfo  {journal} {Astrophys. Space Sci. Proc.}\ }\textbf {\bibinfo
  {volume} {40}},\ \bibinfo {pages} {217} (\bibinfo {year} {2015})},\ \Eprint
  {https://arxiv.org/abs/1406.4206} {arXiv:1406.4206 [gr-qc]} \BibitemShut
  {NoStop}%
\bibitem [{\citenamefont {Mark}\ \emph {et~al.}(2015)\citenamefont {Mark},
  \citenamefont {Yang}, \citenamefont {Zimmerman},\ and\ \citenamefont
  {Chen}}]{Mark:2014aja}%
  \BibitemOpen
  \bibfield  {author} {\bibinfo {author} {\bibfnamefont {Z.}~\bibnamefont
  {Mark}}, \bibinfo {author} {\bibfnamefont {H.}~\bibnamefont {Yang}}, \bibinfo
  {author} {\bibfnamefont {A.}~\bibnamefont {Zimmerman}},\ and\ \bibinfo
  {author} {\bibfnamefont {Y.}~\bibnamefont {Chen}},\ }\bibfield  {title}
  {\bibinfo {title} {{Quasinormal modes of weakly charged Kerr-Newman
  spacetimes}},\ }\href {https://doi.org/10.1103/PhysRevD.91.044025} {\bibfield
   {journal} {\bibinfo  {journal} {Phys.Rev.}\ }\textbf {\bibinfo {volume}
  {D91}},\ \bibinfo {pages} {044025} (\bibinfo {year} {2015})},\ \Eprint
  {https://arxiv.org/abs/1409.5800} {arXiv:1409.5800 [gr-qc]} \BibitemShut
  {NoStop}%
\bibitem [{\citenamefont {Yang}\ \emph
  {et~al.}(2015{\natexlab{a}})\citenamefont {Yang}, \citenamefont {Zimmerman},\
  and\ \citenamefont {Lehner}}]{Yang:2014tla}%
  \BibitemOpen
  \bibfield  {author} {\bibinfo {author} {\bibfnamefont {H.}~\bibnamefont
  {Yang}}, \bibinfo {author} {\bibfnamefont {A.}~\bibnamefont {Zimmerman}},\
  and\ \bibinfo {author} {\bibfnamefont {L.}~\bibnamefont {Lehner}},\
  }\bibfield  {title} {\bibinfo {title} {{Turbulent Black Holes}},\ }\href
  {https://doi.org/10.1103/PhysRevLett.114.081101} {\bibfield  {journal}
  {\bibinfo  {journal} {Phys. Rev. Lett.}\ }\textbf {\bibinfo {volume} {114}},\
  \bibinfo {pages} {081101} (\bibinfo {year} {2015}{\natexlab{a}})},\ \Eprint
  {https://arxiv.org/abs/1402.4859} {arXiv:1402.4859 [gr-qc]} \BibitemShut
  {NoStop}%
\bibitem [{\citenamefont {Yang}\ \emph
  {et~al.}(2015{\natexlab{b}})\citenamefont {Yang}, \citenamefont {Zhang},
  \citenamefont {Green},\ and\ \citenamefont {Lehner}}]{Yang:2015jja}%
  \BibitemOpen
  \bibfield  {author} {\bibinfo {author} {\bibfnamefont {H.}~\bibnamefont
  {Yang}}, \bibinfo {author} {\bibfnamefont {F.}~\bibnamefont {Zhang}},
  \bibinfo {author} {\bibfnamefont {S.~R.}\ \bibnamefont {Green}},\ and\
  \bibinfo {author} {\bibfnamefont {L.}~\bibnamefont {Lehner}},\ }\bibfield
  {title} {\bibinfo {title} {{Coupled Oscillator Model for Nonlinear
  Gravitational Perturbations}},\ }\href
  {https://doi.org/10.1103/PhysRevD.91.084007} {\bibfield  {journal} {\bibinfo
  {journal} {Phys. Rev.}\ }\textbf {\bibinfo {volume} {D91}},\ \bibinfo {pages}
  {084007} (\bibinfo {year} {2015}{\natexlab{b}})},\ \Eprint
  {https://arxiv.org/abs/1502.08051} {arXiv:1502.08051 [gr-qc]} \BibitemShut
  {NoStop}%
\bibitem [{\citenamefont {Teukolsky}(1973)}]{Teukolsky:1973ha}%
  \BibitemOpen
  \bibfield  {author} {\bibinfo {author} {\bibfnamefont {S.~A.}\ \bibnamefont
  {Teukolsky}},\ }\bibfield  {title} {\bibinfo {title} {{Perturbations of a
  rotating black hole. 1. Fundamental equations for gravitational
  electromagnetic and neutrino field perturbations}},\ }\href
  {https://doi.org/10.1086/152444} {\bibfield  {journal} {\bibinfo  {journal}
  {Astrophys. J.}\ }\textbf {\bibinfo {volume} {185}},\ \bibinfo {pages} {635}
  (\bibinfo {year} {1973})}\BibitemShut {NoStop}%
\bibitem [{\citenamefont {Newman}\ and\ \citenamefont
  {Penrose}(1962)}]{Newman:1961qr}%
  \BibitemOpen
  \bibfield  {author} {\bibinfo {author} {\bibfnamefont {E.}~\bibnamefont
  {Newman}}\ and\ \bibinfo {author} {\bibfnamefont {R.}~\bibnamefont
  {Penrose}},\ }\bibfield  {title} {\bibinfo {title} {{An Approach to
  gravitational radiation by a method of spin coefficients}},\ }\href
  {https://doi.org/10.1063/1.1724257} {\bibfield  {journal} {\bibinfo
  {journal} {J. Math. Phys.}\ }\textbf {\bibinfo {volume} {3}},\ \bibinfo
  {pages} {566} (\bibinfo {year} {1962})}\BibitemShut {NoStop}%
\bibitem [{\citenamefont {Li}\ \emph {et~al.}()\citenamefont {Li},
  \citenamefont {Wagle}, \citenamefont {Chen},\ and\ \citenamefont
  {Yunes}}]{Li2022InPrep}%
  \BibitemOpen
  \bibfield  {author} {\bibinfo {author} {\bibfnamefont {D.}~\bibnamefont
  {Li}}, \bibinfo {author} {\bibfnamefont {P.}~\bibnamefont {Wagle}}, \bibinfo
  {author} {\bibfnamefont {Y.}~\bibnamefont {Chen}},\ and\ \bibinfo {author}
  {\bibfnamefont {N.}~\bibnamefont {Yunes}},\ }\href@noop {} {\bibinfo {title}
  {{Perturbations of spinning black holes beyond general relativity: Modified
  Teukolsky equation}}}\BibitemShut {NoStop}%
\bibitem [{\citenamefont {Chrzanowski}(1975)}]{Chrzanowski:1975wv}%
  \BibitemOpen
  \bibfield  {author} {\bibinfo {author} {\bibfnamefont {P.~L.}\ \bibnamefont
  {Chrzanowski}},\ }\bibfield  {title} {\bibinfo {title} {{Vector Potential and
  Metric Perturbations of a Rotating Black Hole}},\ }\href
  {https://doi.org/10.1103/PhysRevD.11.2042} {\bibfield  {journal} {\bibinfo
  {journal} {Phys. Rev. D}\ }\textbf {\bibinfo {volume} {11}},\ \bibinfo
  {pages} {2042} (\bibinfo {year} {1975})}\BibitemShut {NoStop}%
\bibitem [{\citenamefont {Wald}(1978)}]{Wald:1978vm}%
  \BibitemOpen
  \bibfield  {author} {\bibinfo {author} {\bibfnamefont {R.~M.}\ \bibnamefont
  {Wald}},\ }\bibfield  {title} {\bibinfo {title} {{Construction of Solutions
  of Gravitational, Electromagnetic, Or Other Perturbation Equations from
  Solutions of Decoupled Equations}},\ }\href
  {https://doi.org/10.1103/PhysRevLett.41.203} {\bibfield  {journal} {\bibinfo
  {journal} {Phys. Rev. Lett.}\ }\textbf {\bibinfo {volume} {41}},\ \bibinfo
  {pages} {203} (\bibinfo {year} {1978})}\BibitemShut {NoStop}%
\bibitem [{\citenamefont {Ori}(2003)}]{Ori:2002uv}%
  \BibitemOpen
  \bibfield  {author} {\bibinfo {author} {\bibfnamefont {A.}~\bibnamefont
  {Ori}},\ }\bibfield  {title} {\bibinfo {title} {{Reconstruction of
  inhomogeneous metric perturbations and electromagnetic four potential in Kerr
  space-time}},\ }\href {https://doi.org/10.1103/PhysRevD.67.124010} {\bibfield
   {journal} {\bibinfo  {journal} {Phys. Rev. D}\ }\textbf {\bibinfo {volume}
  {67}},\ \bibinfo {pages} {124010} (\bibinfo {year} {2003})},\ \Eprint
  {https://arxiv.org/abs/gr-qc/0207045} {arXiv:gr-qc/0207045} \BibitemShut
  {NoStop}%
\bibitem [{\citenamefont {Keidl}\ \emph {et~al.}(2010)\citenamefont {Keidl},
  \citenamefont {Shah}, \citenamefont {Friedman}, \citenamefont {Kim},\ and\
  \citenamefont {Price}}]{Keidl:2010pm}%
  \BibitemOpen
  \bibfield  {author} {\bibinfo {author} {\bibfnamefont {T.~S.}\ \bibnamefont
  {Keidl}}, \bibinfo {author} {\bibfnamefont {A.~G.}\ \bibnamefont {Shah}},
  \bibinfo {author} {\bibfnamefont {J.~L.}\ \bibnamefont {Friedman}}, \bibinfo
  {author} {\bibfnamefont {D.-H.}\ \bibnamefont {Kim}},\ and\ \bibinfo {author}
  {\bibfnamefont {L.~R.}\ \bibnamefont {Price}},\ }\bibfield  {title} {\bibinfo
  {title} {{Gravitational Self-force in a Radiation Gauge}},\ }\href
  {https://doi.org/10.1103/PhysRevD.82.124012} {\bibfield  {journal} {\bibinfo
  {journal} {Phys. Rev. D}\ }\textbf {\bibinfo {volume} {82}},\ \bibinfo
  {pages} {124012} (\bibinfo {year} {2010})},\ \Eprint
  {https://arxiv.org/abs/1004.2276} {arXiv:1004.2276 [gr-qc]} \BibitemShut
  {NoStop}%
\bibitem [{\citenamefont {Stein}(2014)}]{Stein:2014xba}%
  \BibitemOpen
  \bibfield  {author} {\bibinfo {author} {\bibfnamefont {L.~C.}\ \bibnamefont
  {Stein}},\ }\bibfield  {title} {\bibinfo {title} {{Rapidly rotating black
  holes in dynamical Chern-Simons gravity: Decoupling limit solutions and
  breakdown}},\ }\href {https://doi.org/10.1103/PhysRevD.90.044061} {\bibfield
  {journal} {\bibinfo  {journal} {Phys. Rev. D}\ }\textbf {\bibinfo {volume}
  {90}},\ \bibinfo {pages} {044061} (\bibinfo {year} {2014})},\ \Eprint
  {https://arxiv.org/abs/1407.2350} {arXiv:1407.2350 [gr-qc]} \BibitemShut
  {NoStop}%
\bibitem [{\citenamefont {Yagi}\ \emph {et~al.}(2012)\citenamefont {Yagi},
  \citenamefont {Yunes},\ and\ \citenamefont {Tanaka}}]{Yagi:2012ya}%
  \BibitemOpen
  \bibfield  {author} {\bibinfo {author} {\bibfnamefont {K.}~\bibnamefont
  {Yagi}}, \bibinfo {author} {\bibfnamefont {N.}~\bibnamefont {Yunes}},\ and\
  \bibinfo {author} {\bibfnamefont {T.}~\bibnamefont {Tanaka}},\ }\bibfield
  {title} {\bibinfo {title} {{Slowly Rotating Black Holes in Dynamical
  Chern-Simons Gravity: Deformation Quadratic in the Spin}},\ }\href
  {https://doi.org/10.1103/PhysRevD.86.044037} {\bibfield  {journal} {\bibinfo
  {journal} {Phys. Rev. D}\ }\textbf {\bibinfo {volume} {86}},\ \bibinfo
  {pages} {044037} (\bibinfo {year} {2012})},\ \bibinfo {note} {[Erratum:
  Phys.Rev.D 89, 049902 (2014)]},\ \Eprint {https://arxiv.org/abs/1206.6130}
  {arXiv:1206.6130 [gr-qc]} \BibitemShut {NoStop}%
\bibitem [{\citenamefont {Pani}\ and\ \citenamefont
  {Cardoso}(2009)}]{Pani:2009wy}%
  \BibitemOpen
  \bibfield  {author} {\bibinfo {author} {\bibfnamefont {P.}~\bibnamefont
  {Pani}}\ and\ \bibinfo {author} {\bibfnamefont {V.}~\bibnamefont {Cardoso}},\
  }\bibfield  {title} {\bibinfo {title} {{Are black holes in alternative
  theories serious astrophysical candidates? The Case for
  Einstein-Dilaton-Gauss-Bonnet black holes}},\ }\href
  {https://doi.org/10.1103/PhysRevD.79.084031} {\bibfield  {journal} {\bibinfo
  {journal} {Phys. Rev. D}\ }\textbf {\bibinfo {volume} {79}},\ \bibinfo
  {pages} {084031} (\bibinfo {year} {2009})},\ \Eprint
  {https://arxiv.org/abs/0902.1569} {arXiv:0902.1569 [gr-qc]} \BibitemShut
  {NoStop}%
\bibitem [{\citenamefont {Antoniou}\ \emph {et~al.}(2018)\citenamefont
  {Antoniou}, \citenamefont {Bakopoulos},\ and\ \citenamefont
  {Kanti}}]{Antoniou:2017hxj}%
  \BibitemOpen
  \bibfield  {author} {\bibinfo {author} {\bibfnamefont {G.}~\bibnamefont
  {Antoniou}}, \bibinfo {author} {\bibfnamefont {A.}~\bibnamefont
  {Bakopoulos}},\ and\ \bibinfo {author} {\bibfnamefont {P.}~\bibnamefont
  {Kanti}},\ }\bibfield  {title} {\bibinfo {title} {{Black-Hole Solutions with
  Scalar Hair in Einstein-Scalar-Gauss-Bonnet Theories}},\ }\href
  {https://doi.org/10.1103/PhysRevD.97.084037} {\bibfield  {journal} {\bibinfo
  {journal} {Phys. Rev. D}\ }\textbf {\bibinfo {volume} {97}},\ \bibinfo
  {pages} {084037} (\bibinfo {year} {2018})},\ \Eprint
  {https://arxiv.org/abs/1711.07431} {arXiv:1711.07431 [hep-th]} \BibitemShut
  {NoStop}%
\bibitem [{\citenamefont {Kinnersley}(1969)}]{Kinnersley:1969zza}%
  \BibitemOpen
  \bibfield  {author} {\bibinfo {author} {\bibfnamefont {W.}~\bibnamefont
  {Kinnersley}},\ }\bibfield  {title} {\bibinfo {title} {{Type D Vacuum
  Metrics}},\ }\href {https://doi.org/10.1063/1.1664958} {\bibfield  {journal}
  {\bibinfo  {journal} {J. Math. Phys.}\ }\textbf {\bibinfo {volume} {10}},\
  \bibinfo {pages} {1195} (\bibinfo {year} {1969})}\BibitemShut {NoStop}%
\bibitem [{\citenamefont {Stephani}\ \emph {et~al.}(2003)\citenamefont
  {Stephani}, \citenamefont {Kramer}, \citenamefont {MacCallum}, \citenamefont
  {Hoenselaers},\ and\ \citenamefont {Herlt}}]{Stephani:2003tm}%
  \BibitemOpen
  \bibfield  {author} {\bibinfo {author} {\bibfnamefont {H.}~\bibnamefont
  {Stephani}}, \bibinfo {author} {\bibfnamefont {D.}~\bibnamefont {Kramer}},
  \bibinfo {author} {\bibfnamefont {M.~A.~H.}\ \bibnamefont {MacCallum}},
  \bibinfo {author} {\bibfnamefont {C.}~\bibnamefont {Hoenselaers}},\ and\
  \bibinfo {author} {\bibfnamefont {E.}~\bibnamefont {Herlt}},\ }\href
  {https://doi.org/10.1017/CBO9780511535185} {\emph {\bibinfo {title} {{Exact
  solutions of Einstein's field equations}}}},\ Cambridge Monographs on
  Mathematical Physics\ (\bibinfo  {publisher} {Cambridge Univ. Press},\
  \bibinfo {address} {Cambridge},\ \bibinfo {year} {2003})\BibitemShut
  {NoStop}%
\bibitem [{\citenamefont {Chandrasekhar}(1984)}]{Chandrasekhar:1984siy}%
  \BibitemOpen
  \bibfield  {author} {\bibinfo {author} {\bibfnamefont {S.}~\bibnamefont
  {Chandrasekhar}},\ }\bibfield  {title} {\bibinfo {title} {{The Mathematical
  Theory of Black Holes}},\ }\href
  {https://doi.org/10.1007/978-94-009-6469-3_2} {\bibfield  {journal} {\bibinfo
   {journal} {Fundam. Theor. Phys.}\ }\textbf {\bibinfo {volume} {9}},\
  \bibinfo {pages} {5} (\bibinfo {year} {1984})}\BibitemShut {NoStop}%
\bibitem [{\citenamefont {Dias}\ \emph {et~al.}(2015)\citenamefont {Dias},
  \citenamefont {Godazgar},\ and\ \citenamefont {Santos}}]{Dias:2015wqa}%
  \BibitemOpen
  \bibfield  {author} {\bibinfo {author} {\bibfnamefont {O.~J.~C.}\
  \bibnamefont {Dias}}, \bibinfo {author} {\bibfnamefont {M.}~\bibnamefont
  {Godazgar}},\ and\ \bibinfo {author} {\bibfnamefont {J.~E.}\ \bibnamefont
  {Santos}},\ }\bibfield  {title} {\bibinfo {title} {{Linear Mode Stability of
  the Kerr-Newman Black Hole and Its Quasinormal Modes}},\ }\href
  {https://doi.org/10.1103/PhysRevLett.114.151101} {\bibfield  {journal}
  {\bibinfo  {journal} {Phys. Rev. Lett.}\ }\textbf {\bibinfo {volume} {114}},\
  \bibinfo {pages} {151101} (\bibinfo {year} {2015})},\ \Eprint
  {https://arxiv.org/abs/1501.04625} {arXiv:1501.04625 [gr-qc]} \BibitemShut
  {NoStop}%
\bibitem [{\citenamefont {Chitre}(1976)}]{Chitre:1976bb}%
  \BibitemOpen
  \bibfield  {author} {\bibinfo {author} {\bibfnamefont {D.~M.}\ \bibnamefont
  {Chitre}},\ }\bibfield  {title} {\bibinfo {title} {{Perturbations of Charged
  Black Holes}},\ }\href {https://doi.org/10.1103/PhysRevD.13.2713} {\bibfield
  {journal} {\bibinfo  {journal} {Phys. Rev. D}\ }\textbf {\bibinfo {volume}
  {13}},\ \bibinfo {pages} {2713} (\bibinfo {year} {1976})}\BibitemShut
  {NoStop}%
\bibitem [{\citenamefont {Leaver}(1986)}]{Leaver:1986gd}%
  \BibitemOpen
  \bibfield  {author} {\bibinfo {author} {\bibfnamefont {E.~W.}\ \bibnamefont
  {Leaver}},\ }\bibfield  {title} {\bibinfo {title} {{Spectral decomposition of
  the perturbation response of the Schwarzschild geometry}},\ }\href
  {https://doi.org/10.1103/PhysRevD.34.384} {\bibfield  {journal} {\bibinfo
  {journal} {Phys. Rev. D}\ }\textbf {\bibinfo {volume} {34}},\ \bibinfo
  {pages} {384} (\bibinfo {year} {1986})}\BibitemShut {NoStop}%
\bibitem [{\citenamefont {Leaver}(1985)}]{Leaver:1985ax}%
  \BibitemOpen
  \bibfield  {author} {\bibinfo {author} {\bibfnamefont {E.~W.}\ \bibnamefont
  {Leaver}},\ }\bibfield  {title} {\bibinfo {title} {{An Analytic
  representation for the quasi normal modes of Kerr black holes}},\ }\href
  {https://doi.org/10.1098/rspa.1985.0119} {\bibfield  {journal} {\bibinfo
  {journal} {Proc. Roy. Soc. Lond. A}\ }\textbf {\bibinfo {volume} {402}},\
  \bibinfo {pages} {285} (\bibinfo {year} {1985})}\BibitemShut {NoStop}%
\bibitem [{\citenamefont {Dolan}\ \emph {et~al.}(2021)\citenamefont {Dolan},
  \citenamefont {Kavanagh},\ and\ \citenamefont {Wardell}}]{Dolan:2021ijg}%
  \BibitemOpen
  \bibfield  {author} {\bibinfo {author} {\bibfnamefont {S.~R.}\ \bibnamefont
  {Dolan}}, \bibinfo {author} {\bibfnamefont {C.}~\bibnamefont {Kavanagh}},\
  and\ \bibinfo {author} {\bibfnamefont {B.}~\bibnamefont {Wardell}},\
  }\bibfield  {title} {\bibinfo {title} {{Gravitational perturbations of
  rotating black holes in Lorenz gauge}},\ }\href@noop {} {\  (\bibinfo {year}
  {2021})},\ \Eprint {https://arxiv.org/abs/2108.06344} {arXiv:2108.06344
  [gr-qc]} \BibitemShut {NoStop}%
\bibitem [{\citenamefont {Wald}(1973)}]{Wald:1973a}%
  \BibitemOpen
  \bibfield  {author} {\bibinfo {author} {\bibfnamefont {R.~M.}\ \bibnamefont
  {Wald}},\ }\bibfield  {title} {\bibinfo {title} {On perturbations of a kerr
  black hole},\ }\href {https://doi.org/10.1063/1.1666203} {\bibfield
  {journal} {\bibinfo  {journal} {Journal of Mathematical Physics}\ }\textbf
  {\bibinfo {volume} {14}},\ \bibinfo {pages} {1453} (\bibinfo {year}
  {1973})},\ \Eprint {https://arxiv.org/abs/https://doi.org/10.1063/1.1666203}
  {https://doi.org/10.1063/1.1666203} \BibitemShut {NoStop}%
\bibitem [{\citenamefont {Geroch}\ \emph {et~al.}(1973)\citenamefont {Geroch},
  \citenamefont {Held},\ and\ \citenamefont {Penrose}}]{Geroch:1973am}%
  \BibitemOpen
  \bibfield  {author} {\bibinfo {author} {\bibfnamefont {R.~P.}\ \bibnamefont
  {Geroch}}, \bibinfo {author} {\bibfnamefont {A.}~\bibnamefont {Held}},\ and\
  \bibinfo {author} {\bibfnamefont {R.}~\bibnamefont {Penrose}},\ }\bibfield
  {title} {\bibinfo {title} {{A space-time calculus based on pairs of null
  directions}},\ }\href {https://doi.org/10.1063/1.1666410} {\bibfield
  {journal} {\bibinfo  {journal} {J. Math. Phys.}\ }\textbf {\bibinfo {volume}
  {14}},\ \bibinfo {pages} {874} (\bibinfo {year} {1973})}\BibitemShut
  {NoStop}%
\bibitem [{\citenamefont {Toomani}\ \emph {et~al.}(2022)\citenamefont
  {Toomani}, \citenamefont {Zimmerman}, \citenamefont {Spiers}, \citenamefont
  {Hollands}, \citenamefont {Pound},\ and\ \citenamefont
  {Green}}]{Toomani:2021jlo}%
  \BibitemOpen
  \bibfield  {author} {\bibinfo {author} {\bibfnamefont {V.}~\bibnamefont
  {Toomani}}, \bibinfo {author} {\bibfnamefont {P.}~\bibnamefont {Zimmerman}},
  \bibinfo {author} {\bibfnamefont {A.}~\bibnamefont {Spiers}}, \bibinfo
  {author} {\bibfnamefont {S.}~\bibnamefont {Hollands}}, \bibinfo {author}
  {\bibfnamefont {A.}~\bibnamefont {Pound}},\ and\ \bibinfo {author}
  {\bibfnamefont {S.~R.}\ \bibnamefont {Green}},\ }\bibfield  {title} {\bibinfo
  {title} {{New metric reconstruction scheme for gravitational self-force
  calculations}},\ }\href {https://doi.org/10.1088/1361-6382/ac37a5} {\bibfield
   {journal} {\bibinfo  {journal} {Class. Quant. Grav.}\ }\textbf {\bibinfo
  {volume} {39}},\ \bibinfo {pages} {015019} (\bibinfo {year} {2022})},\
  \Eprint {https://arxiv.org/abs/2108.04273} {arXiv:2108.04273 [gr-qc]}
  \BibitemShut {NoStop}%
\bibitem [{\citenamefont {Nichols}\ \emph {et~al.}(2012)\citenamefont
  {Nichols}, \citenamefont {Zimmerman}, \citenamefont {Chen}, \citenamefont
  {Lovelace}, \citenamefont {Matthews}, \citenamefont {Owen}, \citenamefont
  {Zhang},\ and\ \citenamefont {Thorne}}]{Nichols:2012jn}%
  \BibitemOpen
  \bibfield  {author} {\bibinfo {author} {\bibfnamefont {D.~A.}\ \bibnamefont
  {Nichols}}, \bibinfo {author} {\bibfnamefont {A.}~\bibnamefont {Zimmerman}},
  \bibinfo {author} {\bibfnamefont {Y.}~\bibnamefont {Chen}}, \bibinfo {author}
  {\bibfnamefont {G.}~\bibnamefont {Lovelace}}, \bibinfo {author}
  {\bibfnamefont {K.~D.}\ \bibnamefont {Matthews}}, \bibinfo {author}
  {\bibfnamefont {R.}~\bibnamefont {Owen}}, \bibinfo {author} {\bibfnamefont
  {F.}~\bibnamefont {Zhang}},\ and\ \bibinfo {author} {\bibfnamefont {K.~S.}\
  \bibnamefont {Thorne}},\ }\bibfield  {title} {\bibinfo {title} {{Visualizing
  Spacetime Curvature via Frame-Drag Vortexes and Tidal Tendexes III.
  Quasinormal Pulsations of Schwarzschild and Kerr Black Holes}},\ }\href
  {https://doi.org/10.1103/PhysRevD.86.104028} {\bibfield  {journal} {\bibinfo
  {journal} {Phys. Rev. D}\ }\textbf {\bibinfo {volume} {86}},\ \bibinfo
  {pages} {104028} (\bibinfo {year} {2012})},\ \Eprint
  {https://arxiv.org/abs/1208.3038} {arXiv:1208.3038 [gr-qc]} \BibitemShut
  {NoStop}%
\bibitem [{\citenamefont {Cook}\ and\ \citenamefont
  {Zalutskiy}(2014)}]{Cook:2014cta}%
  \BibitemOpen
  \bibfield  {author} {\bibinfo {author} {\bibfnamefont {G.~B.}\ \bibnamefont
  {Cook}}\ and\ \bibinfo {author} {\bibfnamefont {M.}~\bibnamefont
  {Zalutskiy}},\ }\bibfield  {title} {\bibinfo {title} {{Gravitational
  perturbations of the Kerr geometry: High-accuracy study}},\ }\href
  {https://doi.org/10.1103/PhysRevD.90.124021} {\bibfield  {journal} {\bibinfo
  {journal} {Phys. Rev. D}\ }\textbf {\bibinfo {volume} {90}},\ \bibinfo
  {pages} {124021} (\bibinfo {year} {2014})},\ \Eprint
  {https://arxiv.org/abs/1410.7698} {arXiv:1410.7698 [gr-qc]} \BibitemShut
  {NoStop}%
\bibitem [{\citenamefont {Zenginoglu}(2011)}]{Zenginoglu:2011jz}%
  \BibitemOpen
  \bibfield  {author} {\bibinfo {author} {\bibfnamefont {A.}~\bibnamefont
  {Zenginoglu}},\ }\bibfield  {title} {\bibinfo {title} {{A Geometric framework
  for black hole perturbations}},\ }\href
  {https://doi.org/10.1103/PhysRevD.83.127502} {\bibfield  {journal} {\bibinfo
  {journal} {Phys. Rev. D}\ }\textbf {\bibinfo {volume} {83}},\ \bibinfo
  {pages} {127502} (\bibinfo {year} {2011})},\ \Eprint
  {https://arxiv.org/abs/1102.2451} {arXiv:1102.2451 [gr-qc]} \BibitemShut
  {NoStop}%
\bibitem [{\citenamefont {Panosso~Macedo}(2020)}]{PanossoMacedo:2019npm}%
  \BibitemOpen
  \bibfield  {author} {\bibinfo {author} {\bibfnamefont {R.}~\bibnamefont
  {Panosso~Macedo}},\ }\bibfield  {title} {\bibinfo {title} {{Hyperboloidal
  framework for the Kerr spacetime}},\ }\href
  {https://doi.org/10.1088/1361-6382/ab6e3e} {\bibfield  {journal} {\bibinfo
  {journal} {Class. Quant. Grav.}\ }\textbf {\bibinfo {volume} {37}},\ \bibinfo
  {pages} {065019} (\bibinfo {year} {2020})},\ \Eprint
  {https://arxiv.org/abs/1910.13452} {arXiv:1910.13452 [gr-qc]} \BibitemShut
  {NoStop}%
\bibitem [{\citenamefont {Ripley}(2022)}]{Ripley:2022ypi}%
  \BibitemOpen
  \bibfield  {author} {\bibinfo {author} {\bibfnamefont {J.~L.}\ \bibnamefont
  {Ripley}},\ }\bibfield  {title} {\bibinfo {title} {{Computing the quasinormal
  modes and eigenfunctions for the Teukolsky equation using horizon
  penetrating, hyperboloidally compactified coordinates}},\ }\href@noop {} {\
  (\bibinfo {year} {2022})},\ \Eprint {https://arxiv.org/abs/2202.03837}
  {arXiv:2202.03837 [gr-qc]} \BibitemShut {NoStop}%
\bibitem [{\citenamefont {Cardoso}\ \emph {et~al.}(2019)\citenamefont
  {Cardoso}, \citenamefont {Kimura}, \citenamefont {Maselli}, \citenamefont
  {Berti}, \citenamefont {Macedo},\ and\ \citenamefont
  {McManus}}]{Cardoso:2019mqo}%
  \BibitemOpen
  \bibfield  {author} {\bibinfo {author} {\bibfnamefont {V.}~\bibnamefont
  {Cardoso}}, \bibinfo {author} {\bibfnamefont {M.}~\bibnamefont {Kimura}},
  \bibinfo {author} {\bibfnamefont {A.}~\bibnamefont {Maselli}}, \bibinfo
  {author} {\bibfnamefont {E.}~\bibnamefont {Berti}}, \bibinfo {author}
  {\bibfnamefont {C.~F.~B.}\ \bibnamefont {Macedo}},\ and\ \bibinfo {author}
  {\bibfnamefont {R.}~\bibnamefont {McManus}},\ }\bibfield  {title} {\bibinfo
  {title} {{Parametrized black hole quasinormal ringdown: Decoupled equations
  for nonrotating black holes}},\ }\href
  {https://doi.org/10.1103/PhysRevD.99.104077} {\bibfield  {journal} {\bibinfo
  {journal} {Phys. Rev. D}\ }\textbf {\bibinfo {volume} {99}},\ \bibinfo
  {pages} {104077} (\bibinfo {year} {2019})},\ \Eprint
  {https://arxiv.org/abs/1901.01265} {arXiv:1901.01265 [gr-qc]} \BibitemShut
  {NoStop}%
\bibitem [{\citenamefont {McManus}\ \emph {et~al.}(2019)\citenamefont
  {McManus}, \citenamefont {Berti}, \citenamefont {Macedo}, \citenamefont
  {Kimura}, \citenamefont {Maselli},\ and\ \citenamefont
  {Cardoso}}]{McManus:2019ulj}%
  \BibitemOpen
  \bibfield  {author} {\bibinfo {author} {\bibfnamefont {R.}~\bibnamefont
  {McManus}}, \bibinfo {author} {\bibfnamefont {E.}~\bibnamefont {Berti}},
  \bibinfo {author} {\bibfnamefont {C.~F.~B.}\ \bibnamefont {Macedo}}, \bibinfo
  {author} {\bibfnamefont {M.}~\bibnamefont {Kimura}}, \bibinfo {author}
  {\bibfnamefont {A.}~\bibnamefont {Maselli}},\ and\ \bibinfo {author}
  {\bibfnamefont {V.}~\bibnamefont {Cardoso}},\ }\bibfield  {title} {\bibinfo
  {title} {{Parametrized black hole quasinormal ringdown. II. Coupled equations
  and quadratic corrections for nonrotating black holes}},\ }\href
  {https://doi.org/10.1103/PhysRevD.100.044061} {\bibfield  {journal} {\bibinfo
   {journal} {Phys. Rev. D}\ }\textbf {\bibinfo {volume} {100}},\ \bibinfo
  {pages} {044061} (\bibinfo {year} {2019})},\ \Eprint
  {https://arxiv.org/abs/1906.05155} {arXiv:1906.05155 [gr-qc]} \BibitemShut
  {NoStop}%
\bibitem [{\citenamefont {V\"olkel}\ \emph {et~al.}(2022)\citenamefont
  {V\"olkel}, \citenamefont {Franchini},\ and\ \citenamefont
  {Barausse}}]{Volkel:2022aca}%
  \BibitemOpen
  \bibfield  {author} {\bibinfo {author} {\bibfnamefont {S.~H.}\ \bibnamefont
  {V\"olkel}}, \bibinfo {author} {\bibfnamefont {N.}~\bibnamefont
  {Franchini}},\ and\ \bibinfo {author} {\bibfnamefont {E.}~\bibnamefont
  {Barausse}},\ }\bibfield  {title} {\bibinfo {title} {{Theory-agnostic
  reconstruction of potential and couplings from quasinormal modes}},\ }\href
  {https://doi.org/10.1103/PhysRevD.105.084046} {\bibfield  {journal} {\bibinfo
   {journal} {Phys. Rev. D}\ }\textbf {\bibinfo {volume} {105}},\ \bibinfo
  {pages} {084046} (\bibinfo {year} {2022})},\ \Eprint
  {https://arxiv.org/abs/2202.08655} {arXiv:2202.08655 [gr-qc]} \BibitemShut
  {NoStop}%
\bibitem [{\citenamefont {Vigeland}\ \emph {et~al.}(2011)\citenamefont
  {Vigeland}, \citenamefont {Yunes},\ and\ \citenamefont
  {Stein}}]{Vigeland:2011ji}%
  \BibitemOpen
  \bibfield  {author} {\bibinfo {author} {\bibfnamefont {S.}~\bibnamefont
  {Vigeland}}, \bibinfo {author} {\bibfnamefont {N.}~\bibnamefont {Yunes}},\
  and\ \bibinfo {author} {\bibfnamefont {L.}~\bibnamefont {Stein}},\ }\bibfield
   {title} {\bibinfo {title} {{Bumpy Black Holes in Alternate Theories of
  Gravity}},\ }\href {https://doi.org/10.1103/PhysRevD.83.104027} {\bibfield
  {journal} {\bibinfo  {journal} {Phys. Rev. D}\ }\textbf {\bibinfo {volume}
  {83}},\ \bibinfo {pages} {104027} (\bibinfo {year} {2011})},\ \Eprint
  {https://arxiv.org/abs/1102.3706} {arXiv:1102.3706 [gr-qc]} \BibitemShut
  {NoStop}%
\bibitem [{\citenamefont {Misner}\ \emph {et~al.}(1973)\citenamefont {Misner},
  \citenamefont {Thorne},\ and\ \citenamefont {Wheeler}}]{Misner:1973prb}%
  \BibitemOpen
  \bibfield  {author} {\bibinfo {author} {\bibfnamefont {C.~W.}\ \bibnamefont
  {Misner}}, \bibinfo {author} {\bibfnamefont {K.~S.}\ \bibnamefont {Thorne}},\
  and\ \bibinfo {author} {\bibfnamefont {J.~A.}\ \bibnamefont {Wheeler}},\
  }\href@noop {} {\emph {\bibinfo {title} {{Gravitation}}}}\ (\bibinfo
  {publisher} {W. H. Freeman},\ \bibinfo {address} {San Francisco},\ \bibinfo
  {year} {1973})\BibitemShut {NoStop}%
\end{thebibliography}%

\end{document}